\documentclass[11pt, oneside]{article}   	
\usepackage{geometry}                		
\usepackage{graphicx}				

\usepackage{amssymb}
\usepackage{amsmath}
\usepackage[utf8]{inputenc}

\title{The Stable Marriage Problem: an Interdisciplinary Review\\
from the Physicist's Perspective}

\author{\centerline{Enrico Maria Fenoaltea*, Izat B. Baybusinov*,
Jianyang Zhao**, Lei Zhou**, Yi-Cheng Zhang*}\\
\\
*Department of Physics, University of Fribourg, 1700 Fribourg, Switzerland\\
\centerline{**Faculty of Computer and Software Engineering, Huaiyin Institute of Technology, Huaian, China}}

\usepackage[sorting=none,style=numeric-comp,backend=biber]{biblatex}
\begin{document}
\maketitle

    \begin{abstract}
    We present a fascinating model that has lately caught attention among physicists working in complexity related fields. Though it originated from mathematics and later from economics, the model is very enlightening in many aspects that we shall highlight in this review. It is called The Stable Marriage Problem (though the marriage metaphor can be generalized to many other contexts), and it consists of matching men and women, considering preference-lists where individuals express their preference over the members of the opposite gender. This problem appeared for the first time in 1962 in the seminal paper of Gale and Shapley and has aroused interest in many fields of science, including economics, game theory, computer science, etc. Recently it has also attracted many physicists who, using the powerful tools of statistical mechanics, have also approached it as an optimization problem. Here we present a complete overview of the Stable Marriage Problem emphasizing its multidisciplinary aspect, and reviewing the key results in the disciplines that it has influenced most. We focus, in particular, in the old and recent results achieved by physicists, finally introducing two new promising models inspired by the philosophy of the Stable Marriage Problem. Moreover, we present an innovative reinterpretation of the problem, useful to highlight the revolutionary role of information in the contemporary economy.
    \end{abstract}
    
    \maketitle
    
    \tableofcontents
    
    \section{Introduction}

\subsection{SMP Overview}

SMP concerns a system where two different sets of $N$ elements must be matched. In the literature it is often said that the two sets of elements are composed of $N$ men and $N$ women and all individuals must have a partner of the opposite sex. Both men and women have preference-lists in which each person expresses his/her preference over the members of the opposite gender. The final matching must be stable, which means that there are no man and woman who both prefer marriage with each other rather than stay with their current partner.\\
This problem was introduced in 1962 in the seminal paper of Gale and Shapley \cite{10.2307/2312726}, and has attracted researchers in several areas, including mathematics, economics, game theory, computer science, physics etc.\\

More importantly for our discussion, SMP can be studied as a combinatorial optimization problem where, instead of finding a stable solution, the problem consists of finding the solution that maximizes the \textit{global} happiness (the higher the ranking of your partner in your preference-list, the higher your \textit{personal} happiness). In this case, SMP is also called \textit{Random Matching Problem} \cite{mezard1987solution}.\\
To show why this problem has aroused the interest of many physicists let us consider the example of a similar problem introduced by Gaspard Monge already in 1781 \cite{monge1781memoire}. He wrote:\\
\\
"Let us suppose that we have a certain number of mines and the same number of deposits. We want to associate each mine to one deposit only (where the production of the considered mine will be transported and stored). How can we perform this matching in such a way that the total transport cost is minimum?"\\
\\
The cost of transportation from a mine and a deposit is given by a function that depends on their distance. Obviously, the greater the distance, the greater the cost. The goal is therefore to minimize the total distance, that is the sum of the individual distances between mine and deposit. This problem is usually called \textit{optimal transport problem} \cite{bogachev2012monge, villani2008optimal}. The location of the mines and deposits are assigned, so the problem is fixed in every detail. Although this problem seems simple from its formulation, it turns out to be very difficult to solve. If the number of mines is $ N $, then there are $ N! $ ways of matching those mines with the deposits. If $ N $ is large enough, it is evident that a \textit {brute force} approach is not feasible (at least with the computers available today). A solution to this problem was found only two centuries later, thanks to Kuhn's algorithm \cite{kuhn1955hungarian}. \\
Since the position of mines and deposits is fixed, in this formulation of the problem there are no randomness and no disorder. Once the matching that minimizes the cost is found, the problem is solved. \\
The problem, however, can be considered from another point of view: suppose that mines and deposits are two sets of random points, and we require to find the optimal matching so that a certain function is minimized (in our example the cost or distance function). The specific solution, in this case, is not interesting. What interests most is the \textit{average properties} of the optimal solution, in particular, the value of the minimum cost averaged over all the possible configurations of the distances. Furthermore, if we assume that the distances between the points (and therefore the costs) are not correlated, we find exactly the formulation of the Random Matching Problem, in fact, in its original version the preference lists of men and women are random and independent between each other (the version where the distances are correlated is called \textit{Euclidean Matching Problem} \cite{mezard1988euclidean}).\\
\\
Another example of a problem that can be treated similarly is the very famous \textit {Traveling Salesman Problem} (TSP). Its original formulation is as follows: given a set of cities, and known the distances between each pair of them, it is required to find the shortest distance route that a travelling salesman must follow to visit all the cities once and only once, returning to the city of departure \cite{junger1995traveling}. Also in this case, since the position of the cities is assigned, the problem is fixed in every detail. Hence, having found the path that minimizes the distance, the problem is solved. However, the TSP, unlike the random matching problem, is shown to be NP-complete \cite{christofides1976worst}, i.e. there are no efficient algorithms that solve the problem in a reasonable time. \\
Similarly to the previous problem, however, we can consider cities as a set of random points (with correlated distances) and now we have to minimize a cost function that depends on distances. As before, a specific solution to the problem is not important, but it is interesting to study the minimum value of the cost function \textit{averaged} over all the possible configurations that the distances between the points can assume.\\   
\\
At this point the physicists come into play: when, instead of a given instance of an optimization problem, all the possible configuration of that problem are considered, according to a suitable probability distribution, ideas, methods, and powerful mathematical tools that physicists have developed in statistical mechanics of systems with disorder and frustration can be applied and be very efficient. In particular, the cost function to be minimized is associated with the Hamiltonian of the system and, by introducing a fictitious temperature, we can find the minimum of the Hamiltonian using the typical techniques of statistical mechanics. \\ 
Indeed, we will see that SMP (and not only) can be solved by exploiting techniques from the theory of spin glasses such as the \textit{replica method}.\\
\\
Therefore SMP can be treated both as a \textit{game theory problem}, in which one looks for the stability of the system and it is a useful approach above all in economic models, and as an \textit{optimization problem}, in which one look for a globally optimal solution, and is a more familiar approach to problems encountered in statistical physics.\\
For these reasons, this problem has aroused interest in many fields of science. It is also characterized by its mathematical elegance and its particular adaptability to real-world models. Indeed, the SMP became so popular that in 2012 Lloyd S. Shapley and Alvin E. Roth won the Nobel Memorial Prize in Economic Sciences for their theoretical works on Stable Marriage Problem and the practice of market design \cite{henderson2012marriage}.\\
\\
SMP is particularly known for being suitable for describing most of the two-side markets, such as buyers and sellers, workers and firms, students and universities, doctors and hospitals etc. Of course, none of these two-side markets can be fully described by an idealized model like the classic version of SMP, but it can be modified ad hoc to satisfactorily describe the complexity of the real world. For example, as we shall see in detail, the general SMP model can be modified by making the realistic assumption that the agents of the system do not have complete information. We will devote an entire section to this topic, as the assumption of partial information will lead us to interesting conclusions on the two-side markets in the modern economy. In particular, it will lead us to introduce the new theory of \textit {information economy} in which the classical paradigms of the standard economy are upset. As we will see, the applications of the stable marriage problem are not reduced only to economic and physics systems, but also address biological or technological systems.\\

\subsection{Historical Backgrounds}

More than twenty years ago the Fribourg team, led by Yi-Cheng Zhang, began research activities around the topic of \textit{Stable Marriage Problem} (SMP). Fribourg group, generation after generation, made some key contributions like partial information, information economy, and, recently, it developed new models like Negotiation problem and Seating Problem (we shall cover all these topics in this review). \\

This interest starts thanks to Sergei Maslov, that was a close collaborator of the physicist Per Bak at Brookhaven National Lab, and now he is professor of physics and biology at Urbana Champaign. Each summer a small group of original thinkers used to gather at Swiss Alps, and they challenged each other to find the most beautiful ideas in sciences, wherever they may come. Sergei, who did beautiful work on Bak-Sneppen model \cite{paczuski1996avalanche,maslov1996infinite}, one day introduced the SMP by Gale-Shapley. Fribourg team was fascinated by this elegant, simple, yet not easy problems and its endless ramifications.\\
It is hard to imagine to get a PhD from Fribourg without knowing SMP!\\
\\
In light of all that we described here and in the previous paragraph, we can say that the beauty of the model, its practical utility and its interdisciplinary aspect are sufficient elements to write a review on this topic.  We will tell the story of the SMP, retracing with a pedagogical approach the studies that have allowed researchers to obtain the most interesting results, always keeping an eye on the possible future scenarios in which research on this model could lead.\\
Moreover, this review cannot come more timely. Indeed, new directions, as we already mentioned, are suddenly popping up again in \textit{information economy} where asymmetrical information is ubiquitous and many related research subjects need new tools and new methodology.\\
\\
To review a subject across many disciplines is never easy. Nevertheless, we tried our best to bring the original ideas up to date and to relate all these disciplines to each other whenever possible. We hope that more researchers in physics, especially the younger generation, beyond learning new tools, can share the excitement of this subject of complexity sciences profiting from the state of the art of this problem, and hopefully making breakthroughs.

\subsection{Outline of the Review}
The organization of this paper is structured as follows: in section 2 we will make a brief introduction to SMP in its most general form. In section 3 we will address the main work of physicists regarding SMP. In particular, we will use a thermodynamic formalism, familiar to physicists, to find the main properties of the problem both in the classical case and in its more interesting variants. \\
In section 4, after a brief introduction to computational complexity theory, we will deal with the main algorithms that made the history of SMP and some of its generalizations. Of particular importance are the analysis of the Gale-Shapley algorithm and the Hungarian algorithm. \\
in section 5 we will study the main "out of the box" applications of SMP, such as the study of the dynamics of microbe communities in biology. Also, we will mention some works complementary to what is dealt with in section 4, mainly carried out by mathematicians. \\
In section 6, we will study SMP in the environment in which it has become popular, that is, among economists. We will introduce some notions of game theory and then we will retrace the main applications that allowed Alvin Roth and Lloyd Shapley to win the Nobel Prize in 2012. Finally, we will study some extensions of the SMP that are important in the economic field. \\
In section 7 we will present a reinterpretation by the authors of this review, of the original SMP model. The fundamental concept will be the information present in the system. We will see that, if re-analyzed from this new point of view, SMP can have interesting implications for economic theories. In particular, we will show how this reinterpretation of the model is directly linked to the new information economy theory by Yi-Cheng Zhang. \\
Finally, In section 8, we will study some of the most recent research performed by the authors of this review related to the SMP. We will introduce two entirely new models inspired by the simplicity and elegance of the SMP, and we will show that they will lead to absolutely non-trivial results.
\newpage

\section{General Introduction to the Stable Marriage Problem}
In this section we will introduce the main notions of the SMP to lay the foundations for a more in-depth study. We will introduce general notation and terminology and show the core results of the SMP. In later sections, we will explore the details of each topic present in this section and more.

\subsection{The Model}
SMP  addresses the situation where two sets, $ X $ and $ Y $, with the same number of elements, must be arranged in pairs $ (m \in X, w \in Y) $ taking into account the opinions of each element with respect to those of the others. In general, in the literature, the two sets are associated with men and women, and couples are called marriages, hence the problem's name.\\
Each player has an ordered list of individuals of the opposite sex that reflects his/her preferences and that we will call \textit{preference-list}. In particular, the first item on a woman (man) list represents the man (woman) of her (his) dreams; on the contrary, the last element of the list of a woman (man) represents her (his) worst-case partner.\\
An instance of the SMP is defined by a set of $ N $ men and $ N $ women, and by the preference-lists of each individual.\\
The decisions of both sexes consist in choosing a partner. The cost $ x_ {i, j} = H_i (w_j) $ associated to man $ m_i $ if he marries the woman $ w_j $ is given by the ranking of $ w_j $ in the preference list of $ m_i $. Similarly we define the cost $ y_ {i, j} = F_i (m_j) $ associated to woman $ w_i $ married to the man $ m_j $. Of course, the lower the cost for an individual, the higher his/her happiness.\\
In the simplest version of the model, the preference lists are random and independent of each other, i.e. there is no correlation between the players' preference lists. \\
One state $M$ of the problem is the assignment of couples in which all elements are married to one and only one element of the opposite sex.

\subsection{Stable Solutions}
A problem state is \textit {unstable} if a man $ m_i $ and a woman $ w_j $ are not married to each other in that state, but would rather be married than stay with their current partners.\\
In real life, this would be enough for the divorce in the marriages to which $m_i$ and $w_j$ belong. In this way they can form a new marriage for themselves, both saving cost (i.e. gaining happiness).\\
So unstable couples lead to the reconfiguration of states if the players are considered acting rationally and selfishly. A \textit{stable} state will therefore be a state that does not contain \textit{any} unstable pair, as no player will find anything better to do than stay with his/her current partner. In other words, such a state will be stable to the individual actions of the players, and we will call that states "stable solutions".\\
\\
To fix ideas, we make an easy example with $ N = 3 $, i.e. three men and three women. Consider the preference-lists in the table:

\begin{center}
\begin{tabular}{|c|c|c|c|}
\hline
\multicolumn{4}{|c|}{}\\
\multicolumn{4}{|c|}{\textbf{\Large SMP Instance}}\\
\multicolumn{4}{|c|}{}\\
\hline
\multicolumn{1}{|c|}{\textbf{Man}} &\textbf{Man's preference-list} &\textbf{Woman} &\multicolumn{1}{c|}{\textbf{Woman's preference-list}}\\
\hline
$m_1$ &  $(w_1,w_2,w_3)$   &  $w_1$  & $(m_3,m_1,m_2)$     \\
$m_2$ &  $(w_1,w_3,w_2)$   &  $w_2$  & $(m_1,m_2,m_3)$     \\
$m_3$ &  $(w_2,w_3,w_1)$   &  $w_3$  & $(m_3,m_2,m_1)$      \\
\hline
\end{tabular}
\end{center}

The cost for each person is assigned in this way: if $m_1$ marries $w_1$ the cost will be equal to 1 since woman 1 is at the top of his list, i.e. $x_{1,1}=1$. Vice versa, if $w_1$ marries $m_1$ the associated cost will be equal to 2 since he occupies the second position in her preference-list, i.e. $y_{1,1}=2$. Similarly if $m_2$ marries $w_3$ the costs will be $x_{2,3}=2$ and $y_{3,2}=2$. And so on.\\
A matching $ M $ is a set of $ N $ pairs: $ M = \{(m_i, w_j); i,j = 1, ..., N \} $. There are $ N! $ Possible matchings in a system of size $ N $, so in our case there are $ 3! = 6 $ possible ways of pairing the three men with the three women. \\
The problem is to find a matching $ M = \{(m_i, w_i) \} $ that is stable, i.e. where there is no pair $ (m_i, w_j) $ such that $ y_{j,i} <y_{j,j} $ and $ x_{i,j} <x_{i,i} $.\\
In our example it is evident that the matching $ M = \{(m_1,w_1), (m_2,w_3), (m_3,w_2) \} $ is stable since there is no man $m_i$ and no woman $w_j$ to who they are not married, but that they would both prefer to marry rather than stay with their respective current partners $ w_p $ and $ m_q $. On the contrary, the matching $ M = \{(m_1,w_2), (m_2,w_1), (m_3,w_3) \} $ is not stable: in fact the pair $ (m_1,w_1) $ constitutes an unstable pair since both $ m_1 $ and $ w_1 $ would rather marry each other than stay with their current partners $ w_2 $ and $m_2 $. 

\begin{figure}
\centering
\includegraphics[width=0.3\textwidth,scale=0.3]{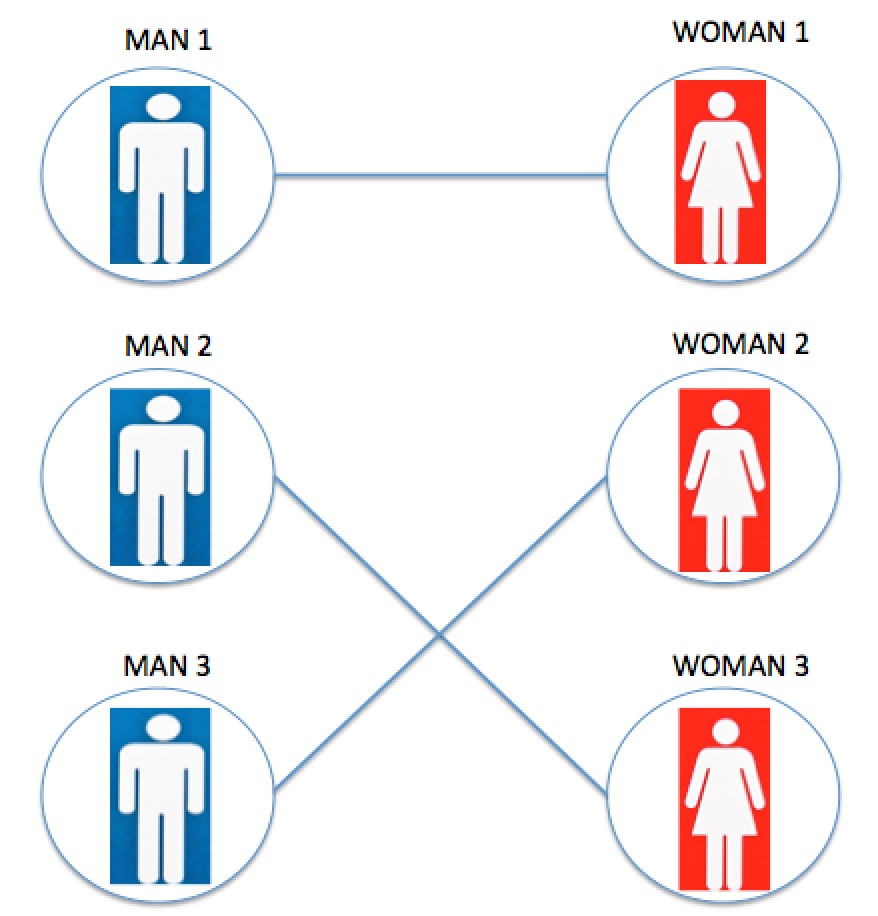}\quad\includegraphics[width=0.3\textwidth,scale=0.3]{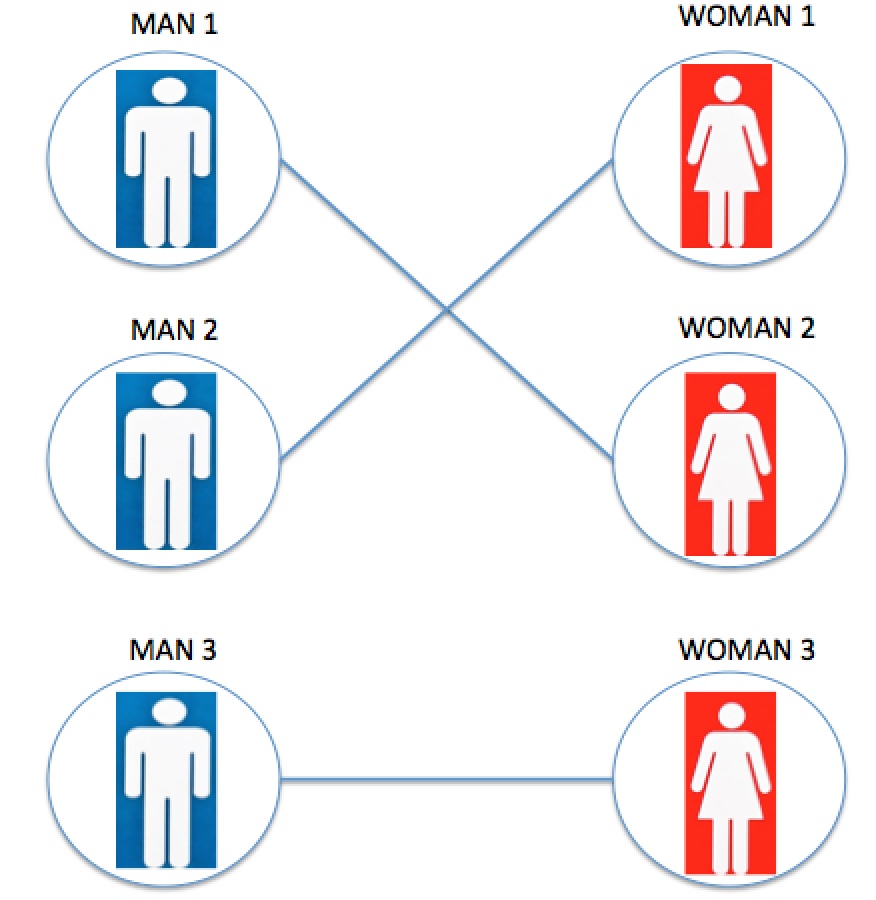}
\caption{\textbf{SMP solutions}: example of a stable (on the left) and unstable (on the right) solutions.}
\end{figure}

\subsection{Gale-Shapley Algorithm}
When $ N $ is large enough, it is not easy to find stable solutions (note that there may be more than one). The first, and most famous, algorithm capable of finding a particular stable solution was proposed by Gale and Shapley in 1962  \cite{10.2307/2312726}. It is called \textit{Gale-Shapley (GS) Algorithm}. \\
There are two versions of the GS algorithm: the \textit{men-oriented} and the \textit{women-oriented} versions. The mechanisms of both versions are equivalent then we will show only the men-oriented version for simplicity
\\
Given an instance of the problem, the algorithm follows these steps:
\begin{enumerate}
\item Begin with every man and woman being free.
\item While there exists a free  man $m$, he  proposes to his most preferred  woman whom he has not proposed to yet as follows:
\begin{itemize}
\item $m_i$ proposes to $w_j$, that is the best woman in his list to whom he has not yet proposed.
\item If $w_j$ is free: $m_i$ and $w_j$ get engaged.
\item If $w_j$ is currently engaged to another man $m_p$, and she prefers $m_p$ to $m_i$, i.e. $y_{j,p}<y_{j,i}$, then she rejects the proposal from $m_i$ and he remains free.
\item If $w_j$ is currently engaged to another man $m_p$, and she prefers $m_i$ to $m_p$, i.e. $y_{j,p}>y_{j,i}$, then she breaks off the engagement with $m_p$ and gets engaged to $m_i$. Consequently $m_p$ returns free.
\end{itemize}
\item The algorithm continues in this way until every man is married.
\end{enumerate}

The women-oriented version is equivalent, but the women are the proposers.\\
\\
It is possible to show that the algorithm ends in a stable solution and that the solution found by the men-oriented version is the best possible stable solution for men (we will show the proofs in section 4). It means that each man, in this particular solution, will receive the best partner among all the possible partners that he could receive in the other stable solutions. In this case, the solution is said to be \textit {men-optimal}. Similarly, the women-oriented version finds a \textit {women-optimal} solution (also stable). \\
Furthermore, the men-optimal solution is also the worst possible solution for women and, likewise, the women-optimal one is the worst possible for men. In summary: whoever takes the initiative gets the best result. \\
\\
Referring to the example of the previous paragraph, the GS (men-oriented) algorithm develops as follows: at the beginning all individuals are free. $ m_1 $ makes a proposal to $ w_1 $ (as she is at the top of his list). $ w_1 $ is free and therefore accepts. $ m_2 $ proposes to $ w_1 $, but she is already married to $ m_1 $ and $ y_ {1,1} <y_ {1,2} $, so $ m_2 $'s offer is rejected. At this point he proposes to $ w_3 $, which is free and therefore accepts. Finally, $ m_3 $ proposes to $ w_2 $ that is free and they get married. So the men-optimal matching is $ M = \{(m_1, w_1); (m_2, w_3); (m_3, w_2) \} $.\\
Similarly, the women-optimal solution is found to be $ M = \{(m_1, w_1); (m_2, w_2); (m_3, w_3) \} $.\\
Note that the total cost of men in the men-optimal solution is $ X = 1 + 2 + 2 = 5$ while that of women is $ Y = 2 + 3 + 2 = 7$, so men get a better result. however, the situation is reversed in the women-optimal solution where $ X = 6 $ and $ Y = 5 $. \\
\\
So men get the best result in the men-optimal solution and the worst in the women-optimal one, all other solutions are in between these two extremes. We will show in section 4 the algorithms to get all the possible stable states given an instance of the problem. \\
The particular case in which the men-optimal and the women-optimal solutions coincide corresponds to the situation in which there is only one stable state.

\subsection{Global Minimum Solution}
The purpose of SMP is, as we mentioned above, to find stable states in which stability affects the behaviour of individuals: in a stable state, there are no two individuals who are both better off getting divorced to form a new marriage between themselves. In other words, everyone acts selfishly thinking about their benefit and not the global one. This type of stability is formally called \textit {Nash equilibrium}, and we will learn more about this concept in the part dedicated to economics.\\ 
\\ 
Another type of interesting state, especially from a physicist's point of view, is the solution that maximizes the total benefit, or, equivalently, that minimizes the total cost. The total cost is defined as the sum of the costs of the individuals, i.e. $ X + Y $. \\
In this way, the concept of stability is no longer important, but SMP becomes an \textit {optimization problem}, in which one looks for the global minimum (or \textit {Ground State} as we will explain later). As we mentioned in the introduction, the version of the SMP where the goal is optimization is also called \textit {assignment problem}. \\ 
In section 3 we will show the analytical results related to the global minimum; in section 4 we will show the main algorithm that allows finding the solution that minimizes the total cost given one instance of the problem. In this paragraph, we will limit ourselves to showing some of the most significant characteristics of the global minimum through the case of the previous example. \\
 \\
 Referring to the SMP instance in the table in paragraph 2.2, it is easy to verify that the matching with the minimum cost is $ M = \{(m_1, w_2); (m_2, w_1); (m_3, w_3) \} $. \\
The total cost in this case is $ X + Y = 10$. In fact, $ X = 2 + 1 + 2 = 5$ and $ Y = 3 + 1 + 1=5 $. This solution is the one with the lowest possible cost. For example, the men/women-optimal solutions seen above have a total cost of $ 12 $ and $ 11 $ respectively. \\ 
\\ 
We also note that this \textit {solution} is not stable: there are two individuals, $ m_1 $ and $ w_1 $, who are not married to each other but who both would rather marry than stay with their current partners, respectively $ w_2 $ and $ m_2 $. In fact, in this matching, $ x_ {1,2}> x_ {1,1} $ and at the same time $ y_ {1,2}> y_ {1,1} $. \\ 
Thus, with this small example, we can state that, in general, \textit {the global minimum of SMP is not stable}. \\
At this point it is possible to understand better why, in stable solutions, we have said that individuals act selfishly: in the previous example, $ m_1 $ and $ w_1 $ would divorce with their respective partners, increasing their happiness, but decreasing overall happiness. ($ X + Y = 10 \to X + Y = 12 $). \\
 \\
Therefore, if everyone acted on their own, the system would end up in a stable state, but the total cost would not be the minimum. It happens because individuals have a local view of the system, i.e. they only know their preference list. To obtain the global minimum solution one needs a \textit {matchmaker} who has global knowledge of the system and who has an interest in minimizing the total cost. \\
In section 7 we will deepen in detail the reflections on the role of the matchmaker in society. 

\subsection{SMP Variants}
The version of SMP that we have covered in this section is the simplest one. There are many SMP extensions and, in this Review, we will try to give an overview of all its ramifications.\\
In particular, in the next section, in addition to analyzing in more detail the properties of stable solutions and the global minimum, we will study other particular solutions such as the optimal stable solution or the solution that minimizes inequalities between men and women. We will also study variants of the SMP where the number of men is different from the number of women, or where the preference-lists are correlated and incomplete.\\
In the following, we will study generalized versions of SMP in which the matching is not one-to-one (a man with a woman) but can be many-to-one or many-to-many (more men with more women). We will also study the extensions in which the matching is monopartite or multipartite, i.e. the is no distinction between men and women (monopartite case), or there are more than two categories of individuals (multipartite case, i.e. not only men-women but men-women-children or more). \\
Finally, we will show how to adapt SMP to real world, particularly in sections 5, 6 and 7.
\newpage

\section{Stable Marriage Problem and Physics}

In this section we shall show how SMP can be studied through statistical physics, obtaining non-trivial results both from a mathematical and an interpretation points of view. In the following pages, we shall show in detail how physicists,  with the help of the powerful mathematical tools developed in the statistical mechanics, obtained the most significant features of the SMP.

\subsection{Physics Interpretation of SMP}
As we saw in the previous section, one can study both stability and optimization in SMP. The latter has always been familiar to physicists: there are many different types of optimization problems one may encounter in physics. In these problems, it is often necessary to optimize some physical quantity such as distance, velocity, time, mass, acceleration, force, energy etc. In particular SMP can be studied as a \textit {combinatorial optimization problem}. It is known that equilibrium statistical mechanics and combinatorial optimization have common roots \cite{kumar2015mathematical}. For example, the understanding of relevant physical problems, such as three-dimensional lattice statistics or two-dimensional quantum statistical mechanics problems, are solving counting problems over non-planar lattices, that has purely combinatorial origin \cite{kumar2015mathematical}. The other example concerns phase transitions: they are phenomena which are not limited to physical systems but are typical of many combinatorial problems, like the percolation transition in random graphs \cite{fagnanimathematical}. \\
Combinatorial problems are usually written as Constraint Satisfaction Problems that, in general, concern finding a zero-energy ground state of an appropriate energy function and its analysis amounts at performing a zero temperature statistical physics study. \\
\\
So the connection between physics and SMP optimization is natural. On the other hand, the search for stable states of the system is a better-known problem in game theory and economics. Recently, however, the concept of stability (in the sense of Nash) inspired many physicists, in particular for those who dealt with complex systems and interdisciplinary physics. So various systems (physical, social, economic, biological etc.) have been studied in this context by physicists \cite{lage2006marriage}. \\
\\
Therefore, in the SMP,  where there are two classes of $ N $ agents (i.e. men and women), the main goals are the following:
\begin{itemize}
\item Finding the solution that maximizes the total happiness of the system: \textit{finding the global minimum solution (or Ground State)}.
\item Finding the states that is stable with respect to the decisions of individuals: \textit{finding the stable solutions}.
\end{itemize}
Both problems have algorithmic solutions in polynomial times and we will study these algorithms in detail in section 4. Furthermore, the properties of these solutions have been deeply studied in recent years and here we will analyze them in detail. \\
At this point, we can ask interesting questions about the connection between these two solutions: how much are the Stable solutions different from the global minimum? How stable is a solution to external perturbations? Or, how do real system achieve stable states? Regarding this last question, for example, Lage-Castellanos and Mulet in \cite{lage2006marriage} have proposed a microscopic dynamic that brings the system to a stationary state that can be studied analytically. The first two questions will be analyzed in more detail at the end this section, helping to clarify how optimally matched agents still tend to act selfishly. \\
From an interpretative point of view, the global minimum represents a model in which there is a matchmaker that forces the agents to match in the optimal solution; the stable solution instead is considered as the natural state in which society evolves if one assumes infinitely rational agents. \\
\\ 
So now it remains to analyze the physicists' approach to SMP through the most significant studies that have been carried out in recent years. This will be the main focus of this section.

\subsection{Standard SMP}
Let us recapitulate the main features of the SMP and set the notation. The classic problem is to match $N$ men and $N$ women so that the system is stable. In this case, the system has size $N$. Suppose that the men and women in the two sets have complete information (so that each one knows everyone). Based on this information each man expresses a preference-list of the desired women in descending order so that at the top of his list there is the woman of his dreams, while at the bottom there is the woman to marry only in the worst case in which all the other women have refused him. Women do the same thing. In the simplest case, the preference-lists are random and independent. Note that each individual must have a partner. We have already defined the concept of stability in section 2, but it is worth to repeat it. First, let us define the concept of "blocking pair": given a matching $ M $, a pair $ (m, w) $, where $ m $ is a man and $ w $ a woman, is a blocking pair if $ m $ and $ w $ are not partners in $ M $, but both would prefer to marry each other rather than stay with their current partners.\\
At this point, the definition of stability is as follows:  \textit{a matching $M$ is stable if it has no blocking pairs.}\\
\\
For simplicity, we assume that the happiness of an individual depends on the rank of his/her partner in his/her preference-list. In general, the rank is associated with a cost function: if the first choice is satisfied, then the cost will be 1 (the lowest possible); instead, if only the last choice is satisfied, then the cost will be $N$ (the highest possible). Sometimes it is convenient to normalize this values so that the cost is in the interval $\{0,1\}$ (the lower the cost, the better the rank of the partner).\\
The complexity of the problem arises from the fact that it can happen that two men (or women) put the same woman (or man) at the top of their list. In this way, conflicts arise, making the SMP a non-trivial problem.\\
\\
We denote by $y(w,m)$ the position of man $ m $ in the list of woman $ w $. Similarly, $x(m,w)$ denotes the position of the woman $ w $ in the list of the man $ m $. The rank in the preference-list can be associated with energy in statistical physics \cite{omero1997scaling}, for this reason, we will refer to $x$ and $ y $ as "energies". \\
In general, in statistical mechanics, energy is the quantity that must be minimized; in the case of the classic SMP, the goal is to find a stable solution that does not necessarily correspond to the solution with the lowest total cost. Indeed, each individual in the system tries to minimize his/her \textit{own} cost function without worrying about the total cost of the system. This situation is a typical situation of game theory in which one looks for equilibrium in the sense of "Nash" rather than a global equilibrium.\\
The analogy between energy and cost function will be clearer when we will study the ground state solution, that is not necessarily stable.\\

\subsubsection{Average Number of Stable Solutions}
We have seen in the previous section how the Gale-Shapley algorithm (GS) is very efficient in finding particular stable solutions. It can also be interpreted as a theorem (in fact, the literature often refer to it as GS theorem) \cite{gusfield1987three, dubins1981machiavelli, 10.2307/2312726}, and it turned out to be fundamental to answer the first and most significant question about the SMP: does stability exist? The GS theorem answers "yes" by guaranteeing the existence of at least one stable solution. \\
Since at least one stable matching exists, it becomes natural to ask what is the average number of stable solutions for a given $ N $. In this paragraph, we will analyze this question in detail by using physics methods such as the mean-field approximation.\\
\\ 
In the Knuth monograph \cite{donald1999art}, among the various open problems proposed, there was that of estimating the expectation value of the number of stable matchings for random and independent preference-lists. He indicated that the key to solving this problem should be found through the integral formula that a given matching is stable. Taking advantage of the symmetry, i.e. that each of the possible $ N! $ matchings has the same probability $ P $ of being stable, and assuming that the cost for each marriage is a random variable between 0 and 1, Knuth \cite{knuth1997stable} has shown that:
\begin{equation}
P=\int_{0}^{1} d^Nx\int_{0}^{1} d^Ny \prod_{i \ne j} (1-x_iy_j) \;, 
\end{equation}
where $ x_i = x (i, i) $ is the cost for the man $ m_i $ of marrying the woman $ w_i $, while $ y_j = y (j, j) $ is the cost for the woman $ w_j $ in marrying the man $ m_j $.\\
The formula (1) is intuitively justified with the following reasoning: a certain matching is unstable if there is a man $ m_i $ and a woman $ w_j $ who are not married to each other and whom mutual cost $ x (i, j) $ and $ y (j, i) $ are both lower than those of their existing marriages, $ x_i $ and $ y_j $, or $ x (i, i)> x (i, j) $ and $ y (j, j)> y (j, i) $. So the probability that the man $ m_i $ and the woman $ w_j $ prefer to stay with their current partners is $ p_ {ij} = 1-x_iy_j $ and consequently the probability that the whole system is stable is given by (1). \\
This integral can only be done exactly for very small $ N $. Boris Pittel, in his 1988 article \cite{pittel1989average}, derived with complicated probabilistic reasoning the following asymptotic formula valid for $N \to \infty$:
\begin{equation}
P_{Pittel} \approx \frac{logN}{e\Gamma(N)} \;,
\end{equation}
where $ \Gamma (N) $ is the gamma function of Euler for which $ \Gamma(N + 1) = N! $.\\
Subsequently, Dzierzawa and Omero \cite{dzierzawa2000statistics} showed that the formula derived from Pittel had large discrepancies with the numerical simulations on the average number of stable matchings. in the rest of this paragraph, we will follow their reasoning to find an asymptotic formula of $ P $ that does not have such discrepancies. \\
\\
Since the major contribution to the integral in (1) comes from regions where the products $ x_iy_j $ are small, it is justifiable to replace the term $ 1-x_iy_j $ with $ e^ {- x_iy_j} $. Defining the total energy (cost) of men and the total energy (cost) of women as:
\[ X= \sum_{i=1}^N x_i \;, \]
\[ Y= \sum_{j=1}^N y_j \;, \] 
and neglecting the constraint $ i \ne j $, we get the approximation:
\begin{equation}
P=\int_{0}^{1} d^Nx\int_{0}^{1} d^Ny \, e^{-XY} \;.
\end{equation}
From a physical point of view, equation (3) can also be seen as the sum of a partition of two species of particles confined in the interval $\{0,1\}$ whose interactions are the product of the coordinates of their centres of mass  \cite{dzierzawa2000statistics}.\\
At this point, it is convenient to make the transition to the coordinates $ X $ and $ Y $, but to do this it is necessary to know their probability distributions $\rho(X)$ and $\rho(Y)$ for small $X$ and $Y$. The probability distribution $ \rho (X) $ is formally given by
\begin{equation}
\rho(X)=\int_{0}^{1} d^Nx \, \delta(X-\sum_{i=1}^N x_i) \;.
\end{equation}
According to the central limit theorem it converges to a normal distribution with mean $ N / 2 $ and variance $ N / 12 $. However, this is only true in the central part of the distribution while the distribution tails are very different. It therefore becomes useful to introduce new coordinates $ u_i $ such that $ x_i = u_i-u_ {i-1} $ for $ i = 1, .., N $ e with $ u_0 = 0 $ and $ u_N = X $. Neglecting the constraint $ x_i <1 $ for every $ i $, we can write
\[ \rho(X)=\frac{1}{\Gamma(N)} \int_{0}^{X} du_1...\int_{0}^{X} du_{N-1}=\frac{X^{N-1}}{\Gamma(N)} \;, \]
where the Euler gamma function is because the variables $ u_i $ must be placed in ascending order so that it holds $x_i \ge 0$.\\
To take into account the constraint $ x_i <1 $ just note that the $ u_i $ are random variables distributed in the interval $ (0, X) $ and therefore we obtain a poisson distribution for the $ x_i $:
\[p(x)= \frac{N}{X} e^{-Nx/X} \;. \]
Then the probability that all $ x_i $ are less than 1 is:
\[ p(x_i<1, i=1,..,N)=\left(\int_{0}^{1} dx \, p(x)\right)^N= (1-e^{-N/X})^N \;. \]
So putting it all together we have:
\begin{equation}
\rho(X)=\frac{X^{N-1}}{\Gamma(N)}(1-e^{-N/X})^N \;.
\end{equation}
Equation (5) is valid only for $ X <N / logN $ because of the term $(1-e^{-N/X})^N$ and for reasons that will be clearer in the next paragraph when we talk about total average energy.\\

\begin{figure}[!h]
\begin{center}
\includegraphics[width=0.7\textwidth,scale=0.7]{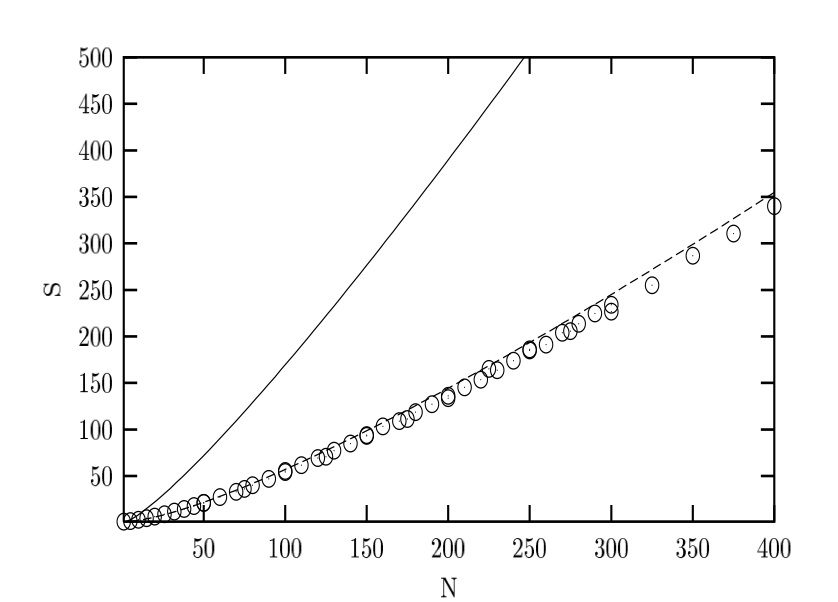}
\caption{\textbf{Average number of stable matchings}: each data point represents an average over 200 random realizations of the preference lists. The dashed curve is the result of equation (8) and the solid curve is the asymptotic formula derived by Pittel. Figure reprinted from \cite{dzierzawa2000statistics}, with permission from Elsevier}
\end{center}
\end{figure}

At this point we can rewrite (3) as:
\begin{equation}
P=\int_{0}^{N} dX\int_{0}^{N} dY \rho(X) \rho(Y) e^{-XY}  \;. 
\end{equation}
Inserting the equation (5) in (6) and doing the change of variable $ t = XY $ we obtain:
\begin{equation}
P=\frac{1}{\Gamma(N)} (logN-2log(logN)) \;,
\end{equation}
valid for $ logN <X <N / logN $. \\
\\
Now, since there are $ N! = \Gamma(N + 1) $ possible matchings, the average number of stable matching S will be equal to $ P \Gamma (N + 1) $, namely:
\begin{equation}
S \approx \frac{N}{e}(logN-2log(logN)) \;.
\end{equation}
Equation (2) and (8) are plotted in figure 2, and using the Gusfiel and Irving algorithm \cite{gusfield1987three, krishnamoorthy1991stable} for the determination of all stable marriages, the numerical simulations are also plotted: remarkably, equation (8) reproduces empirical results much better than equation (2).

\subsubsection{Energy of the Agents: a Mean Field Approximation}
Here we will analyze in more detail the statistical properties of the agents' energy following, in particular, the reasoning in \cite{omero1997scaling, dzierzawa2000statistics, laureti2003matching}. We will calculate the statistical properties of the agents' energies in the two extreme men/women optimal solutions, and then generalize the results to any stable solution, calculating the correlation between the energy of men and women.
\subsubsection*{Gale-Shapley Energies of Men and Women}
We already showed that the men-oriented GS algorithm provides an optimal stable solution for men, that is, for proposers. It is very interesting that although men generally receive more rejections and only one positive response, they achieve a much better result than women, in terms of total energy. On the other hand, women have the privilege of rejecting all their suitors except to those who like them the most, but even with this dynamic they reach the worst possible stable solution. The lesson is that people who take the initiative (men in this case) are rewarded. In this paragraph we will quantify this statement by calculating the average energy of men and women in the GS dynamic. \\
\\
Let us imagine that, as before, the cost of each marriage is a random variable between 0 and 1. First of all, we count the average number of proposals that a man has to make to find a definitive partner. The proposals define an intrinsic time in the algorithm \cite{omero1997scaling}. Since we are considering independent and random preference-lists, if at time $ t_k $ the $ k^{th} $ woman gets married, then the probability that the next proposal is made to one of the $ N-k $ free women is $ 1-k/N $. On average, men need $ N / k = <t_{k + 1} -t_k> $ proposals to marry one more woman (we changed the variable to $N+1-k \to k$). If we call $ r $ the average number of proposals of each man in the men-oriented GS algorithm, then the total number of proposals is equal to $ Nr = t_N $ and therefore holds:
\begin{equation}
r=\sum_{k=1}^N \frac{1}{k} \approx log(N)+C \;,
\end{equation}
where $ C = 0.5772 ... $ is the Euler constant and the corrections are of the order $ O ((log (N)) ^ 2 / N) $. \\
\\
At the same time, each woman receives on average $ r $ proposals. By accepting only the best offer, which is the smallest of $ r $ random numbers between 0 and 1, a woman gets $ y = 1 / (r + 1) $ as her best value. However, we must keep in mind that the $ r $ number of proposals that a woman receives is not fixed but is distributed according to a binomial:
$$ b(r)={R \choose r} p^{r} (1-p)^{R-r} \;, $$
where $ R $ is the total number of proposals that men make during the execution of the GS algorithm and $ p = 1 / N $ is the probability that an offer is made to a particular woman. Hence, averaging we get:
\begin{equation} 
<y>=b(r) \frac{1}{r+1} \approx \frac{1}{r} (1-e^{-r}) \approx \frac{1}{r} \;.
\end{equation}
Now, after each proposal, a man's total energy grows on average by $ 1 / N $ because he has to scroll down one position on his preference list. So the total energy of men $ X $ is on average equal to the number of proposals $ r $ that each man makes:
\begin{equation}
X=Nr \frac{1}{N} \approx  log(N)+C \;,
\end{equation}
while the total energy for women $Y$ is: 
\begin{equation}
Y=N<y> \approx \frac{N}{log(N)+C} \;.
\end{equation}
In figure 3 are showed men's energy $X$ and women's energy $Y$ against $N$. Note that in the men-oriented GS the energy of women is always much larger than that of men. Putting together (11) and (12) we get the relation:
\begin{equation}
XY=N \;,
\end{equation}
for GS dynamics. We will see in the next paragraph that this relation is valid on average for each SMP solution. Furthermore, equations (11) and (12) agree with the cut-offs $ logN <X <N / logN $ performed to obtain the average number of stable solutions (8). \\
All this remains valid even if the roles are reversed, i.e. in the women-oriented version of the GS algorithm.

\begin{figure}[!h]
\begin{center}
\includegraphics[width=0.7\textwidth,scale=0.7]{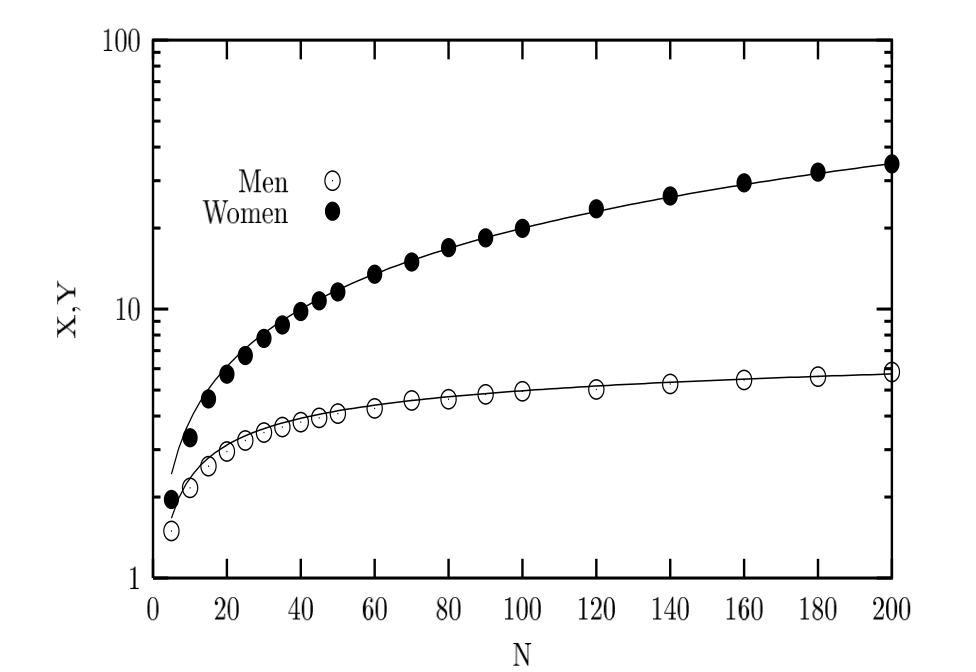}
\caption{\textbf{GS energies}: men-oriented GS energy of men and women as a function of the size of the system in a y-log scale. The energy corresponds to the ranking and therefore with values between 1 and $ N $. Figure reprinted from \cite{dzierzawa2000statistics}, with permission from Elsevier}
\end{center}
\end{figure}
\subsection*{Average Energy of Men and Women in the Mean-Field Approximation}
\begin{figure}[!h]
\begin{center}
\includegraphics[width=0.7\textwidth,scale=0.7]{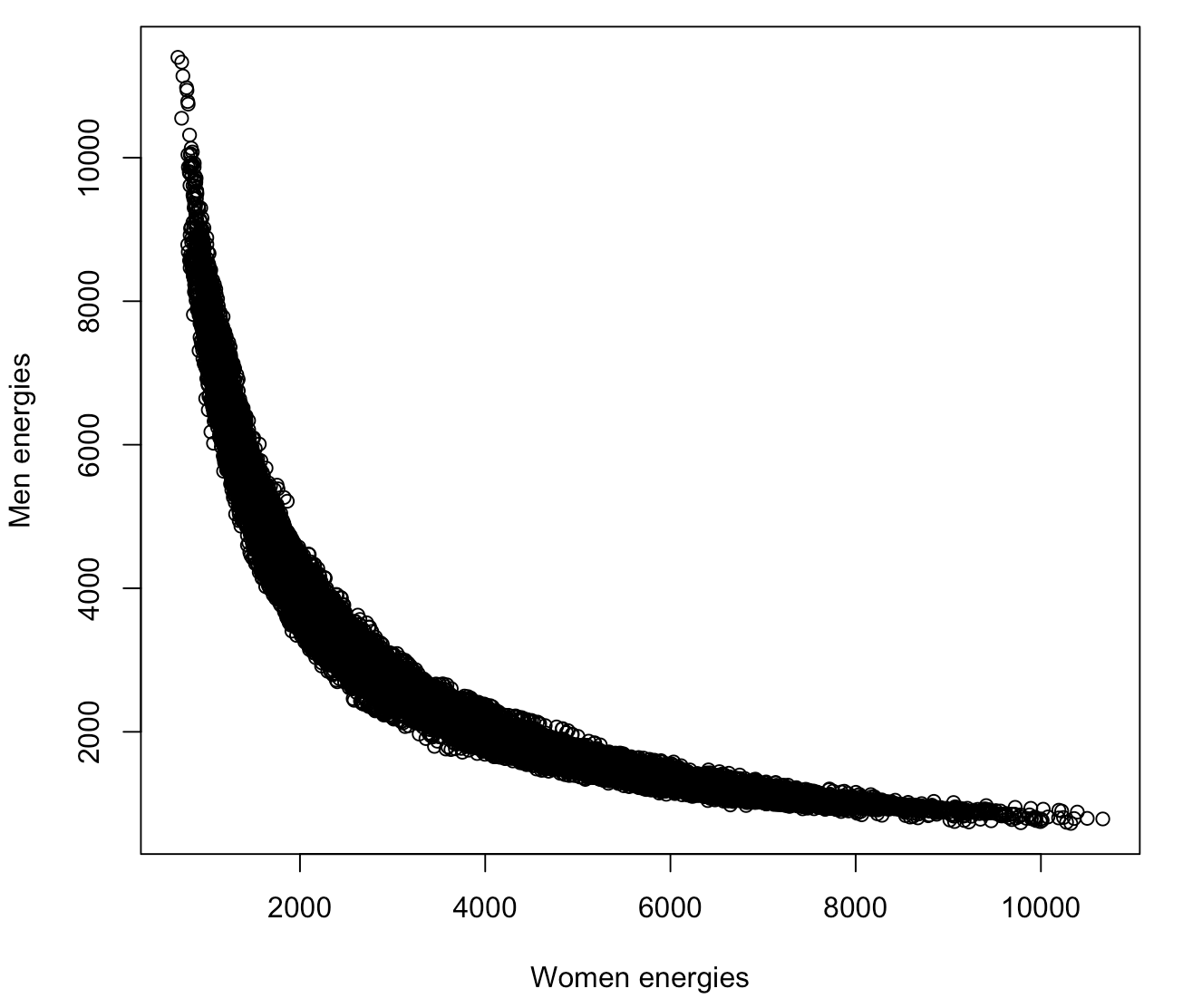}
\caption{\textbf{Men and Women energies}: each point in the figure represents the sum of the energies of women given the sum of the energies of men obtained in the same solution. The figure was performed by plotting all the solutions of a game with $ N = 200 $, for 200 simulations with randomly generated preference lists.}
\end{center}
\end{figure}

Equation (13) is also valid in general for any other stable state (not necessarily men/women-optimal). Following a mean-field approach, as in \cite{omero1997scaling}, the goal is evaluating the energy of men, given that of women in a stable state. We assume that there is a stable solution in which the woman $ w $ has energy $ y_w $. Without loss of generality, we consider the energy to correspond to the ranking (therefore with values between 1 and $ N $). To find such stable matching, Dzierzawa, Marsili, and Zhang consider a situation where women "know" that they can achieve such stable matching. So the best strategy for women is to reject all proposals from men lower than $ y_w $. At the same time, the best strategy for men remains that of the GS algorithm. \\
\\
Now consider a man $ m $ with preference list $ x (m, w) = w $ for $ w = 1, ..., N $. His proposal to the woman $ w^{th} $ will randomly fall between 1 and $ N $ relative to the ranking of the woman $ w $, and this proposal will be accepted with probability $ y_w / N $. Obviously if $ w $ refuses, then the man $ m $ will propose to the woman $ w + 1^{th} $. The probability that this man marries with his choice $ k^{th} $ is therefore equal to:
\begin{equation}
P_x(k)=(\frac{y_k}{N}) \prod_{w=1}^{k-1} (1-\frac{y_w}{N}) \;. 
\end{equation}
Under the assumption that $ y_w $ are independent and random variables with mean $ Y $, we can write:
\begin{equation}
P_x(k)=(\frac{Y}{N})(1-\frac{Y}{N})^{k-1} \;.
\end{equation}
Hence, the average energy of men can be calculated as $ X = \sum_ {k = 1}^{\infty} kP_x (k) = N / Y $. So we have:
\begin{equation}
XY=N \;,
\end{equation}
valid now for any stable solution (figure 4). In \cite{omero1997scaling} the authors showed that, based on the assumption of independent $y_w$, there is only a weak correlation and that therefore (16) is exact for $ N \to \infty $. Equation (16) is also confirmed by numerical simulations: figure 4 shows the trend of the total energy of women as a function of the total energy of men for each solution with $ N = 200$. As one can easily verify, it holds that $XY/200^2 \approx 200$.\\
In \cite{dzierzawa2000statistics}, Dzierzawa and Omero obtained the same results starting from the probability distribution (5).

\subsection{Modified Stable Marriage Problems}
The standard SMP has many interesting properties and has many application outlets. However, it remains an ideal model in which all the agents of the system have total information, the preference-lists are random and independent and the two sets have the same size $ N $. \\
All these conditions are not realistic and must be revisited to make them more suitable for models in the real world.\\
\\
In the following, we will deal with the possible variants of the SMP studied by physicists. In particular, we will find the stable solution that minimizes the difference between men and women's energy (to ensure fairness); we will analyze the presence of correlation in the preference's lists, that can be interpreted as the presence of competition or Euclidean distances; we will also deal with the more realistic case in which the number of men differs with that of women.\\
All these variants have been studied in various fields of science such as mathematics and computer science. In these pages, however, we shall show the results obtained by physicists who study these variants through familiar and easily interpretable concepts (such as mean-field approximation), often obtaining interesting results also from an analytical point of view.

\subsubsection{Equitable SMP} 
In 1989 Gusfield and Irving \cite{gusfield1989stable} proposed the "equitable stable marriage problem" (ESMP), which requires finding the stable SMP solution that minimizes "the distance" between the energy of men and of women to avoid discrimination between the two sides. Since for many real world applications ESMP is more appropriate than the classic SMP, it has attracted a lot of attention \cite{kato1993complexity, iwama2010approximation, gelain2010local, morge2011privacy, everaere2012casanova, everaere2013minimal}. \\
What we want to minimize in ESMP is the so-called \textit{sex-equality cost} $ D_{sec} $, defined in this way:
\begin{equation}
D_{sec}(M)=|\sum_{i=1}^{N}x_i-\sum_{i=1}^{N}y_i| \;,
\end{equation} 
where $ x_i $ and $ y_i $ are still the costs of marriages of man $ i $ and woman $ i $ in matching $ M $, respectively. \\
Unfortunately, as we will see in more detail in the next section, no algorithm solves the $ D_{sec} $ search in polynomial time: in \cite{kato1993complexity} Kato has shown that this problem is strongly NP-hard. However, in this paragraph, we will deal with a statistical analysis of the ESMP solution. We will rely on the results obtained in the previous paragraphs to calculate the energy of individuals in the system in the state of sex-equality. For this purpose, we will follow the arguments of  Zhang and  Laureti  in \cite{laureti2003matching}.

\subsubsection*{ESMP Individual Energy Distribution}
Since the average energy of men and women obeys the relationship (16), in the solution of the ESMP the average energy per person is:
\begin{equation}
<E>=\sqrt N \;,
\end{equation}
where the cost of the marriage corresponds to the ranking of the partner in their preference list.\\
\\
We now focus on the individual agents of the system. We shall show how to obtain the individual energy distribution $ p (x)_{ESMP} $ in the sex-fair solution. We can follow the same calculation of section 3.2.1 by adding the constraint that $ X = Y $, where now $ X $ and $ Y $ again represent the total energy of men and women respectively. We re-define $ x_i $ and $ y_j $ as the costs (between 0 and 1) of the marriages of man $ i $ and woman $ j $ respectively in the sex-fair solution. With the equality constraint, the probability $ P $ that a matching is stable can be rewritten as:
\begin{equation}
P_{ESMP}=\int_{0}^{1} d^Nx\int_{0}^{1} d^Ny \prod_{i \ne j} (1-x_iy_j) \delta(\sum_{i=1}^{N}x_i-\sum_{i=1}^{N}y_i)  \;, 
\end{equation}
and rewriting the probability distribution of $ X $ and $ Y $ as in (5), we can rewrite equation (19) as:
\begin{equation}
P_{ESMP}\approx \int_{0}^{1} d^Nx e^{-X^2} \rho_N(X)^2  \;,
\end{equation}
Where the subscript $ N $ in $ \rho_N(X) $ indicates that the probability distribution refers to a system of size $ N $.\\
Now, note that the probability that the individual $ i $ has energy $ x $ is equal to (probability that the total energy is $ X $) $\times$ (probability that the total energy without the individual $ i $ is $ X-x $), with the condition of stability and sex equity. Hence, the single energy distribution can be written as the average on the ensemble of all the stable and sex-fair solutions of $ \delta(x-x_N) $:
\begin{equation}
p(x)_{ESMP}=<\delta(x-x_N)>=\frac{1}{P}\int_{0}^{N} dX e^{-X^2} \rho_{N-1}(X-x) \rho_N(X) \;. 
\end{equation}
The integral in (21) can be calculated through the saddle point method around $ X = \sqrt N $. Therefore we can finally write (for $ N>>x$):
\begin{equation}
p(x)_{ESMP} \approx (N-1)\frac{(\sqrt N - x)^{N-2}}{N^{(N-1)/2}} \to \sqrt N e^{-x\sqrt N} \;.
\end{equation}
The simulation results are shown in figure 5.

\begin{figure}[!h]
\begin{center}
\includegraphics[width=0.9\textwidth,scale=0.9]{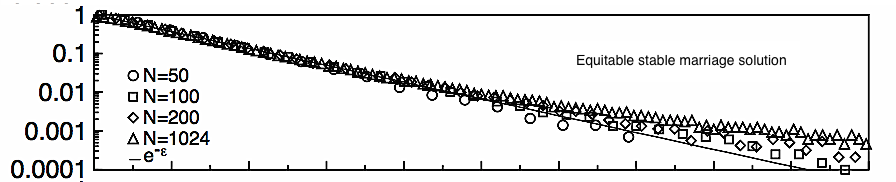}
\caption{\textbf{Individual's energy distribution}: probability distribution of average individual energies $\epsilon_i$, rescaled according to equation (22). Figure adapted from \cite{laureti2003matching}, with permission from Elsevier.}
\end{center}
\end{figure}

\subsubsection{SMP with Competition in the System}
The ideal model in which preference lists are completely independent is unrealistic. In most "two-side markets" there are correlations between the preferences of agents in the system \cite{abdulkadirouglu2005new, haruvy2006dynamics, mongell1991sorority}: think about the products-consumers market, where there are certainly better quality products that are more likely than others to be at the top of consumer preference-lists.\\
In general, in most real systems the rankings in the preference-lists are influenced by intrinsic properties such as beauty, quality, intelligence etc.\\
Continuing with the metaphor of marriage, Introducing the correlation between the preference lists means introducing the concept of beauty of individuals \cite{caldarelli2001beauty} and of competition between them \cite{kong2018competition}. Indeed, the more beautiful an individual is, the more likely he/she is to be at the top of the preference-lists of other individuals, so there will be more competition between agents to mate with him/her.

 \subsubsection*{Quantify the Competition in the System}
Consider the extreme case in which individuals build their preference-lists based solely on "objective beauty", that is, the case in which all men and women have identical preference-lists. In the GS dynamic (say men-oriented), all men propose to the woman they all prefer. This woman will only accept the man at the top of her list, while all the other $ N-1 $ men will be rejected.\\
In \cite{kong2018competition} the authors define the men who are rejected after the first step as "initial competitors". They use the number of initial competitors as a measure of competition in the system. So, when the competition is maximum, the number of initial competitors is $ N-1 $. When the lists are independent and random, one can reason in this way: the probability that a woman does not receive any proposal is equal to $(1-1/N)^N \to 1/e $ (for $N \to \infty$). Since the number of men who have been refused is equal to the number of women who have not received proposals, then the average number of initial competitors is equal to $ N/e $. Hence, in situations of intermediate competition, the number of initial competitors is in the range $ [N/e, N-1] $. \\
To take these intermediate situations into account, we say that each person $ i $ assigns a value $ V_ {ij} $ to all individuals $ j $ of the opposite sex:
\begin{equation}
V_{ij}=\omega F_j+(1-w)\eta_{ij} \;,
\end{equation}
where $ F_j \in [0,1] $ represents the intrinsic beauty of the individual $ j $, $ \eta_ {ij} \in [0,1] $ is a random number representing the subjectivity of personal preferences and $ \omega $ is a universal parameter for men and women that weighs the role of objective beauty.\\
The preference-lists rank the scores $V$ in descending order. Note that, when $ \omega = 0 $, the number of initial competitors is minimal and therefore the competition is also minimal; when $ \omega = 1 $, all the lists are equal and therefore the competition is maximum.

\subsubsection*{GS Energy of the Agents with Correlated Lists}
\begin{figure}[!h]
\begin{center}
\includegraphics[width=0.7\textwidth,scale=0.7]{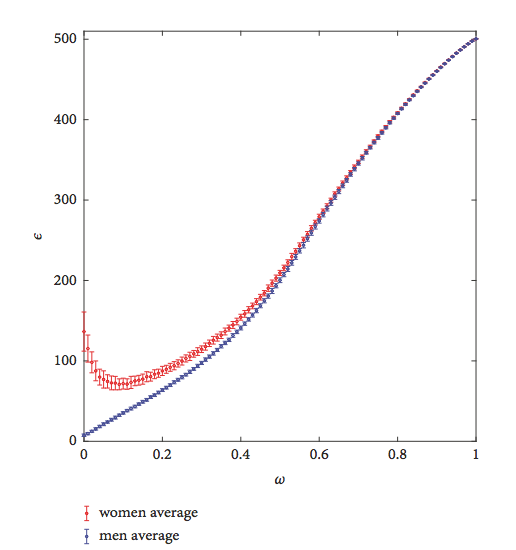}
\caption{\textbf{GS energies with correlated lists}: average GS energy, over 100 realization, as a function of $\omega$ for $N=1000$. Figure reprinted from \cite{kong2018competition}.}
\end{center}
\end{figure}

\begin{figure}[!h]
\begin{center}
\includegraphics[width=0.7\textwidth,scale=0.7]{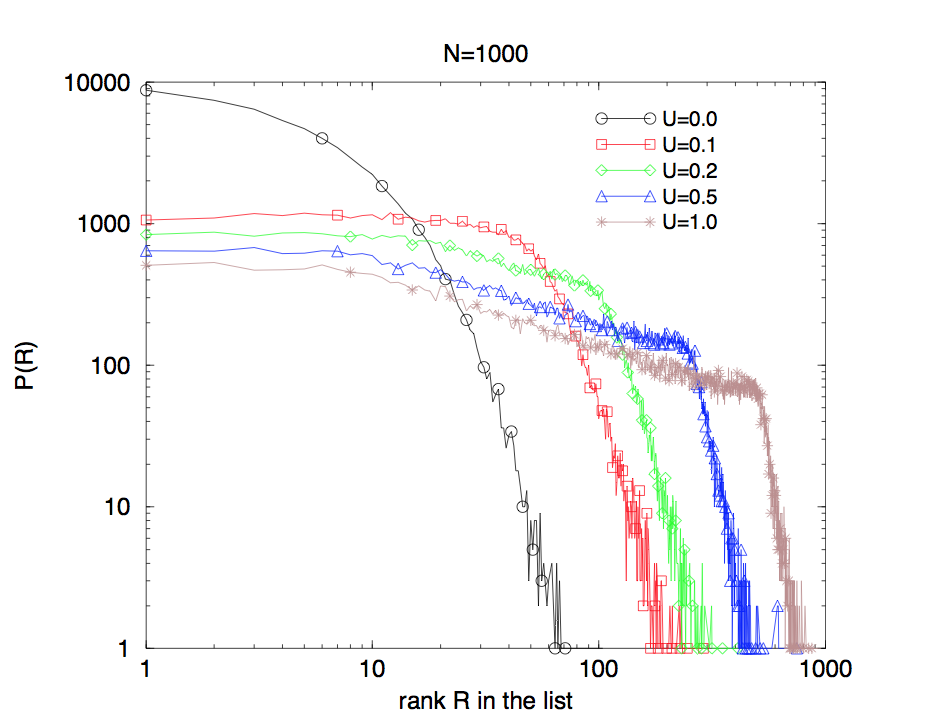}
\caption{\textbf{Ranking's probability distribution}: probability $P(R)$ that a man is married with a woman that is ranked $R$ in his preference list. The value $U$ is analogous to $\omega$ in the above model. Figure reprinted from \cite{caldarelli2001beauty}, with permission from Elsevier.}
\end{center}
\end{figure}
Building the preference-lists with the above process, it is possible to start the men-oriented GS dynamic and see what happens to the energies $\epsilon $ of men and women as a function of competition in the system, i.e. as a function of $ \omega $. The results are shown in figure 6:  the average energy of men (of the proposers) grows monotonically with the growth of the competition $\omega $. It happens because, with increasing competition between proposers, on average more and more proposals are needed to marry permanently.\\
Regarding women, however, the curve has a minimum. It happens because, by adding a little competition in the system, women receive more proposals than the case $ w = 0 $ and this allows them to make better choices.\\
Exceeded a certain critical threshold of competition, the same thing happens to men: high competition implies that many women will have to marry men at the bottom of their preference-lists, increasing total energy. \\
In general, at the point of minimum energy for women, the increase in men's energy is lower than the decrease in energy for women, therefore the total energy of the system also has a minimum.\\
This is a significant result because it means that adding the right amount of competition can increase the overall happiness of the system. \\
Another interesting result is that the gap between men's energy and women's energy decreases with increasing competition.\\
\\
Regarding the energy of individuals, the beauty-energy relationship was analyzed again in \cite{kong2018competition}. In particular, they studied the average energy as a function of $ \omega $ for individuals with different values of $ F $. Not surprisingly, they found that men's energy grows monotonically for any $ F $, obviously the bigger the $ F $ the lower the energy.\\
All of this is also true for women. However, when $ F $ is large enough, the average energy of individuals has a minimum at a certain critical amount of competition.\\
\\
In \cite{caldarelli2001beauty}, using a slightly different model (yet equivalent for the interpretation of the results), they analyzed the probability $ P (R) $ that an individual has a partner ranked $ R $ in his/her preference-list, for different levels of competition in the system in the GS dynamic. The results are shown in figure 7: when the competition in the system is zero, it is easier to make more people happy, so $ P(R) $ is a picked distribution around the lowest value. On the other hand, when the competition between agents is maximum ($ \omega = 1 $), a uniform distribution is expected. It is easy to show that, when the preference-lists are identical, the rankings of the men's wives from the lowest to the highest are $ 1,2,3 ... N $. So men's satisfaction univocally depends on their ranking in the list of women and, since they are drawn by uniform distribution, this produces a step function for $ P (R) $ when there is partial competition.

\subsubsection*{Number of Stable Solution}
\begin{figure}[!h]
\begin{center}
\includegraphics[width=0.6\textwidth,scale=0.6]{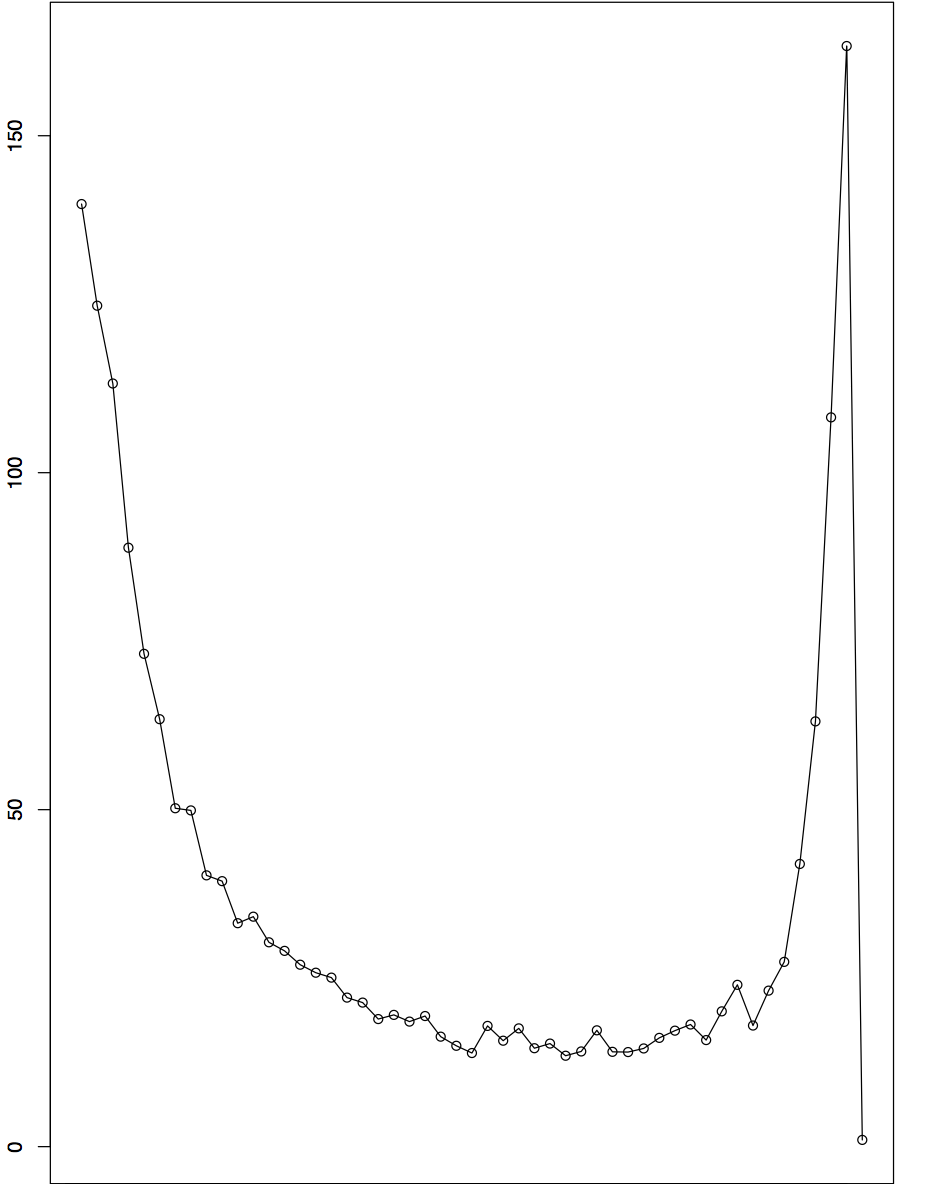}
\caption{\textbf{Average number of stable solution with correlated lists}: number of stable solutions as a function of $\omega \in [0,1]$. Each point is the average over 200 simulations for a system with $N=200$.}
\end{center}
\end{figure}
We now deal with the average number of stable solutions as a function of $ \omega $. We have already seen that when $ \omega = 0 $ we get the case of random and independent lists, so the average number of stable solutions is given by (8). On the other hand, when $ \omega = 1 $, i.e. the lists are all identical, it is easy to show that there is only one stable matching possible.\\
We expect that, with increasing competition, the average number of stable solutions decreases monotonically, given a fixed size $ N $. Indeed, one of us performed numerical simulations showing that the average number of stable solutions has a very high peak when $ \omega $ is close to 1. This is due to the symmetry of the stable configurations when the competition is close to the maximum:  in this situation, most of the stable solutions have the same total energy but are different from each other. The results are shown in figure 8.

\subsubsection*{Other Measures of Correlation}
The correlation between the lists can be inserted not only as in equation (23), but there are several possible generalizations to obtain realistic situations in which the preferences among the agents have certain conformity. Caldarelli and Capocci in \cite{caldarelli2001beauty} introduced the concept of distance in the SMP. They considered a spatial distribution of the agents of the system, making marriages that occur at great distances between two individuals more expensive. They defined the score $ S_ {ij} $ that the individual $ i $ assigns to the individual $ j $ as:
\begin{equation}
S_{ij}=\alpha d(i,j)+\eta_{ij} \;,
\end{equation}
where $ d(i, j) $ is the Euclidean distance between the individual $ i $ and $ j $, $ \eta_{ij} \in [0,1] $ is again a random number representing the subjectivity of personal preferences and $ \alpha \in [0,\infty] $ weighs the contribution of distance.\\
Since $ \eta_ {ij} $ is a random number between 0 and 1, they normalized the distance by dividing it by the largest possible value it can have (obviously dependent on $ N $). This way, when $ \alpha = 0 $, we have random and independent lists; when $ \alpha \to \infty $, we have that preference lists are dominated by each individual's first neighbours. This fact is important because, even when $ \alpha $ is very large, no individual will have the same preference-lists since each has different first neighbours. The results of Caldarelli and Capocci show that in GS dynamics, when the correlation due to distance is present, more advantageous matchings are obtained for the proposers compared to the classic case. \\
\\
Other correlation measures have been analyzed in \cite{celik2007marriage}. The authors also studied the case where only one of the two sexes has correlations between the lists. \\
A curious result regards the study on relaxation time carried out in \cite{nyczka2011stable} by Nyczka and Cislo. Instead of starting from the GS dynamics, they used a Montecarlo dynamics: starting from a random matching, they counted the number of switches between two individuals needed to achieve a stable configuration (simulating for $ N $ not too large). Their results show that when even little competition is added to the system (according to the model described in the previous paragraphs), relaxation time (i.e. the number of switches) decreases drastically even for small $ N $. In our opinion, it is an interesting result that can have different applications: for example, assume that individuals in the system can reproduce only when they are in a stable state, then it is evident that evolution prefers a system where competition exists between agents as it decreases the time to reach a stable state among individuals who, in turn, are more likely to reproduce at a young age.

\subsubsection{SMP with Unequal Number of Men and Women}
So far we focused on the case of matching two sets with \textit{same} number of agents. The reality, however, is that most two-sided markets are seldom symmetrical. Most real systems are matching between two classes with \textit{different} size. \\
Here, we shall study the GS model in the case of a different number of men and women. We will see how this difference modifies the classical results regarding the energies of the agents.\\
Dzierzawa and Omero in \cite{dzierzawa2000statistics} performed numerical simulations in the case of $ N + 1 $ men and $ N $ women. In this case, of course, a man is forced to remain single. The authors showed that, with the men-oriented GS algorithm, women are much happier than men, and this is different from what occurs in the symmetrical case. This is very interesting as it shows that the GS model is strongly sensitive to the asymmetry of the system (even of a single element). \\
\\
Consider the classic SMP scenario with $ N $ men and $ M $ women. When $ N = M $, we get the conventional SMP. When $ N \ne M $, there will be some individuals who will remain single. How to define the energy for these people? There are several ways to do this, here we shall adopt the convention used in \cite{shi2018instability} where the energy of a single is defined as $ 1 + N $ for men and $ 1 + M $ for women. For $ N \le M $, all men will be married; while for $ N> M, $ there will be $ M $ married men and $ N-M $ single men. 

\subsubsection*{Energy of the Agents in the Men-Oriented GS}
We consider the energy of the agents in the men-oriented GS model when there are more women than men. The reasoning will be similar to what we did in the symmetrical case. \\
Remember that the total energy of men is the same as the number of their proposals. Now let us focus on the number of marriages that form during the GS process. When this number is equal to $ N $ it means that all men are married and the process ends. When the number of couples is $ C $, then there will be $ M-C $ unmarried women. Hence, the probability that a proposal is made to an unmarried woman is $ (M-C) / M $. In this case, the number of couples increases by one. So, on average, $ M / (M-C) $ proposals will have to be made before a new pair is formed. So the total number of proposals $ L $ is
\begin{equation}
L=\sum_{C=0}^{N-1}\frac{M}{M-C}=M\left( \sum_{i=1}^{M} \frac{1}{i}-\sum_{j=1}^{M-N} \frac{1}{j} \right) \;.
\end{equation}
So the average energy of men is \cite{shi2018instability}:
\begin{equation}
\overline{E_m}=\frac{L}{N}=\frac{M}{N}log\frac{M}{M-N} \;.
\end{equation}

\begin{figure}[!h]
\begin{center}
\includegraphics[width=0.7\textwidth,scale=0.7]{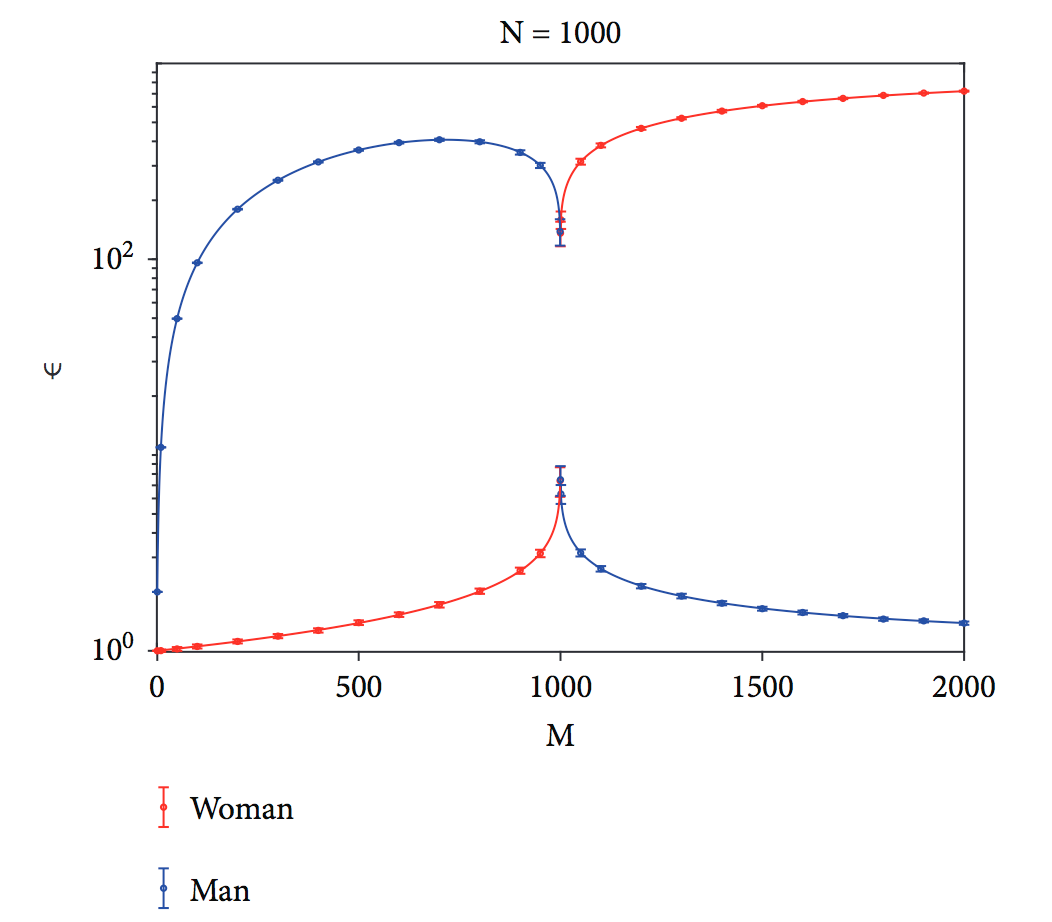}
\caption{\textbf{Agents' energies with different number of men and women}: average energy of $N$ men and $M$ women as a function of $M$. $N$ is 1000. Each point represents an average over 100 realizations. The solid lines represents the analytical solution. Figure reprinted from \cite{shi2018instability}.}
\end{center}
\end{figure}

We now calculate the average energy of women $ \overline {E_w} $ in the men-oriented GS dynamics, always with $ M> N $. Similarly to what was done in the symmetrical case, we consider the final stable solution in which all the partners of the women are already determined. Let us say that a certain woman is married to a man with ranking $ x $ in her list. All men with rankings better than $ x $ have not sent any proposals to her. Consequently, these men should have a better partner than the woman that we are considering. We can write the probability that a woman is married to a man of ranking $ x $ as
\begin{equation}
P_x= \frac{E_m(x)}{M}\cdot \prod_{i=1}^{x-1}\left(1-\frac{E_m(i)}{M} \right) \;.
\end{equation} 
Single women energy is $ N + 1$, so we can write
\begin{equation}
P_{N+1}=\prod_{i=1}^{N} \left(1-\frac{E_m(i)}{M} \right) \;.
\end{equation}
Now, summing equations (27) and (28), and replacing $ E_m (i) $ with its average value $ \overline{E_m} $, we get the average energy of women evaluating the following quantity $ \overline {E_w} = \sum x \cdot P_x $. We obtain
\begin{equation}
\overline{E_w}=\frac{\overline{E_m}}{M}\sum_{i=1}^{N}i \cdot \left(1-\frac{\overline{E_m}}{M} \right)^{i-1}+(N+1)\cdot \left(1-\frac{\overline{E_m}}{M} \right)^N \;.
\end{equation}
Performing the summation and using the approximation $(1-(\overline{E_m}/M))^N \sim (M-N)/M$, after some algebra we get  \cite{shi2018instability}
\begin{equation}
\overline{E_w}=\frac{N}{\overline{E_m}}+\frac{M-N}{M}\;.
\end{equation}

Now, by analogous calculations, it is possible to obtain the average energies of women and men in the men-oriented GS with $ N> M $, obtaining respectively  \cite{shi2018instability}
\begin{equation}
\overline{E_w}=\frac{N}{M}log \left(\frac{N}{N-M} \right)\;,
\end{equation}
\begin{equation}
\overline{E_m}=\frac{M}{\overline{E_w}}+\frac{N-M}{M}\;.
\end{equation}

Figure 9 shows the results of the numerical simulations compared to the analytical solution.

\subsection{Ground State Energy}
Forget for a moment the notion of stability: in this section, we seek for the solution that minimizes the total energy of the system, or even that maximizes total happiness, without taking into account the stability between the agents. That is, we want to study how to obtain the minimum energy $ E_{TOT}^*$ given by
\begin{equation}
E_{TOT}^*=\sum_{i=1}^{N} x_i + \sum_{i=1}^{N} y_i=X+Y \;, 
\end{equation}
and we want to study its general properties. To do this we will frame the SMP in a class of optimization problems and we will exploit the techniques used by physicists to solve these problems.\\
First, we want to introduce the main concepts as well as the definition of \textit{optimization problem}. For a more detailed description on can refer, for example, to \cite{papadimitriou1998combinatorial, christos1994papadimitriou, moore2011nature}. The instance of an optimization problem consists of two mathematical elements:
\begin{itemize}
\item \textit{Space of solution} $\tau \ne \emptyset$, $\tau \to \mathbb{R}$.
\item \textit{Cost function} (in our case the energy) : $E$.
\end{itemize}
The goal is to find the \textit{globally optimal solution}, that is the solution $ C_0 \in \tau $ which minimizes the cost function, i.e. $ E (C_0) = min (E (C \in \tau)) $. Note that in general the existence of $ C_0 $ is not guaranteed, but this is not the case with SMP. In fact, the matching problem belongs to an optimization problems class called \textit{combinatorial optimization} in which there is \textit {always} a solution but the number of possible configurations is often extremely large and therefore an approach of brute force is computationally infeasible. \\
\\
We introduce the analytical methods that have been developed to find the minimum solution of the matching problem. As we will see in the next section, given an instance of the problem, to find the value of $E_{TOT}^* $ there are specific efficient algorithms with a polynomial execution time. However, to study the general properties of a given optimization problem, it is often useful to consider the complexity of the problem and its solution \textit{on average}, for large system sizes. Indeed, the probabilistic approach to combinatorial optimization problems helps the understanding of their mathematical properties. \\
In this context, statistical physics has played an important role. Many techniques developed by physicists for the study of disordered systems and phase transitions are very effective in the study of random optimization problems. In particular, we will deal with the \textit {replica method}, that is a powerful technique to analytically find the average ground state of the matching problem in the thermodynamic limit ($ N \to \infty $). \\

\subsubsection{Statistical Physics and Optimization}
In recent years, statistical mechanics has been a very useful tool for solving and understanding many optimization problems. We will make a general recapitulation of the basic notions of statistical mechanics and its link with optimization.\\
As we have already mentioned, each optimization problem consists of 1) a set $ \tau $ of possible configurations $ C \in \tau $ and 2) of a cost function $ H (C) $ to be minimized. $ H (C) $ can be interpreted as the energy of a physical system and $ C $ as the numerical quantities that determine its state. Thinking about the "equilibrium" system, one can try to derive its thermodynamics by counting the configurations that have a certain energy $ E $
\begin{equation}
N(E) = \sum_ {C \in \tau} \delta(H (C) -E) \;,
\end{equation}
where $ \delta (x-x^*) $ is Dirac's delta function, which is 1 if $ x = x^* $ and zero otherwise. The optimization problem can therefore be solved finding the minimum energy $ E_{min} $ for which there is at least one configuration, $ N(E_{min}) \ge 1 $. Since the entropy of the system can be defined as the logarithm of $ N(E) $, then one can try to derive all the thermodynamic quantities of this system. Any physical system with a large number of degrees of freedom is, for convenience, treated with statistical mechanics, that is, one tries to deduce the thermodynamic quantities starting from the knowledge of the microscopic properties of the system. \\

\subsubsection*{Statistical Mechanics Approach} 
Here we present some basic notions of statistical mechanics. There is a very vast literature on the subject: for a detailed discussion, one can refer, for example, to \cite{reif1965statistical, huang1967statistical, ellis2006entropy,}. \\
Consider a volume $ V $ that confines $ n $ particles. We want to associate a probability measure $ P(C) $ to each configuration $ C \in \tau_{n} $ of the system of $ n $ particles. One can imagine having an \textit{ensamble}, that is, a large number of copies of the physical system which evolve independently of each other. When all these copies are in equilibrium, the probability $ P (C) $ can be interpreted as the probability of drawing from the ensemble a copy that is exactly in the configuration $ C \in \tau $. At this point, we define the entropy $ S $ of the system as:
\begin{equation}
S=-\sum_{C \in \tau_n} P(C) ln(P(C)) \;.
\end{equation}
This quantity, according to the second principle of thermodynamics, can never decrease (for an isolated system at equilibrium). Given the Hamiltonian $ H(C) $, and setting the average energy to $ \epsilon = <H (C)> = \sum_{C \in \tau_n} P(C) H(C) $, the goal is to find the probability distribution $ P(C) $ that maximizes the entropy. This distribution is the Boltzmann distribution:

\begin{equation}
P(C)=\frac{1}{Z_n(\beta)}e^{-\beta H(C)} \;,
\end{equation}

\begin{equation}
Z_n(\beta)=\sum_{C \in \tau_n} e^{-\beta H(C)} = \int dE N(E) e^{-\beta E} \;,
\end{equation}
where the parameter $ \beta $ was introduced using the Lagrange multipliers method and corresponds to the inverse of the temperature, that is $ \beta = \frac {1}{T} $. At this point, the following identities apply:

\begin{equation}
\epsilon=<H(C)>=-\frac{\partial ln(Z_n(\beta))}{\partial \beta} \mid_{\beta=\beta (\epsilon)} \;,
\end{equation}

\begin{equation}
S(\epsilon)=\beta (\epsilon) \epsilon+ ln(Z_n(\beta(\epsilon)))  \;,
\end{equation}

\begin{equation}
\frac{\partial S(\epsilon)}{\partial \epsilon}=\beta=\frac{1}{T} \;.
\end{equation}

So the \textit{canonical partition function} $ Z_n(\beta) $ contains all the information about the system. The free energy $F(\beta)$ is:

\begin{equation}
F(\beta)=-\frac{1}{\beta} ln(Z_n(\beta)) \;.
\end{equation}

The rest of thermodynamics can be deduced from the free energy (entropy, specific heat, etc.). \\
Now, the application of the central limit theorem becomes essential when, keeping the density $ \rho = n/V $ constant, the volume $ V $ is increased. We want to reach the so-called \textit{thermodynamic limit}. All extensive quantities should grow linearly with the number of particles $ n $. So we expect that, when $ n \to \infty $, the energy per particle $ e = \epsilon / n $ is a finite value. The basic idea is that when the volume is infinite, most systems of the canonical ensemble have energy $\epsilon = ne $ and only these states contribute to the entropy. \\
\\ 
Now, by lowering the temperature, the energy of the system decreases. When the temperature is very close to zero, the probability that a system has certain energy is different from zero only for energies that are very close to the minimum energy, that is to the ground state. It is clear the link between optimization problems and statistical mechanics: the ground state coincides with the optimal configuration of our optimization problem. The use of statistical mechanics to solve optimization problems is very advantageous. We can compare, both qualitatively and quantitatively, different problems through the same physics language.

\subsubsection*{Frustrated Systems}
Now let us see what are the main properties that characterize frustrated and disordered systems. In this way, we will be able to interpret the intractability of most of the complex optimization problems. \\
Consider $ n $ objects that in pairs can attract each other (negative energy) or repel each other (positive energy). For example, think of $ n $ individuals who can love or hate each other in pairs. The goal is to divide individuals into two groups achieving maximum total happiness. Formally it can be written using $ n $ variables $ \sigma_i $ to represent a configuration such that $ \sigma_i = \pm 1 $, depending on the group to which the individual $ i $ belongs. We also define the quantity $ J_{ij} = + 1 $ if individuals $ i $ and $ j $ love each other and $ J_{ij} = - 1 $ if they hate each other. We note that the quantity $ 1-J_{ij} \sigma_i \sigma_j $ is equal to zero only if two people who love each other are in the same group or if two people who hate each other are in different groups, otherwise this quantity is equal to 2. We can therefore interpret this quantity as the degree of \textit {unhappiness} in the system.\\
This example represents a classic optimization problem in which the quantity to be minimized is the following Hamiltonian:

\begin{equation}
H(\sigma)=- \sum_{1 \le i \le j \le n} J_{ij} \sigma_i \sigma_j \;.
\end{equation}

One can think of solving this problem iteratively: at each turn, each individual is asked if he wants to change the group he belongs to. This individual decides to change only if this operation leads him to have greater happiness, even at the expense of the others. People damaged by this operation may want, in their turn, to restore the previous configuration. Let us see what happens in complicated situations, for example where three individuals, $ i $, $ j $ and $ z $, hate each other or if $ i $ loves $ j $ and $ j $ loves $ z $, but $ i $ and $ z $ hate each other. In this situation, a vicious circle is triggered in which individuals chase each other or avoid each other endlessly without ever reaching a configuration that is satisfactory for everyone. It is not possible to make everyone happy, so one has to choose between different equivalent configurations, as our system has multiple ways to be ordered. This last feature defines the \textit{disordered} systems. One of the main causes of the emergence of the disorder is the presence of \textit{frustrations}: we are in the presence of a frustration when $ J_ {ij} J_ {jz} J_ {zi} = - 1 $. The higher the number of frustrations, the more difficult it is to find a configuration that minimizes the Hamiltonian in (42). \\
\\
The shape of this Hamiltonian is equal to that obtained by Sherrington and Kirkpatrick in \cite{sherrington1975solvable} in the study of spin glasses. This model shows that for large $ n $ the Hamiltonian is generally characterized by a huge number of local minimums separated from each other by high energy barriers. An approximate way to estimate the height of these barriers is as follows: we can take $ J_ {ij} $ and $ \sigma_i $ randomly. Since energy fluctuates around its average value by an amount proportional to $ n $, we can assume that this is also the order of magnitude of these energy barriers. When the temperature becomes low enough, the dynamics we have described above (flipping one spin at a time) tends to become trapped in one of these local minimums. It will only be able to escape when it manages to overcome energy barriers of the order of $ O (n) $. These barriers, therefore, diverge in the thermodynamic limit. This is called the "spin-glass phase" in the Sherrington and Kirkpatrick model (SK model). Since the occupancy numbers at this stage behave as if they were frozen, only small variations in energy are allowed. The minima of this Hamiltonian are called \textit {metastable state}, as the time in which the system remains in such states could be so long that they appear as a state of equilibrium.\\
\\
In general, the time averages will be different from those of the ensemble and will depend on the initial situation (unless you wait for the system to enter and exit several times from these metastable states).\\
Furthermore, if the number of metastable states grows exponentially with the size of the system, when the temperature drops too much, the convergence to equilibrium also slows down drastically. So, for $ n \to \infty $, the system is no longer ergodic (remember that the property of ergodicity states that, sooner or later, the system will surely pass through any configuration and therefore the time average and the ensemble average are equivalents). Given these considerations, it is quite unlikely that the method of flipping the spins one at a time (also called \textit{Metropolis} algorithm \cite{beichl2000metropolis}) will be able to derive the energy of the ground state. \\
\\
The metastable states are not all the same. The most populated states will be the deepest ones, i.e. with the least free energy (consistently with the principles of thermodynamics). So the idea is to count how many metastable states there are for each free energy value \cite{parisi2003course}: denote by $ \nu = 1, ..., M $ the metastable states. Since in the thermodynamic limit the states are divided by energy barriers, we can imagine breaking down the partition function in this way
\begin{equation}
Z_n(\beta)=\sum_{\nu} \sum_{C \in \nu} e^{-\beta H(C)} \approx \sum_{\nu}e^{\beta F_{\nu}(\beta)} \approx \int_{F_{min}}^{F_{max}} N(F,V) e^{-\beta F}\, dF \;,
\end{equation}

where $ N (F, V) $ represents the number of metastable states with energy $ F $; the logarithm of this quantity is called \textit {complexity}. We can reasonably assume that the number of such states is non-zero only within the range $ [F_ {min}, F_ {max}] $. Since complexity, in analogy with entropy, is an extensive quantity, then in the thermodynamic limit we have

\begin{equation}
log (N(F)) \equiv \Sigma_{n}(F) \approx n\Sigma_{n}(F/n)= n\Sigma_{n}(f) \;,
\end{equation}
where $ f = F / n $ and we define $ F_ {min} = nf_{min} $ and $ F_{max} = nf_{max} $. So the complexity grows exponentially with the size of the system. Taking the maximum of the exponent in the integral in equation (43), we obtain the density of free energy from which to derive all the thermodynamics:
\begin{equation}
\beta f(\beta)= min(\beta f -\Sigma(f)) ; f \in [f_{min}, f_{max}] \;.
\end{equation}

At this point two scenarios are possible. In the case of high temperatures, the minimum of equation (45) is within the range $ [f_ {min}, f_ {max}] $ and the solution can be found from the relation
\begin{equation}
\frac{\partial \Sigma (f)}{\partial \beta} |_{f=f^*(\beta)}=\beta \;,
\end{equation}
from which one obtains
\begin{equation}
f(\beta)=f^*(\beta)-\Sigma(f^*(\beta)) \;.
\end{equation}
Therefore the number of minimums in which the system will occur will be of the order of $ e^{n\Sigma(f^*(\beta))} $. \\
In the case of low temperatures, on the other hand, the minimum is at the extremes of the range $ [f_ {min}, f_ {max}] $, so we have $ f^* = f_{min} $. In this case, given that complexity is an increasing function of $ f $, we have that the complexity contribution to total free energy is close to zero.\\
\\
Hence, the form of the complexity function is very important for understanding the characteristics of a system. It is also directly linked to optimization: imagine having
\begin{equation}
e^{-n\psi \beta}=\sum_{\nu}e^{-n \beta f_{\nu}} \approx \int_{f_{min}}^{f_{max}} e^{-n(\beta f - \Sigma (f))}\, df \;.
\end{equation}
Estimating this integral through the saddle point method, we see that the $ \psi $ function is linked to complexity in the same way that entropy is linked to energy:
\begin{equation}
\beta \psi (\beta)= min(\beta f - \Sigma (f))= \beta f^*(\beta) - \Sigma(f^*(\beta)) \;.
\end{equation}
Note that cancelling the first derivative of $ \psi $ corresponds to finding the free energy value for which the complexity is zero:
\begin{equation}
\frac{\partial \psi (\beta)}{\partial \beta} |_{\beta=\beta_0}=0 \to \Sigma(f^*(\beta_0))=\Sigma(f_{min})=0  \;.
\end{equation}
The latter condition means that there have been no states with free energy less than $ f_{min} $. Furthermore, since both $ \psi $ and $ \Sigma $ depend on temperature and chemical potential, when we send the temperature to zero, that is $ \beta \to \infty $, for each state $ \nu $ and for $ C \in \nu $, free energy must be
\begin{equation}
nf_{\nu}=min(H(C)) \to f_{min}=min(f_{\nu})=\frac{1}{n}min(H(C)) \;,
\end{equation}
where $ f_{min} $ is the energy of the ground state. So we have just seen that complexity is very useful for finding the solution to any optimization problem.\\
\\
This methodology was used first for the studies of spin glass system. For more details one can refer to \cite{castellani2005spin, mezard1987spin, nishimori2001statistical, parisi1983order, de1978stability, parisi1988statistical}. The importance of studying spin glasses goes beyond physics applications: for example, Mezard and Parisi in \cite{mezard1985replicas} discussed the applications of the analytical techniques used in spin glasses for some optimization problems. In particular they studied the \textit{Traveling salesman problem} and the \textit{Matching problem} using \textit{replica method}. It turns out to be a powerful tool to solve many optimization problems with frustrations and, as we shall show in the next paragraph, it helps us to find (analytically) the ground state of the SMP.
 
\subsubsection{Average Ground State with Replica Method}
Let us go back to the SMP. We want to find the solution that minimizes energy regardless of stability. As we will see in section 7, a system can achieve the maximum global happiness with the help of a matchmaker, who has all the information available in the system. \\
As we already mentioned, combinatorial optimization can be formulated as a problem of statistical mechanics \cite{kirkpatrick1981lectures}. In this way, it is possible to cope with frustrations and great complexity. Indeed, on can introduce an artificial temperature and a Boltzmann weight for every possible configuration; the cost function corresponds to the energy of the system and, in the language of statistical mechanics, one studies the properties of the system at low temperatures such as ground state configuration and its energy. \\
Here, we will see how the analytical methods developed in the mean-field theory of spin glasses can be used to solve the SMP ground state problem. In particular, we will exploit the replica approach.\\
 Mezard and Parisi in 1985 \cite{mezard1985replicas} wrote a pioneering article. They solved the ground state problem both in the monopartite and bipartite matching problem.\\
We will show the calculations in the more general case of monopartite matching problem (i.e. each agent can marry with all the other agents, without distinction between males and females), and then we shall modify only some details to solve the problem also in the SMP (or, equivalently, in the case of the bipartite matching problem), following what Omero, Dzierzawa, Marsili, and Zhang did in \cite{omero1997scaling}.

\subsubsection*{SMP Partition Function}
In the matching problem, there are $2N$ individuals who can mate with each other two by two. The observable of this system is the total happiness, i.e. total energy: $ H(C) $, where $ \{C \} $ is the set of all possible configurations. In the limit $ N \to \infty $, we can formulate this problem according to statistical mechanics: the goal is to study the partition function of the system $ Z (\beta) = \sum_{C} e^{-\beta H( C)} $. If $ F $ is the free energy we have
\begin{equation}
\lim_{\beta \to \infty} F(\beta)=\lim_{\beta \to \infty} -\frac{ln(Z)}{\beta}=min(H(C)) \;,
\end{equation}
therefore finding the minimum energy corresponds to calculating the logarithm of the partition function.\\
\\
In the matching problem, writing the set of all possible configurations such as $ C = n \in \{0,1\} $, the Hamiltonian (energy) can be written in the following way
\begin{equation}
H(n)=\sum_{i=1}^{2N}\sum_{j=i+1}^{2N} n_{ij} l_{ij} \;,
\end{equation} 
where $\sum_{j=1}^{2N} n_{ij} =1$ and $l_{ij}$ is the cost corresponding to pairing the person $ i $ to the person $ j $ and, in the terminology of SMP, it corresponds to the sum of the ranking of person $i$ in $j$'s list, and the ranking of person $j$ in $i$'s list. Note that the conditions on $ n_{ij} $ fix the number of couples at $ N $, thus obtaining perfect matching (i.e. without singles or people who have more than one partner). \\
The partition function associated with this Hamiltonian is
\begin{equation}
Z(\beta)=\sum_{n} \left[\prod_{i=1}^{2N}\delta\left(1-\sum_{j=1}^{2N} n_{ij} \right)\right]
 \exp\left\lbrace-\beta \sum_{i=1}^{2N}\sum_{j=i+1}^{2N} n_{ij}l_{ij}\right\rbrace \;,
\end{equation}
where the $ \delta $ function enforces the condition $ \sum_ {j = 1}^{2N} n_{ij} = 1 $. By writing the $ \delta $ function as Fourier series $\delta(x)=\int_{0}^{2\pi}\frac{d\lambda}{2\pi} e^{ix\lambda}$, and adding up all the possible values of $ n_{ij} $ (which are 0 and 1), we can rewrite the partition function in the following way

\begin{equation}
Z(\beta)=
\left[\prod_{j=1}^{2N}\int_{0}^{2\pi}\frac{d\lambda_j}{2\pi}\right]
e^{\sum_{j=1}^{2N}i\lambda_j}
\left[\prod_{k=1}^{2N}\prod_{j=k+1}^{2N}\left(1+e^{-i(\lambda_k+\lambda_j) -\beta l_{kj}}\right)\right] \;.
\end{equation}

\subsubsection*{Replica Method}
Now, from equation (52), we know that we must calculate the logarithm of the partition function. On the other hand, we are interested not in a specific instance but in the \textit{configurational average}: that is, the average calculated on the distribution of the coupling factor, in this case, $ l_ {ij} $.\\
Calculating the quantity $ \overline {log (Z)} $ is extremely complicated and therefore, as in the case of spin glasses, to simplify the calculations we use the replica method, namely, we use the identity
\begin{equation}
\overline{log(Z)}=\lim_{n \to 0} \frac{\overline{Z^n}-1}{n} \;.
\end{equation}
So the problem is to calculate the following quantity:
\begin{equation}
\overline{Z^n}=\int \rho(l) Z^n \,dl \;,
\end{equation}
where $ \rho (l) $ is the probability distribution of the coupling cost $ l_{ij} $. Let us now calculate $ \overline{Z^n} $: in equation (55) we have $ 2N $ integrals over the variables $ \lambda $ and therefore we introduce the associated "replicas"  $ \lambda_i^a $ that come from the expansion of the power of $ Z (\beta)^n $, i.e.
\begin{equation}
\overline{Z^n}= 
\left[\prod_{a=1}^n\prod_{j=1}^{2N}\int_{0}^{2\pi}\frac{d\lambda^a_j}{2\pi}\right]
e^{\sum_{a=1}^n\sum_{j=1}^{2N}i\lambda^a_j}
\prod_{k=1}^{2N}\prod_{j=k+1}^{2N}\left[
\int dl \rho(l)\prod_{a=1}^n\left(1+e^{-i(\lambda^a_k+\lambda^a_j) -\beta l_{kj}}\right)\right] \;.
\end{equation}

By defining $g_p(\beta)=\int dl \rho(l)e^{-p\beta l}$, we can rewrite equation (58) in the following way
\begin{equation}
 \overline{Z^n}= 
\left[\prod_{a=1}^n\prod_{j=1}^{2N}\int_{0}^{2\pi}\frac{d\lambda^a_j}{2\pi}\right]
e^{\sum_{a=1}^n\sum_{j=1}^{2N}i\lambda^a_j}
\prod_{k=1}^{2N}\prod_{j=k+1}^{2N}\left[
1+\sum_{p=1}^n g_p(\beta) \sum_{1\leq a_1<...<a_p\leq n}\prod_{r=1}^p e^{-i(\lambda^{a^r}_k+\lambda^{a^r}_j)}
\right] \;.
\end{equation}
At this point, Mezard and Parisi make the following ansatz for the distribution of the coupling variable: $ \rho (l) = \frac{l^{d-1}}{(d-1)!}  e^{-l} $, where $ d $ is a characteristic parameter of the distribution, i.e. $ d = 1 $ means that $ \rho (l) $ has a non-zero probability of obtaining infinitely small values of $ l $, while the case $ d = 2 $ is the case of the SMP in which the distribution of $ l $ is triangular (sum of two uniformly distributed random variables).\\
\\
Since, given $ N $ nodes in a network, the nearest neighbor of a point $ i $ is at a distance of order $ N ^ {-1/d} $, then the energy (the cost function) is expected to scale as $ N ^ {1-1 / d} $ and therefore it is convenient to perform the calculations with the following transformation: $ \beta = \beta_0 N^{1/d} $. This way when $ \beta_0 \to \infty $ one has $g_p(\beta)=\frac{1}{p^d(1+\beta)^d}=\frac{1}{N(p\beta_0)^d}$.\\

Now, exploiting the properties of the Gaussian integrals performing Gaussian transformation, we can introduce the Gaussian parameters $ Q_ {a_1, .. ,a_p} $. At the end of the calculation, we have

\begin{equation}
\overline{Z^n}=\prod_{p=1}^n\prod_{1\leq a_1<...<a_p\leq n}\int dQ_{a_1,..a_p} \exp\left\{
-\sum_{p=1}^n\sum_{1\leq a_1<...<a_p\leq n}\frac{1}{2g_p(\beta)} Q_{a_1,..a_p}^2
+2N\log z\right\} \;,
\end{equation}

where we defined the  \textit{one site partition function} $z$:

\begin{equation}
z=\left[\prod_{a=1}^n\int_{0}^{2\pi}\frac{d\lambda^a}{2\pi}\right]
\exp\left\{
\sum_{a=1}^ni\lambda^a+\sum_{p=1}^n \sum_{1\leq a_1<...<a_p\leq n} Q_{a_1,..a_p} e^{-i\sum_{r=1}^p\lambda^{a^r}}
\right\} \;.
\end{equation}

From this form it remains to compute $z$ and integrate over $Q_{a_1,..,a_p}$.\\
To carry out this last step we use the \textit{saddle point technique}: since $g_p(\beta)$ scales as $1/N$, the integral is of the kind $\int dx e^{N\phi(x)}$ which can be approximated by $e^{N\phi(x_0)}$ where $x_0$ is such that $\frac{d}{dx}\phi(x_0)=0$.\\
In our case the integration is equivalent to solving the closed equations for the variables $ Q_ {a_1, .. a_p} $ obtaining $ Q ^ 0 $ and calculating $ Z $ in the saddle point, that is:

$$Q^0 \quad \text{t.c.} \quad Q^0_{a_1,..a_p}=g_p(\beta) \frac{\partial}{\partial Q_{a_1,..a_p}}|\log z (Q^0) \quad \forall \,\, a_1,..,a_p \;. $$

And so we obtain

\begin{equation}
\overline{Z^n}:=\exp\left\{-N\phi(Q^0)\right\}=\exp\left\{
-\sum_{p=1}^n\sum_{1\leq a_1<...<a_p\leq n}\frac{1}{2g_p(\beta)} (Q^0_{a_1,..a_p})^2
+2N\log z(Q^0)\right\} \;.
\end{equation}

\subsubsection*{Replica Symmetry Ansatz}

To solve the saddle point equation we consider a restricted subspace of the possible solutions of $ Q_0 $, i.e. we assume that all these variables are symmetric and therefore equal with respect to the variables $ {a_1, .., a_p} $: $ Q_ {a_1, .. a_p} = Q_p $. In spin glass theory, this symmetry assumption is not enough to correctly describe the system, but we must introduce symmetry breaking \cite{mezard1984replica}. However, in the case of matching this ansatz is sufficient to solve the problem.\\

With this simplification, and using the replica trick of equation (56), we finally can write:
\begin{equation}
\overline{log(Z)}=\lim_{n\rightarrow 0}\frac{\overline{Z^n}-1}{n}=
\lim_{n\rightarrow 0} \frac{e^{-N\phi(Q^0)}-1}{n}=
\lim_{n\rightarrow 0} -N\frac{\phi(Q^0)}{n} \;,
\end{equation}

where we have defined 

\begin{equation}
\phi(Q^0)=-\frac{1}{2}\sum_{p=1}^n \frac{1}{Ng_p(\beta)} {n\choose p} (Q^0_p)^2+ 
2\log \left[
\left(\frac{\partial}{\partial x}\right)^n\exp\left\{\sum_{p=1}^\infty \frac{Q_p}{p!}x^p\right\}|_{x=0}\right] \;.
\end{equation}

Hence, performing the limit, we found the configurational average of the logarithm of the partition function and we can formulate the closed equations for the matching problem:

\begin{equation}
\overline{log(Z)}=-N\lim_{n\rightarrow 0}\frac{\phi(Q^0)}{n}=-N\left\{
-\frac{1}{2}\sum_{p=1}^\infty\frac{1}{N g_p} (-1)^{p-1}
\frac{(Q_p^0)^2}{p}
	+2\left(
	\int_{\infty}^{-\infty} [e^{-e^{l}}-e^{-G(Q^0,l)}]
	\right)
\right\} \;,
\end{equation}

with:

\begin{equation}
G(Q,l)=\sum_{p=1}^\infty (-1)^{p-1} \frac{Q_p}{p!}e^{pl}\;,
\end{equation}

and $Q^0$ is defined implicitly from $\frac{\partial}{\partial Q_p} \frac{\phi(Q^0)}{n}=0 \quad \forall \, p>0 $, i.e (to be solved numerically):

\begin{equation}
 \frac{1}{Ng_p}(-1)^{p-1} \frac{Q^0_p}{p}=2 \int dl \, \left[ (-1)^{p-1}\frac{e^{pl}}{p!}\right]
e^{-G(Q^0,l)}\;.
\end{equation}

From these equations we can find easily the thermodynamic quantities by computing: 

\begin{equation}
E=-\frac{\partial}{\partial \beta} \overline{log(Z)}
\end{equation}

\begin{equation}
F=-\frac{1}{\beta}\overline{log(Z)}
\end{equation}

Now, by solving the equation (68) numerically in the limit $ \beta \to \infty $ (zero temperature) we obtain the value of the minimum energy, that is the ground state. In the case of $ d = 1 $, where the probability of having infinitely small values of $ l_ {ij} $ is not zero, Mézard and Parisi have obtained a constant value for the minimum energy (in the thermodynamic limit $ N \to \infty $), i.e.
\begin{equation}
E^{d=1}(T=0)=\frac{\pi^2}{12}\approx 0.82 \;.
\end{equation}
This result has been obtained recently in a more rigorous way by mathematicians, in particular in \cite{parisi1998conjecture, linusson2004proof, wastlund2009easy, aldous2001zeta, sharma2002parisi, nair2003proof}.\\
\\
As we have already mentioned above, the case of the SMP is the case in which $ d = 2 $. Indeed, $ l_ {ij} $ follows a triangular distribution as it represents the sum of two uniformly distributed variables, i.e. happiness of man plus that of woman, $ x (i, j) + y (j, i) $. Therefore energy is expected to scale as $ N ^ {1/2} $. So one obtains
\begin{equation}
E^{d=2}(T=0)\approx 1.144 \sqrt{N} \;.
\end{equation}
We must remember that the SMP is a \textit {bipartite matching problem}, i.e. the agents are divided into men and women. The calculations made so far have been performed in the case of \textit {monopartite matching problem}.

\begin{figure}[!h]
\begin{center}
\includegraphics[width=0.7\textwidth,scale=0.7]{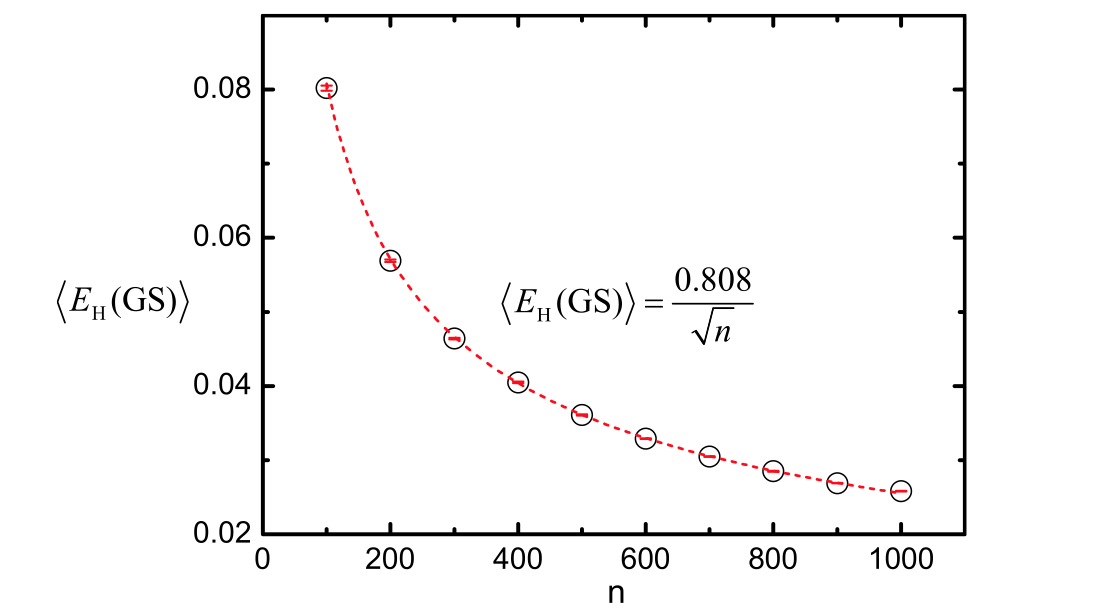}
\caption{\textbf{SMP Ground State energy}: relationship in a bipartite matching problem between the expected average energy and the system size $n$. The dashed line is the theoretical result. The results are averaged over 100 realization. Figure reprinted from \cite{shi2016analysis}, with permission from Elsevier.}
\end{center}
\end{figure}

To consider the bipartite nature of SMP, it is enough to change the energy scaling by a constant. In particular, it is sufficient to multiply the monopartite case by $ 2 ^ {1 / d} $. So, we find that, if $ n = N / 2 $ (i.e. the number of women or men), the expected average energy per person in the ground state is equivalent to \cite{shi2016analysis}
\begin{equation}
<E_H(GS)> \approx \frac{0.808}{\sqrt{n}} \;.
\end{equation}
Figure 10 shows the simulations results obtained with the Hungarian algorithm (which will be described in detail in the next section). The theoretical results are very much in agreement with the simulations. \\
As already mentioned, the minimum energy solution in which stability is preserved is $ 2 \sqrt {N}$ \cite{omero1997scaling}. So, ultimately, the intervention of a matchmaker who forces players to mate to get the ground state and neglecting stability would improve total system happiness by about $ 19 \% $.\\
Other works use the same technique to find the minimal solution in a SMP where bachelors are allowed \cite{nieuwenhuizen1998marriage}.\\
\\
The replica method, as we have seen, is very powerful and is useful for solving many disordered systems \cite{bomze1999maximum, franz2003replica, martin2001statistical, vannimenus1984statistical}. However, even though the solution is exact, it is not a rigorous method. A more exact technique by which the same results can be obtained is the \textit {cavity method}. It too is a powerful mathematical tool born in the field of statistical physics. The detailed explanation of this method in the context of disordered systems is beyond the scope of this review but one can refer to the following literature \cite{martin2005random, martin2004frozen, krauth1989cavity, mezard1987spin, mezard2003cavity} 

\subsection{Ground State vs Stable Solutions}
It is interesting to study the similarities and differences between the ground state and the stable states of the SMP. Comparing these states is useful because it helps us to understand the interplay between acting selfishly and reaching the optimal state for society. Furthermore, it can be useful for choosing the best strategy a matchmaker should adopt when matching two classes of agents who \textit{a priori} could make independent decisions. \\
In the next paragraphs, to make this comparison, we shall compute the ground state (through the Hungarian algorithm) and the optimal stable state, i.e. the stable state with lower energy.
 
\subsubsection{Gap between Ground State and Optimal Stable State}
We have shown that there is an energy gap between the optimal stable state and the global optimum. In particular, in the ground state there is an improvement of about $ 19 \% $ compared to the optimal stable state. In \cite{lage2006marriage}, Lage-Castellanos and Mulet, introduce the following quantity:
\begin{equation}
D(a,b)= \frac{1}{N}\sum_{i=1}^{N} (1-\delta_{(a_i,b_i)})\;.
\end{equation} 
This quantity measures the number of different pairs between the state $ a $ and the state $ b $. We will call it \textit{distance} between two states. We are interested in measuring $D(GS, OSS) $, which is the number of different pairs between the ground state (GS) and the optimal stable state (OSS). The figure shows the results of the simulations. In particular, it holds $ D (GS, OSS) = 0.53 $, that is, about $ 50 \% $ of the pairs present in the optimal stable state differ from those present in the ground state. This suggests that the two states are strongly correlated. Indeed, it can be shown that the average distance between two random matching goes as $ 1-1/N $ \cite{lage2006marriage}.

\begin{figure}[!h]
\begin{center}
\includegraphics[width=0.7\textwidth,scale=0.7]{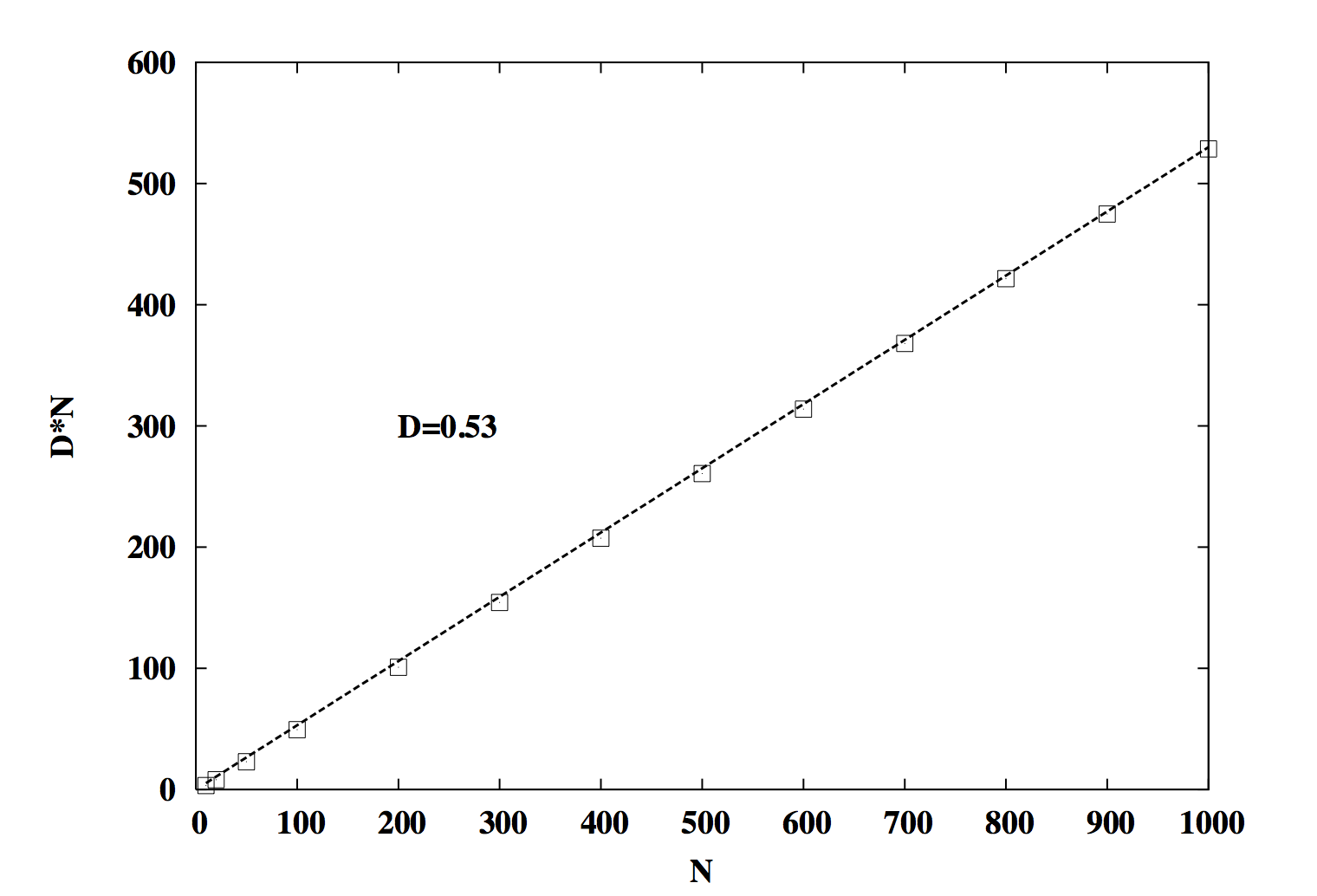}
\caption{\textbf{Distance between GS and OSS}: distance $N \dot D(GS,OSS)$ as a function of $N$. Each point is an average over 1000 realizations. Figure reprinted from \cite{lage2006marriage}, with permission from Elsevier.}
\end{center}
\end{figure}

Finally, it is interesting to see how many individuals improve their situation by passing from the OSS to the GS. By performing numerical simulations, Lage-Castellanos and Mulet found that the $ 24 \% $ of the agents improves their situation while the $ 29 \% $ worsens it. In any case, the improvement of the former is large enough to compensate for the worsening of the latter, hence improving the total happiness of the system.

\subsubsection{Stability of the Ground State}
Now we want to analyze the stability of the GS. A reasonable measure of instability is the number of blocking pairs present in this state \cite{shi2016analysis}. Recall the definition of blocking pair: a matching is stable if there is no man $ i $ and a woman $ \alpha $ who are not married to each other but who both would like to be rather than stay with their current partner. If such a pair were to exist, it would be a blocking pair (BP). \\
With numerical simulations, it is possible to verify that the probability $ P_0 $ that a man $ m_i $ does not form a blocking pair in the ground state is $ P_0 = 0.758 $ (figure 12). In \cite{shi2016analysis} and \cite{omero1997scaling} the authors found analytically that the value of this probability was $ P_0 = 0.59 $. However, this value was based on the wrong approximation that the BPs were independent.
\begin{figure}[!h]
\begin{center}
\includegraphics[width=0.8\textwidth,scale=0.8]{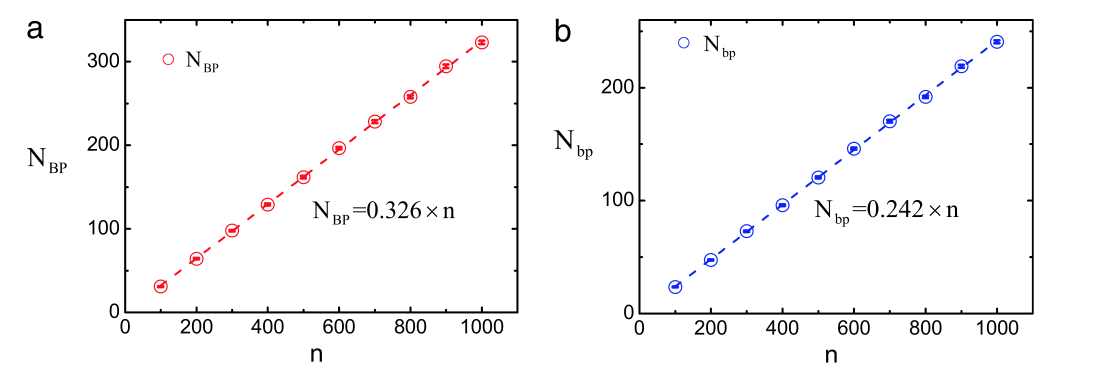}
\caption{\textbf{Number of blocking pairs}: number of blocking pairs as function of the system size $n$. (a) Number of BPs; (b) number of men who form a BP. Each point is an average over 100 realizations. Figure reprinted from \cite{shi2016analysis}, with permission from Elsevier.}
\end{center}
\end{figure}
\begin{figure}[!h]
\begin{center}
\includegraphics[width=0.6\textwidth,scale=0.6]{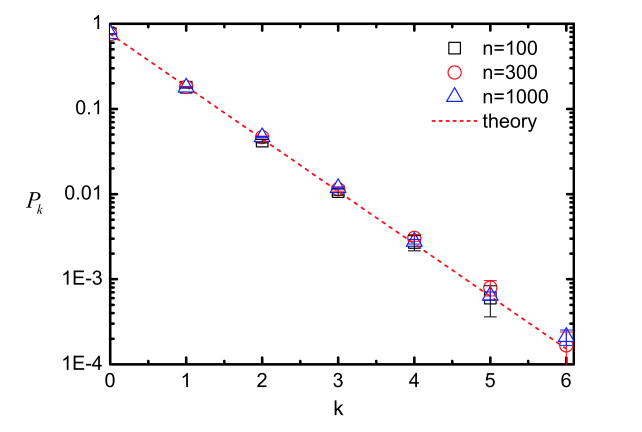}
\caption{\textbf{Probability of forming blocking pairs}: probability that a man forms $k$ Blocking pairs. Figure reprinted from \cite{shi2016analysis}, with permission from Elsevier.}
\end{center}
\end{figure}

Now, we can calculate the probability of a single person forming $ k $ blocking pairs. numerical simulations show that a man has a $ 24.2 \% $ chance of forming a BP on average. Furthermore, among individuals who have already formed a BP, they also have a $ 24.2 \% $ chance of forming a second BP. This process continues until $ k $ blocking pairs are formed. Thus the probability that a single man forms $ k $ BPs is given by the probability that he forms $ k $ BPs by the probability that he does not form a $ k + 1 $ -th BP:
\begin{equation}
P_k=0.758 \cdot (1-0.758)^k \;,
\end{equation}
with $k=0, 1, 2,...$.So in the end we have that the total number of BPs is
\begin{equation}
N_{BPs}=\sum_{k=1}^{\infty}0.758 \cdot 0.242^k \cdot k \cdot n=0.319n \;.
\end{equation} 
This result is in good agreement with numerical simulations.
\newpage

\section{Stable Marriage Problem and Computer Science}
In the previous section, we saw how the SMP is full of non-trivial results if seen through the lens of physicists. In any case, the use of efficient algorithms to verify the analytical results or to find new clues to the problem was of fundamental importance. Furthermore, there are many other problems related to SMP that cannot be solved analytically. For these reasons, we shall study the most important algorithms that have allowed scientists to discover and deepen significant characteristics of the SMP. \\
\\
This section does not intend to be a complete review of all the work on SMP carried out in computer science but only an account of the most important algorithms that have made the history of the problem. We will start with the introduction of some basic concepts of \textit {computational complexity theory} that will be useful for putting the algorithms we will talk about in the right context. \\
Computational complexity theory is that branch of computational theory in which problems and algorithms are evaluated based on the resources necessary to obtain a solution. For an in-depth analysis of this theory it is advisable to refer to the book \textit{Computer and Intractability} \cite{garey1979computers} which has always been a fundamental point of reference.

\subsection{An Introduction to P and NP Problems}
We already mentioned in section 3 that the SMP is a combinatorial problem. But what is meant by \textit{combinatorial problem}? Formally, we can say that a combinatorial problem consists of finding the configurations that accomplish a certain objective within a large configuration space. For example, consider one of the most known combinatorial problems in computer science: \textit {the traveling salesman problem} \cite{lenstra1975some, halim2019combinatorial, beardwood1959shortest}.\\
The problem is the following. Consider $ n $ cities in a Euclidean space, so each city will be characterized by a certain Euclidean distance from all the other $ n-1 $ cities. The goal is to make a complete \textit{tour} of all the cities, passing through them only once, covering the shortest distance possible. In computer science, however, it is known that finding a solution to this problem is a prohibitive task, computationally speaking, even when there are only a few dozen cities. The idea of trying to verify all possible configurations is not feasible: the possible \textit{tours} are $ n! / 2n $ and it takes $ n $ operations to calculate the length of a tour. So already with $ n = 30$ a machine capable of performing one billion operations per second would take billions of years to find a solution. Regardless of how "smart" an algorithm is, the travelling salesman problem remains intractable. The problem is therefore said to be NP-complete. \\
Even in the SMP, regardless of what the goal is, whether it is to find a stable solution or the solution that maximizes total happiness, the number of possible configurations is large. The number of possible matchings is $ n! $, Where $ n $ is the number of men or women. So even in this case, trying to act with brute force by testing all configurations is a failed tactic. In any case, in the SMP there are some objectives (stable solution or maximum happiness) for which it is possible to adopt some tricks that drastically simplify the complexity of the problem. Hence, the problem is called P-complete. \\
In the next paragraphs, we will deepen these two concepts, P and NP, in more detail.

\subsubsection{Worst Case Run Time: P vs NP}
Above we mentioned that the difference between P and NP mainly concerns the number of operations the machine has to perform to find a solution to a problem. How can we strictly define what an operation is and what is processing time? The abstract concept of \textit {Turing machine} is normally used. In its deterministic version, \textit {Deterministic Turing Machine} (DTM), it consists of a set of control states $ s $, a reading and writing tool and memory support. During processing, at each operation, the machine reads an input character $ c $, passes from a state $ s $ to another $ s '$, moves the reading tool by an amount $ q $ and writes another character $ c '$ on the output \cite{karp1972reducibility}. In a deterministic Turing machine the choice of $ s '$, $ c' $ and $ q $ depend only on the current state $ s $ of the machine and on the input data $ c $. When we want to use an algorithm to solve a problem we have to program the DTM and translate the problem into a machine-understandable language through a string $ x $ of input characters. At this point, we define the \textit {run time} $ t (x) $ as the number of operations necessary for a DTM to finish processing and have the solution to the problem. \\
The higher the run time, the more difficult the problem is or the less efficient the algorithm is. With the same input string length $ x $, some cases may be easier than others, so different instances of the same problem could have very different run times. For this reason, the most convenient quantity to consider is the \textit {worst case run time} $ t ^ * (n) = max [t (x)] $. This concept allows us to classify problems: a problem belongs to the $ P $ class if there exists a $ \gamma $ such that $ t ^ * (n) \sim n ^ {\gamma}$. The letter $ P $ stands for \textit{Polynomial} and indicates the class of problems that can be solved in polynomial time with the size of the problem $ n $, regardless of the value of the exponent $ \gamma $. \\
Now, imagine that there is a solution to our problem and that there is an omniscient source able to prove it to us. We say that a problem belongs to the class $ NP $ if the prediction of the omniscient source can be verified in polynomial time with a Turing machine.\\
If a problem can be solved in polynomial time, then the prediction of the omniscient source can certainly be verified in polynomial time. So it is clear that the $ NP $ class contains the $ P $ class. The $ NP $ class can be defined equivalently through the concept of \textit {non deterministic Touring Machine} (NDTM). In simple words, this non-deterministic machine can duplicate itself after each operation and on each copy of itself, it performs a different operation. With such a machine the Travelling Salesman Problem could be solved with a number of operations proportional to the number of cities.\\
We can say that the class $ NP $ constitutes the classes of problems that can be solved in polynomial time with a non-deterministic Touring Machine. In fact, $ NP $ stands for \textit{non deterministic polynomial time}. \\
Therefore \textit{intractable} problems do not belong to the $ P $ class and can be solved with a number of operations that grows more than polynomially with the size of the problem, for example, $t^*(n) \sim e^{\gamma n}$.  

\subsubsection{NP-Completeness}
Determining the relationship between class $ P $ and $ NP $ is a million-dollar problem. While we are writing, no algorithm has been found capable of finding the solution of the Traveling Salesman Problem in polynomial time. Therefore we cannot say that the problem belongs to the $ P $ class, but we cannot even say the opposite because we do not know the necessary algorithm.\\
It is then important to introduce the concept of \textit {reducibility}. A problem $ A $ is reducible to a problem $ B $ if there is a polynomial algorithm that transforms each instance of $ A $ into an instance of $ B $. In this way, the output of a given instance of $ B $ is identical to that of the corresponding instance of $ A $.\\
If the problem $ A $ is reducible to $ B $, then $ B $ must be at least as difficult as $ A $. If $ B $ belongs to the class $ P $ and $ A $ is reducible to $ B $, then also $ A $ belongs to the $ P $ class and therefore can be solved in polynomial time. At this point, the problems that are still intractable can be placed in the $ NP-complete $ (NPC) class: a $ Q $ problem belongs to the $ NPC $ class if it belongs to the $ NP $ class and if every problem of the $ NP $ class it is reducible to $ Q $. By this definition, we mean that if all the problems of the $ NP $ class can be transformed into a $ NP-complete $ problem, then the problems of the $ NPC $ class are the most difficult problems in $ NP $.\\
However, each $ NP-complete $ problem can be reduced to another problem in $ NPC $. There are two possible scenarios \cite{karp1972reducibility}: either the intersection between the classes $ P $ and $ NPC $ is empty and hence $ P \ne NP $; or if there was an algorithm capable of solving a $ NPC $ problem in polynomial time then all the problems in $ NP $ can be solved in polynomial time. Therefore we would have $ P = NP $. Although this last hypothesis is considered improbable, there is still no proof that can confirm it and this problem remains, in fact, a million-dollar problem! \\
To prove that a $ Q $ problem belongs to the $ NPC $ problem class, it is sufficient to show a problem that is $ NP-complete $ reducible to the $ Q $ problem. But one needs to know at least one $ NP-complete $ problem to start the reducibility mechanism in other problems. Explaining such mechanisms is beyond the scope of this paper.\\
\\
It is not necessary, for this review, to go into the details of the computational complexity theory. The excursus carried out so far has served to contextualize, with the correct formalism, the algorithms that we will show in the rest of this section. We shall show SMP algorithms belonging to the $ P $ class and the $ NPC $ class. As already mentioned, the study of computer scientists on the SMP is not all summarized in this section, but we certainly hope to have presented the historically most important and most useful algorithms for understanding the complexity of the SMP.

\subsection{P-Complex Problems in SMP}

\subsubsection{Gale-Shapley Algorithm}
Much effort has been made to build algorithms that quickly find all stable solutions of an SMP of size $ N $. Most of these algorithms are based on the classic Gale-Shapley (GS) algorithm which assigns the role of the proposer to one of the two sets of elements to be paired (men for example), and the role of judges to the other set (women). \\
Here, we see in detail how this algorithm works and we will demonstrate that, as already mentioned, it guarantees the existence of at least one stable solution for any $ N $.\\
\\
The algorithm inputs must be a list of $ N $ men $ \{m_1, m_2, ..., m_N \} $ and a list of $ N $ women $ \{w_1, w_2, ..., w_N \} $, and preference-lists for every man and woman, for example $ P_ {m_i} = (w_6, w_9, ..., w_{13}) $. \\
The \textit{men-oriented} version (that is, with men proposers) of the GS algorithm works in this way: at the beginning of the algorithm, each person is free and gets married during the execution of the algorithm. Once a woman gets married she will never become free again, although she may change partners several times during the run. On the contrary, men can also return free. The following steps are iterated until each man is married: choose a free man $ m $ and he proposes to the first woman $ w $ on his list such that $ w $ has not already refused $ m $. If $ w $ is free then $ m $ and $ w $ get married, if $ w $ is married to a man $ m '$, then she rejects the man she prefers least (between $ m' $ and $ m $) and becomes (or remains) married to the other man. the rejected man becomes, or remains free (see section 1 for a schematic description). \\
When all men are married the algorithm ends and, as we will demonstrate below, the resulting matching is stable.

\subsubsection*{Stability}
We show first of all that the output of the GS algorithm is a perfect matching, that is, every man and every woman are married. We assume absurdly that the resulting matching $M$ is not a perfect matching. So there must be a man $ m $ who is still free at the end of the algorithm. Since the algorithm always chooses a free man who must propose to his favourite woman who has not yet rejected him, this man must have proposed to all $ N $ women and yet he must still be free. This implies that he must either be rejected or left by all women. This means that each woman was married and, as the algorithm is constructed, has always remained married. So there are $ N $ women who have married to $ N-1 $ men (since $ m $ is free), which is a contradiction since it means that at least one man must be married to more than one woman. This can not happen in the algorithm.\\
\\
Now that we have shown that the output is a perfect matching, we demonstrate that such matching is also stable. Assume absurdly that there is an instability in matching $ M $ resulting from the GS algorithm, in particular that there are a man $ m $, a woman $ w $, a man $ m '$ and a woman $ w' $ such that $ m $ is married to $ w '$, $ m' $ is married to $ w $ but that $ P_m (w) <P_m (w ') $ and that $ P_w (m) <P_w (m') $. With this assumption, $ m $ and $ w $ would prefer to get married and to divorce with $ w '$ and $ m' $ respectively. \\
We first observe that $ m $ must have proposed to $ w $, as $ m $ got married to $ w '$ who prefers her less than $ w $. So the only way he came to propose to $ w '$ is after proposing to all the other women he preferred, including $ w $. \\
At the time that $ m $ proposed to $ w $, there were only two possible scenarios.
 \begin{enumerate}
\item If $ w $ had accepted, then the pair $ (m, w) $ would be in $ M $ after the proposal. Since this pair is not in $ M $, $ w $ must have left $ m $ for a man she preferred more. We know that $ m $ got married to $ w '$ and that $ P_w (m) <P_w (m') $, which implies that at some point in the algorithm $ w $ got married to a man she preferred less than her current partner, and this is not possible in the GS algorithm.
\item If $ w $ had refused, she would already have been married to a more preferable $ m''$ man. Since $ w $ got married to $ m '$ she preferred less than $ m $ and therefore less than $ m''$. Then she would have to accept the proposal of a man she preferred less, which is still impossible in the dynamic GS.
\end{enumerate}
We have shown that there can be no instability in $ M $ and, therefore, that $ M $ is a stable matching.

\subsubsection*{Run-Time Analysis}
Let us now analyze the execution time of the algorithm. We want to demonstrate that, at worst, there will be $ N ^ 2 $ proposals during the execution of the GS algorithm.\\
No man makes a proposal to a woman more than once. Since there are $ N $ men and $ N $ women, there will be a maximum of $ N ^ 2 $ proposals. \\
It is very positive because, as mentioned above, a brute force algorithm would take, in the worst case, a time equal to $ N! $ to find a stable solution. The GS algorithm significantly reduces this time.

\subsubsection*{One Side Optimality}
Before revisiting the algorithms that find all the solutions of the SMP for a given $ N $, let us analyze an important aspect of the GS algorithm. It can be shown that the matching $ M $ resulting from the men-oriented version of the GS algorithm is optimal for men and the worst possible for women. A matching is \textit{men-optimal} if each man receives the best partner than all the other stable matching; instead, it is \textit{men-pessimal} if each man receives the worst partner than all other stable matchings.\\
We show men-optimality by contradiction: in the GS dynamic men make proposals in order and assume absurdly that at least one man is rejected by a valid partner (and therefore does not receive the best partner). We also assume that $ m $ undergoes this first refusal from $ w $ in $ M $. this is because $ w $ has chosen a man $ m '$ that she prefers to $ m $. Imagine that $ M '$ is a stable matching in which the pairs $ (m, w) $ and $ (m', w ') $ exist. By assumption, $ m '$ was not rejected by a woman valid in $ M $ before $ m $ was not rejected by $ w $. Therefore $ m' $ prefers $ w $ to $ w '$. But knowing that $ w $ prefers $ m '$ to $ m $, $ m' $ and $ w $ form a blocking pair in $ M '$ and this is not possible.\\
\\
Similarly, we demonstrate that the men-oriented version of the GS algorithm is the worst possible for women: we assume absurdly that $ m $ and $ w $ are a pair in $ M $ and that $ m $ is not the worst possible partner of $ w $. There is, therefore, a stable $ M '$ matching where $ w $ is married to a man $ m' $ better than $ m $. Let us say that $ w '$ is the partner of $ m $ in $ M' $. For the man-optimality $ m $ prefers $ w $ to $ w '$. Hence, we come again to the absurd conclusion that the pair $ (m, w) $ is a blocking pair in $ M' $.\\
\\
If the roles are reversed (women-oriented GS), the GS dynamic finds a stable solution that is optimal for women and very bad for men. So through the GS algorithm, you can find two stable solutions: the men-optimal solution $ M_m $ and the women-optimal solution $ M_w $. All the other solutions are in between these two extremes. This fact is exploited to build algorithms that find all SMP solutions. if the solution is only one then it must be true that $ M_m = M_w $.

\subsubsection{Finding All the Stable Solutions}
The GS algorithm is a useful tool for finding a particular solution to the problem. In general, given the preference-lists, there are more stable solutions. We now present the main algorithms that find all the stable solutions efficiently, i.e. in a faster time than $ N! $ (With brute force).\\
\\
A simple but not so efficient way of finding all the solutions is that proposed by Wirth who uses a trial-and-error and backtracking method \cite{wirth1986algorithms}. The idea of this method is to find solutions to specific problems not by following a fixed calculation rule, but by trial and error. The common pattern is to break down the trial and error process into partial activities. Often these tasks are expressed more naturally in recursive terms and consist of exploring a finite number of secondary activities. Generally, the whole process is a trial process that gradually builds up and scans (prunes) a subactivity tree. The main feature is that the steps towards the total solution are attempted and recorded. In this way, they can be subsequently taken up and cancelled in the records when it is discovered that the passage probably does not lead to the total solution, that is, that the passage leads to a dead end. This action is called backtracking. In the case of SMP, therefore, the idea is to run through the man $m$'s preference list until an acceptable partner is found. If an acceptable partner is found for all $ m_i $, then the solution is registered, otherwise, the marriage is cancelled.\\
\\
A much more efficient algorithm is the one proposed for the first time by  Mc Vitie and Wilson \cite{mcvitie1971stable, wilson1972analysis}, and it is worth explaining it in more detail. This method exploits the GS dynamic and the concept of \textit{breakmarriage}: it consists in breaking the marriage of a selected man $ m_M $, in a stable solution $ M $, and forcing him to marry a worse choice from his list. As a result, the woman $ w_M $ (married to $ m_M $ in $ M $) has the opportunity to get a better marriage. As previously demonstrated, in this way it is possible to start from the men-optimal solution to get to the women-optimal solution. \\
Once a breakmarriage has been carried out, the algorithm restarts again according to the GS dynamic. The process ends either when the woman $ w_M $ receives a proposal from a man better than $ m_M $, or when a man has no more choices, that is, he has been rejected by all the $ N $ women. In this second case, the breakmarriage process fails.\\
It can now be shown that, under certain conditions, a breakmarriage operation that does not fail leads to a new stable solution and that all stable solutions can be found with subsequent applications of the breakmarriage operation, starting from the men-optimal solution. The two conditions necessary to demonstrate these assumptions are that
\begin{enumerate}
\item starting from a stable solution obtained with a breakmarriage operation on man $ m_i $, the subsequent breakmarriage operations must be carried out either on $ m_i $ or on subsequent men (i.e. to whom a breakmarriage has not yet been applied)
\item  if after the breakup of the $ m_i $ marriage the process interferes on a man $ m_j \ne m_i $ to whom the breakmarriage has already been applied, then the process ends and it has failed.
\end{enumerate} 
We, therefore, show that when a breakmarriage operation is applied to a stable solution $ M $, the new matching resulting after this operation will also be stable. First of all, we note that all the couples that have not been affected by the breakmarriage will be stable as they were stable in the previous matching. When breakmarriage affects couples and at the end of the process a man $ m_i $ marries a woman $ w_i $, if that man $ m_i $ prefers the woman $ w_j $, then she should have received a proposal from him and rejected him in favour of a better partner. In the breakmarriage operation, women continue to get better partners, so the woman $ w_j $ will prefer her current partner to man $ m_i $, therefore the matching is stable.\\
\\
With similar reasoning, one can show that starting from the men-optimal solution it is possible to obtain \textit{all} stable solutions once and only once through successive applications of the breakmarriage operation. \\
This method is much more convenient than the previous trial-and-error-based method, and it can be shown to take at least $O(N^3|S|/(log(S^2)))$ time, and no more than $O(N^3|S|)$ time, where S is the set of stable marriages. Knuth \cite{knuth1997stable} proposed a similar algorithm which has the same complexity.\\
\\
An even more efficient algorithm is due to Gusfield, and it is based on the concept of "rotation" \cite{gusfield1987three}. This algorithm is more complicated and less transparent than the previous ones, here we will expose only the main concepts. \\
Let $ M $ be a stable matching, for each man $ m $, be $ w '$ the first woman on the list of $ m $ such that $ m $ prefer his current partner in $ M $ and at the same time $ w' $ prefer $ m $ to her current partner in $ M $. Let now $ m '$ be the man $ w' $ is married to in $ M $. At this point a rotation $ R $ is defined to be an ordered list $ R = \{(m_1, w_1), (m_2, w_2), ..., (m_z, w_z) \} $ such that for every $ i $ from 1 to $ z $, $ m'_i $ is $ m_ {i + 1} $. One can show that except the stable pairs in the women-optimal solution (which are not in a rotation), each stable pair is exactly in a rotation and vice versa each pair in a rotation is stable \cite{irving1986complexity}. It can also be shown that it is possible to find all rotations in a time $ O(N ^ 2 )$. The central theorem for finding all stable solutions is the following \cite{irving1986complexity}:\\
\\
\textit{Let $S$ be the set of all stable marriages for a given problem instance (i.e. given the preference lists), and let $D$ be the corresponding directed graph formed from the set of all rotations. Then there exists a one-to-one correspondence between $S$ and the family of closed subsets in $D$, i.e. each closed subset in $D$ specifies a distinct stable marriage, and all stable marriages are specified in this way.}\\
\\
This theorem tells us that there is a univocal relationship between rotations and stable solutions which is exploited in Gusfield's algorithm and it can be shown that this algorithm finds all stable solutions in a time $O(N^2+N|S|)$. So the basic idea of this algorithm leading to the improved running times is to exploit theorems about the structure of stable marriages to avoid back-up and duplicated work inherent in earlier algorithms.

\subsubsection{Minimum Regret and Egalitarian Solutions}
We have shown that the SMP generally has multiple solutions and that the number of solutions $ S $ grows with $ N $. Thanks to the GS theorem, we have seen that in the set of stable solutions there are two particular solutions (unless the solution is only one): the men-optimal solution and the women-optimal solution, which correspond respectively to the women-pessimal solution and the men-pessimal solution.\\
So the GS algorithm provides two solutions that satisfy only men or only women and hence they are inappropriate for most applications in the real world. It would be appropriate, for example, to find a matching that not only is stable but also fair compared to men and women \cite{roth1992two, gusfield1989stable}; or a stable solution that minimizes the total cost (i.e. the sum of the cost for each agent: $ X + Y $). So it is natural to look for solutions that, in addition to being stable, are also "acceptable" according to certain criteria. In particular, Knuth \cite{knuth10mariages}, Polya et al. \cite{polya2013notes} have proposed three main criteria:
\begin{enumerate}
\item minimization of the \textit{sex-equalness cost}.
\item minimization of the \textit{regret cost}.
\item minimization of the \textit{egalitarian cost}.
\end{enumerate}
We now define the second and the third criteria and summarize the main results on the algorithms to meet these criteria, i.e. we present the main works regarding two stability criteria proposed in \cite{knuth10mariages, polya2013notes}: the egalitarian SMP and the minimum regret SMP. These two problems, although also complicated, were found to be less difficult than the \textit{equitable stable marriage problem} (ESMP), as we will see later. While we have seen that the latter is NP-hard, the two problems that now we will deal with have been solved in polynomial times.
\\
The egalitarian stable marriage problem requires finding the optimal SMP solution that minimizes the total energy of the system while maintaining stability:
\begin{equation}
E_{TOT}=\sum_{i=1}^{N} x_i + \sum_{i=1}^{N} y_i=X+Y \;. 
\end{equation}
The minimum stable total energy can be found by generating all the SMP solutions and comparing their energies. The problem with this approach is that the maximum number of SMP solutions grows exponentially with the size $ N $ of the system. In \cite{irving1987efficient}, Irving, Leather and Gusfield built an algorithm $ O (N ^ 4) $ to solve the problem of the minimum of $ E_ {TOT} $. They took advantage of the aforementioned rotations theorem which states that \textit {in SMP, stable solutions have a unique correspondence with closed subsets of rotations} and have exploited the result that all rotations can be found in a time$O(N^3)$. Later, Feder \cite{feder1992new} improved this algorithm arriving at a $ O(N^3) $ running time.\\
\\
Regarding the minimum regret stable marriage problem, it concerns the minimization of the maximum regret $ R_{max} $ in a stable matching $ M $, where the regret of the man $ m $ (woman $ w $) corresponds to the ranking of the woman (man) married to $ m $ ($ w $) in $ M $ on his (her) preference list:
\begin{equation}
R(M)=max_{(i,j) \in M }(max\{x(i,j),y(j,i)\}) \;.
\end{equation}
The first to discuss the $ R $ minimization was Knuth in \cite{knuth10mariages} where he showed the solution proposed by Alan Selkow in which an algorithm with a running time of $ O(N^4) $ is shown. A more efficient solution, instead, was proposed by Gusfield in \cite{gusfield1987three} in which the breakmarriage operation and its properties are used to avoid that there are duplicate proposals or rejection in the algorithm, to increase the speed of execution time to $ O(N^2) $.   

\subsection{NP-Complex Problems in SMP}
\subsubsection{SMP with Incomplete Lists and Ties}
In section 7 we will interpret the incompleteness of preference-lists as a limitation of individuals to process information. Indeed, the concept of "partial information" was introduced by Zhang in \textit{zhang2001happier} but the SMP with incomplete preference-lists has been studied under another interpretation by mathematicians and computer scientists: the lists of individuals are incomplete not because of lack of information, but because some individuals are "unacceptable" as partners for other individuals. This problem, known in the literature as SMI (Stable Marriage Incomplete), differs from what we present in section 7 in that an individual can marry \textit {only with the individuals present in his preference-list}. In the SMI the "unacceptability" is not symmetrical (the man $ m $ might consider the woman $ w $ unacceptable but, in general, the vice versa is not also valid). \\
The goal of the SMI is therefore to find a stable matching in which individuals accept each other and therefore individuals may remain single.\\
However, All the solutions of an SMI have the same number of pairs \cite{roth1986allocation}, and these solutions are found in a polynomial time.\\
 \\
A more complicated extension of SMI is SMTI: the SMP with incomplete list and ties. The addition of ties implies that two or more individuals may be indifferent to another individual, that is, they are interchangeable in his/her preference-list. \\
In the SMTI it is possible that there are solutions with different numbers of married couples and the goal of the problem is to find the solution with the maximum number of married couples. SMTI is NP-hard \cite{manlove2002hard, munera2015local}. \\
Iwama et al \cite{iwama20062, iwama201425} have constructed interesting approximation algorithms for SMTI. Other approximation algorithms are shown in \cite{irving2008approximation, iwama20071, mcdermid20093, manlove2013algorithmics}. \\
Another way to approach the solution is to use the local search approach as in \cite{gelain2010local, gelain2010local, gelain2011procedural, downey1999parametrized} in which the authors exploited GS dynamics and based a search pattern of the first neighbours to minimize the distance from stability. The algorithm starts from random matching and at each step, it is moved to the closest matching which minimizes the distance from stability, measured as the number of \textit{blockingpairs}. The algorithm ends when a stable solution is found.\\
\\
SMTI is an interesting problem since it shows that despite the two separate problems (only incomplete list or only ties) can be solved in polynomial times, nevertheless, if put together the problem is no longer so simple, that is, it becomes NP-hard.

\subsubsection{Equitable Stable Marriage Problem}
In 1989 Gusfield and Irving \cite{gusfield1989stable} proposed the "equitable stable marriage problem" (ESMP) which requires finding the stable SMP solution that minimizes "the distance" between the energy of men and of women to avoid discrimination between the two sides. Since for many real-world applications, ESMP is more appropriate than the classic SMP, it has attracted a lot of attention \cite{kato1993complexity, iwama2010approximation, tyagilocal, morge2011privacy, everaere2012casanova, everaere2013minimal}. \\
In ESMP we want to minimize is the \textit{sex-equality cost} $ D_{sec} $, defined in this way:
\begin{equation}
D_{sec}(M)=|\sum_{i=1}^{N}x_i-\sum_{i=1}^{N}y_i| \;,
\end{equation} 
where $ x_i $ and $ y_i $ are still the costs of marriages of man $ i $ and woman $ i $ in matching $ M $ respectively. \\
\\
Unfortunately, no algorithm solves the $ D_{sec} $ search in polynomial time: in \cite{kato1993complexity} Kato has shown that this problem is strongly NP-hard. \\
More recent works have used algorithms with approximations and heuristic methods for this problem \cite{aldershof1999refined, roth1990random, ma1996randomized, piette2013swing++, zavidovique2005novel}: for example in \cite{tyagilocal} Gelain et al. have proposed a local search algorithm to find a solution different from the men-optimal and women-optimal solution for small $ N $. Therefore, with a high probability, this solution is the fairest, without however showing that it corresponds to the $ D_ { sec} $. Roth and Vande \cite{roth1990random} have shown that starting from an arbitrary matching and randomly pairing blocking pairs, the result achieved is a stable matching with probability 1. In this way, it could transform an unstable matching into a stable one different from the two extreme solutions men/women optimal and for small $ N $ there is a good chance that such matching is the fairest. Unfortunately, even in this case, there are no certain guarantees to find a fair solution. In \cite{iwama2010approximation} Iwama et al. built an approximation algorithm with a time $ O (n ^{3 + 1 / \epsilon}) $ to obtain a stable solution such that $ D_{sec} \le \epsilon \cdot min\{M_m, M_w \}$, where $ D (M_m) $ and $ D (M_w) $ are the distances between the energies of men and women in the men-optimal and women-optimal solution respectively and $ \epsilon $ is a constant. Later Everaere et al. \cite{everaere2012casanova, everaere2013minimal} have built a heuristic algorithm that manages to achieve a fair solution by allowing both men and women to make proposals repeatedly. The problem with this algorithm, however, is that it could proceed for an indefinite number of iterations even for small $N$.\\
More recently some improvements have been made in \cite{giannakopoulos2015equitable}: a heuristic algorithm has been built based once again on the idea that both parties can propose, this time however in a controlled manner through a non-state dependent and non-periodic function.

\subsection{Algorithms for the Ground State Solution}
In section 3 we introduced the analytical methods that have been developed to study the ground state in the matching problem, here we want to study the algorithms that allow finding this ground state, given any instance of the problem. \\
The main algorithms we will talk about have been developed for the so-called "assignment problem" which, as we will see, has the same structure as the SMP if one looks for optimization.\\
The assignment problem is defined as follows: imagine that there are $ N $ workers $ i $ and $ N $ jobs $ j $ and assume that it is given a score for the skills that each worker has for each job. In this way, each pair $ (i, j) $ will have a certain score. The problem requires finding a matching (each worker is assigned one and only one job) that maximizes the total score. SMP is different in that the assignment problem has no preference-lists and requires no stability whatsoever. However, it is not difficult to rewrite the SMP as the assignment problem: there are $ N $ men $ m $ and $ N $ women $ w $ each with a preference list on individuals of the opposite sex. In this way each pair $ (m_i, w_j) $ will have a certain score given by the sum of the ranking of the woman $ w_j $ in the list of $ m_i $, previously defined as $ x (i, j) = x_i $, and of the ranking of man $ m_i $ in the list of $ w_j $, previously defined as $ y (j, i) = y_j $. Then we ask to minimize $ E_ {TOT} ^ *=X+Y $. Note that we did not mention anything about stability. \\
The two problems are equivalent: returning to the terminology of the assignment problem, build a matrix $ A $ that is $ N \times N $ where the rows are the workers and the columns the jobs. So the element $ A_ {i, j} $ represents the score assigned to the worker-job pair $ (i, j) $. Finding a matching that maximizes the total score corresponds to finding $ N $ elements of $ A $ that are independent, that is, that there is only one in each row and each column, and that the sum is the maximum possible. In SMP terminology it is enough to replace, for example, workers with men, jobs with women and instead of maximizing the sum of the independent elements of $ A $, we must try to minimize it. It is easy to demonstrate that the maximization and minimization processes are equivalent: let $ p = max\{A_ {ij} \} $ if you define the matrix $ A'$ as:
\[A_{ij}'=p-A_{ij} \;, \]
then maximizing $ N $ independent elements of $ A $ corresponds to minimizing $ N $ independent elements of $ A'$. \\
To remain consistent with the literature on the subject, in this paragraph we will refer to the assignment problem and therefore we will talk about maximizing $ N $ independent elements of $ A $, keeping in mind that minimizing is equivalent.\\
\\
The basis of all algorithms to find the solution to the assignment problem is the so-called "Hungarian method" and it is for this reason that in the following paragraphs we will analyze this algorithm in detail, while its variants and its most recent improvements will only be mentioned.

\subsubsection{General Procedure}
The most direct approach to solving the assignment problem would be to use a brute force approach and calculate all possible matching, i.e. all possible combinations of independent elements of $ A $, and choose the one with higher total score $ P_ {max} $. This is only feasible for very small $ N $, as the number of possible matches is $ N! $ and therefore for large $ N $ a brute force approach is unthinkable. \\
The first to find an efficient algorithm for the assignment problem was Kuhn in \cite{kuhn1955hungarian}. His approach is based on two important results \cite{munkres1957algorithms}: 1) Konig's theorem which states that if in an $ A $ matrix, $ m $ is the maximum number of independent elements of $ A $ equal to 0, then there are $ m $ lines (rows or columns) that contain all elements equal to 0 of $ A $; 2) the solution to the problem does not change if the elements $ A_ {ij} $ are replaced with the elements $ B_ {ij} = A_ {ij} -u_i-v_j $, with $ u_i $ and $ v_j $ arbitrary constants.\\
The main steps of the Kuhn algorithm, which is called in the literature as the Hungarian algorithm, can be summarized as follows \cite{flood1956traveling}:
\begin{enumerate}
\item subtract from each element of $ A '$ the smallest of its elements, obtaining a matrix $ A_1' $ which has all non-negative elements and at least one equal to 0.
\item find a minimum set $ L_1 $ of $ N_1 $ lines (rows or columns) containing all the zeros of $ A_1 '$. If $ N_1 = N $, there is a set of $ N $ independent zeros and the elements of $ A '$ in these $ N $ positions correspond to the solution of the problem.
\item if $ N_1 <N $, call $ s_1> 0 $ the smallest element of $ A_1 '$ which is not in $ L_1 $. For each line in $ L_1 $ add $ s_1 $ to all elements of that line and then subtract $ s_1 $ from all elements of $ A_1 '$. Call the new matrix $ A_2 '$.
\item Repeat steps 2 and 3 using $ A_2 '$ instead of $ A_1' $. Since with each application of the third step the sum of the elements of the matrix decreases by $ N (N-N_k) s_k $, then the algorithm certainly ends after a finite number of steps.
\end{enumerate}
To complete the algorithm it is necessary to know a procedure for step 2, that is, to know how to find a minimum set of lines that contains all the zeros and therefore a maximum set of independent zeros (Konig theorem) .\\
 In the following paragraphs, we will see how the Hungarian method solves this problem. \\
For completeness, it is worth knowing that there are several variations of the Hungarian algorithm and other more recently developed algorithms that solve the assignment problem, for example in \cite{jonker1986improving, wright1990speeding, shah2015improvement}.

\subsubsection{The Hungarian Method}
As Kuhn himself wrote in \cite{frank2005kuhn}, the idea of the Hungarian algorithm came to him while reading Konig's book on graph theory in the summer of 1953. If a matrix is composed only of 0 or 1, then Konig's theorem states that the minimum number of rows and columns containing all 1's is equal to the maximum number of independent 1's. The primitive problem for which Konig enunciated his theorem was a particular version of the assignment problem, called simple assignment problem (SAP), in which the score that workers receive for a given job can only be 0 or 1. \\
In a footnote of the Konig's book, the author referred to an article by E. Egervary (written in Hungarian) \cite{egervary1931matrixok} in which a simple computational method was provided to reduce the more general assignment problem (which from now on we will call GAP) to SAP. Kuhn, reading the article in Hungarian, had the idea which he later gave the name of the Hungarian algorithm. \\
We now introduce the theoretical bases of the algorithm, studying the reduction from GAP to SAP.

\subsubsection*{Simple Assignment Problem}
The SAP can be formulated as follows: build a $ Q $ matrix in which the columns represent the jobs and the rows the workers. $ Q_{ij} = 1 $ if worker $ i $ is qualified for the job $ j $, otherwise $ Q_{ij} = 0 $. The problem requires finding the maximum number of jobs that can be assigned to qualified workers, i.e. the maximum number of independent 1's in $ Q $. \\
Following Kuhn's terminology in \cite{kuhn1955hungarian} we will call \textit{complete} any assignment such that it is no longer possible to assign an unassigned worker to an unassigned job for which he is qualified. However, an assignment can be improved through a \textit {transfer}, namely by moving a worker already assigned to another job for which he is qualified. Also, we will call all workers involved in a transfer \textit{essential} and we will also call all jobs to which a non-essential individual has been assigned in the same way. It follows that if we conventionally define the operation of leaving everything unchanged as transfer, then in this case all jobs will be essential. It is not difficult to prove the following theorems \cite{kuhn1955hungarian}:
\begin{itemize}
\item \textit{1}: For a given assignment, if all transfers lead to a complete assignment, for each worker qualified for a job, either the worker or the job is essential, or possibly both.
\item \textit{2}: There is always an assignment that is complete after every possible transfer.
\end{itemize}
A dual aspect of the problem can now be defined. Consider a budget to be given to the value of an individual assigned to a job for which he is qualified. This budget consists of a unit or nothing to be delivered to every worker and every job. A budget is \textit{adequate} if, for each worker qualified for a job, either a worker, job or both were given a unit. The following theorems \cite{kuhn1955hungarian} can be proved: 
\begin{itemize}
\item \textit{3}: The total sum of the budget (adequate) delivered is greater than or equal to the maximum number of jobs that can be assigned to qualified workers.
\item \textit{4}: There is an adequate budget and an assignment such that the total sum of the budget delivered is equal to the number of jobs assigned to qualified workers.
\end{itemize}
Theorem 3 implies that the assignment of theorem 4 is optimal, i.e. it is the SAP solution. So the maximum number of jobs that can be assigned to qualified workers is equal to the minimum possible sum of the adequate budget delivered, and that each assignment is optimal if and only if it is complete after each possible transfer (theorems 1 and 2). \\
Let us see now how to reduce the GAP to the SAP just described.

\subsubsection*{General Assignment Problem}
We described the GAP previously, so we can directly introduce its dual. Consider an \textit{adequate budget}, that is an amount of non-negative integers (for simplicity) $ u_i $ to be delivered to each worker and $ v_j $ to each job, such that the total amount delivered to worker $ i $ and job $ j $ is not less than $ A_{ij} $:
\[ u_i+v_j \ge A_{ij} \;, \]
for every $ i, j = 1, ..., N $. The analogue of Theorem 3 can be stated in this way \cite{kuhn1955hungarian}:
\begin{itemize}
\item \textit{Theorem 3'}: the total sum of the adequate budget delivered is greater than or equal to the total score of each assignment. 
\end{itemize}
Therefore, the assignment that has the total sum of the adequate budget delivered equal to the total score of the assignment is the solution of the GAP. \\
\\
A very convenient way to map the GAP to SAP is as follows: worker $ i $ is qualified for the job $ j $ if and only if $ u_i + v_j = A_{ij} $. The following theorems immediately follow \cite{kuhn1955hungarian}:
\begin{itemize}
\item \textit{5}: If all the $ N $ workers can be assigned to the jobs for which they are qualified in the SAP associated with the corresponding GAP with an adequate budget. Then the sum of the scores is equal to the total budget delivered and therefore the resulting assignment is the solution of the GAP.
\item \textit{6}: If at most $ N_1 <N $ workers can be assigned to the jobs for which they are qualified in the SAP associated with the corresponding GAP with an adequate budget, then a positive integer amount can be subtracted from the total sum of the budget delivered.
\end{itemize}
In particular, the subtracted sum to which theorem 6 refers can be deduced in this way: consider $ p $ essential individuals and $ q $ essential jobs. Then the following budget changes can be made:
\[u'_1=u_1,..., u'_p=u_p, u'_{p+1}=u_{p+1}-1,..., u'_N=u_N-1 \;, \]
\[v'_1=v_1+1,..., v'_q=v_q+1, v'_{q+1}=v_{q+1},..., v'_N=v_N \;. \]
Note that the total sum of the budget has decreased, in fact the budget has increased by $ q $ and decreased by $ N-p $ and, noting that $ p + q = N_1 $, then the subtracted sum is $ N-p-q = N-N_1 > 0 $.\\
Therefore, starting from any adequate budget, either the assignment is optimal and therefore theorem 5 is valid, or theorem 6 is applied, and the budget can be decreased. Since the budget can decrease a finite number of times, then the highest possible sum of the scores among the possible assignments is equal to the minimum of the adequate budget delivered. This minimum can be found by solving a finite SAP associated sequence. Moreover, the maximum number of steps to find the optimal solution is in the order of $ O (N ^ 3) $ \cite{munkres1957algorithms}. \\
The proofs of all theorems set out in this paragraph have been omitted but can all be found in Kuhn's original article. In the same paper, Kuhn shows a practical example of this method that can help understand this abstract framework.\\

\subsection{Algorithms for the SMP Generalizations}
\subsubsection{College Admission Problem}
The first generalization of the SMP that we will deal with is the "college admission problem" (CAP), sometimes also called "resident hospitals problem" (RHP). While in SMP the goal is to find one-to-one stable matching, in CAP the goal is to find many-to-one stable matching. The metaphor that is commonly used for many-to-one matching is that of students that must apply to various colleges and each college has a limited number of students that it can accept. Similarly, one can think of the original problem described in \cite{irving1998matching} in which medical students must be assigned to different hospitals. \\
The original model was first described in \cite{10.2307/2312726} by Galey and Shapley and is as follows: a set of $ N $ students must be assigned to $ M $ colleges and $ q_i $ is the maximum number of students that a college can accept, also called "quota". Each applicant ranks colleges in order of preference, omitting those colleges they would not accept under any circumstances. Similarly, each college builds a ranking of students in the system, omitting those who would not be admitted under any circumstances. For simplicity, in their original article Galey and Shapley did not consider ties. \\
The definition of "instability" is analogous to that of SMP: a matching between colleges and student is said to be unstable if there are two students $ s_1 $ and $ s_2 $ who are assigned respectively to colleges $ c_1 $ and $ c_2 $ despite $ s_1 $ prefers $ c_2 $ and, at the same time, $ c_2 $ prefers $ s_1 $. So exactly as in the SMP, matching is unstable if it can be disturbed by a college and a student acting together to both benefits. The objective of the CAP is to find a stable matching, i.e. that it is not unstable. \\
Another definition analogous to the case of the classic SMP is that of optimality: a stable matching is defined as optimal if each student is accepted by the best college from which they could be accepted in any other stable matching. This definition of optimality, however, refers only to students. Indeed it would be more appropriate to define it as "students-optimality". There is also a definition of "colleges-optimality" even though the CAP is not symmetric to colleges and students while the SMP was symmetric (as it consisted of one-to-one matching). We will deal with this issue in the next paragraphs. \\
As in the case of SMP, there are several variations of the original model \cite{chade2006college, rios2014college, gale1985some, irving2000hospitals, irving2003strong, roth1984evolution} which, however, we will not describe in detail in this review. \\
In this paragraph, we see what are the main algorithms that allow to find a stable solution to the CAP.

\subsubsection*{Deferred-Acceptance Procedure}
In \cite{10.2307/2312726} the authors propose a natural extension of the classic GS algorithm that develops in the following way: for simplicity, it is assumed that if a college does not accept a certain student in any circumstances then that student will not be able to apply for that college. At first, all students apply to colleges that correspond to their first choice. So a college with a quota $ q $ will place their $ q $ (or less) best applicants on their waiting list and will reject the others. The procedure ends when all students are either on a waiting list or have been rejected by all colleges they are allowed to apply.\\
Now, each college will accept all students on its waiting list and the resulting matching turns out to be stable. Therefore, there is also at least one stable solution in the CAP. The demonstration of stability is analogous to the demonstration of stability for SMP.\\
\\
One can also show that this procedure produces not only stable but also "student-optimal" matching. The demonstration was first described by Gale and Shapley in \cite{10.2307/2312726} by induction. A college to which a student could be assigned in a stable matching is called "possible" for the student. Assume that, at a certain point in the development of the algorithm, no student has been removed from a college that is impossible for him. Now assume that college C accepts students $ \{s_1, s_2, ..., s_q \} $ and rejects student $ s' $. We must show that C is impossible for $ s' $. The student $ s_i $ prefers C to all the other colleges, except those who have already refused him and therefore are impossible for him by assumption. Consider a hypothetical matching where $ s' $ is assigned to C. At least one of the students $ \{s_1, s_2, ..., s_q \} $, say $ s_i $, will be assigned to a less preferred college than C. But this matching is unstable because $ s_i $ and C prefer each other and they would disturb the system. So C is impossible for $ s' $. In conclusion,  The deferred-acceptance (DA) procedure rejects \textit{only} those students from the colleges that are not possible in any stable matching and therefore the resulting matching is "student-optimal".

\subsubsection*{NRMP Algorithm} 
Actually, there was already a procedure very similar to the DA that was used in the "National Residence Matching Program" (NRMP) to assign medical students to hospitals. The NRMP method is equivalent to the DA procedure but it works inversely \cite{gale1985some}: the matching obtained is "college-optimal" rather than "student-optimal". The algorithm works in the following way: each college with quota $ q $ admits $ q $ preferred students. Students who are admitted to multiple colleges choose only their favourite and therefore are deleted from the list of colleges that they have not chosen. At this point, colleges that have fewer students than their quota choose the other preferred students until they reach their quota or until they have exhausted their preference-lists. Once again, students accepted into more than one college accept only their favourite college, and so on. The procedure ends when no college can admit other students (either because they have reached their quota or because they have exhausted their preference list). The resulting matching is stable and it is "college-optimal".\\
\\
Based on these two procedures (DA and NRMP), Gale and Sotomayor \cite{gale1985some} demonstrate the following results:
\begin{enumerate}
\item Although, for students, the NRMP procedure is worse than the DA procedure, it is true that all students admitted to a college with the DA procedure will also be admitted to that college in the NRMP procedure. Vice versa, even if- for colleges- the DA is worse than the NRMP, the number of students $ n (C) $ admitted in college C will be the same both in the DA procedure and in the NRMP procedure.
\item If a college extends the list of eligible students, then this will make the resulting matching (both for DA and NRMP) better for students but worse for colleges.
\item Student-optimal matching is \textit{Pareto optimal}: that is, there are no matching (stable or unstable) that are better for \textit{all} students than student-optimal matching. But this is not true concerning college-optimal matching.
\end{enumerate}

\subsubsection{Stable Roommate Problem}
So far, we have considered the case in which the elements to be joined belong to two disjointed sets (men and women). In other words, we have dealt with a "bipartite matching problem".\\
Now let us imagine we have a system of $N$ individuals (no matter if male or female) that we have to fix in pairs in $ N/2 $ rooms: this type of problem is commonly called \textit{stable roommates problem} (SR) \cite{irving1985efficient, pittel1993stable}. This problem is the non-bipartite version of the SMP, it was first described in the papers of Galey, Shapley and Wilson \cite{wilson1972analysis, 10.2307/2312726} and is formally defined in this way: there are $ N $ people, each of whom has a preference list of the other $ N-1 $ people. The goal is to find a matching in which stability is defined in a similar way to the case of stable marriage, that is, there must not be two individuals who are not in the room together but who both would prefer to be (i.e. no blocking pairs). As for the SMP, the practical applications of this problem are manifold \cite{irving2007cycle, kujansuu1999stable, roth2005pairwise}. \\
The main difference with the SMP is that, while the latter always has at least one stable solution, the SR may have no solution: as we will see later, Irving \cite{irving1985efficient} has proposed an algorithm that recognizes whether a stable solution exists and, if so, how to find it. \\
Also for this problem, there are several variants: it has been demonstrated, for example, that finding a matching that is "almost stable", that is, that presents the minimum number of blocking pairs (defined in a similar way to the SMP), is NP-hard and difficult to approximate \cite{abraham2005almost}; it has been shown that if ties are admitted in the preference lists, the problems of determining whether a stable solution exists or not is NP-complete even if complete lists are admitted \cite{ronn1990np}; on the other hand, it has been shown that the same problem can be solved in polynomial times under different stability conditions, such as super stability or strong stability \cite{scott2005study, irving2002stable}. \\
There are also extensions similar to those seen for the SMP in which, for example, the rooms must contain more than two individuals (multi-dimensional SR \cite{tongmulti}).\\
\\
In the following, we will show in more detail how the Irving algorithm works to emphasize the main differences with the SMP, in particular showing that there can be no stable solution.

\subsubsection*{An Algorithm for the Stable Roommates Problems}
As already mentioned, if the problem is not bipartite, there may be no stable solutions. This was first demonstrated by Galey and Shapley. They showed an example with $ N = 4 $ individuals with the following preference lists:
\[m_1=\{2,3,4\}\] 
\[m_2=\{3,1,4\}\] 
\[m_3=\{1,2,4\}\] 
\[m_4=\{any\} \;,\]
and, as it can be easily proved, any individual placed in the room with individual 4 will constitute a blocking pair for the system. \\
On the other hand, Knuth \cite{knuth10mariages} showed that, as in the SMP, multiple solutions can exist; this was shown through an example in which the system was composed of $ N = 8 $ individuals with 3 possible stable solutions. In the same article, Knuth requested the need to find a polynomial algorithm (at worst) to generate a stable solution, in case it exists. We will show how Irving in \cite{irving1985efficient} solved Knuth's request by building a $ O (N ^ 2) $ algorithm.\\
\\
Irving's algorithm can be divided into two parts. The first part consists of a series of proposals that people make in sequence. In particular, if an individual $ m_x $ receives a proposal from an individual $ m_y $, then $ m_x $ can reject it if he is already in the room with a partner he prefers, or he can accept it if he prefers $ m_y $ to his current partner. As in the case of the GS algorithm, each individual makes proposals starting from their first choice until accepted by someone. This phase of the algorithm ends when each individual has a roommate or if an individual is rejected by all the other individuals in the system. \\
It can be demonstrated that if $ m_y $ refuses $ m_x $ during the sequence of proposals, they cannot be partners in stable matching. As a consequence of this, if an individual $ m_x $ proposes to $ m_y $ then $ 1) $ $ m_x $ cannot have a better partner than $ m_y $ and $ 2) $ $ m_y $ cannot have a partner worse than $ m_x $.\\
Furthermore, if this first phase of the algorithm ends with an individual who has been rejected by all the others, then there is no stable solution. \\
If, on the other hand, it ends with everyone having partner, then the preference-list of the individual $ m_y $ who has accepted the proposal of $ m_x $ can be reduced by eliminating from it all those who are under $ m_x $, and all those who have a partner they prefer to $ m_y $. In this way, $ m_y $ is first in the new list of $ m_x $ which in turn is last in the new list of $ m_y $. Another important result is that if the reduced preference lists contain only one element then they determine the stable solution. \\
If the number of items in the reduced preference lists is greater than one, then the algorithm goes to the second step. In this second phase, the lists will be further reduced and this procedure will be repeated either until everyone has a partner specified in their list reduced to a single element, or until someone is rejected by everyone. In this last case, there is no stable solution. \\
The reduction of the lists in this second phase is based on the presence of \textit {cyclical sequences of people} $ z_1, ..., z_r $ defined in this way:
\begin{itemize}
\item for $ j = 1, ..., r-1 $ the second individual in the reduced list of $ z_j $ is the first individual in the reduced list of $ z_{j + 1} $.
\item the second individual on the $ z_r $'s list is the first individual on the $ z_1 $'s list.
\end{itemize}
Once such a cyclic sequence has been identified, preference-lists can be further reduced with the same rules as in the first phase. \\
The main theorem that Irving has proved is the following: given a cyclic sequence defined as above, call $ a_j $ the first person on the reduced list of $ z_j $. Then $ 1) $ in any stable solution contained in these reduced preference lists $ z_j $ and $ a_j $ either are partners for every $ j $ or for no $ j $. $ 2) $ if there is a stable solution where $ z_j $ and $ a_j $ are partners, then another exists where they are not. \\
Finally, we have that:
\begin{enumerate}
\item if the problem admits a stable solution, then this stable solution is contained in the reduced preference lists.
\item if one or more reduced preference lists contain zero elements, then there is no stable solution.
\item if the reduced preference lists contain only one person, then they determine a stable solution.
\end{enumerate}
In this way, Irving has built an algorithm that in addition to deciding whether a stable solution exists or not, it finds one in the positive case. Furthermore, he has shown that at worst this algorithm has a polynomial execution time $O(N^2)$. 

\subsubsection{The Family Problem}
Throughout most of this review paper, we have considered only "bipartite matching like" problems, in which the system was always composed of two separate sets to be coupled, which were represented, for example, by men and women, colleges and students or workers and firms. Previously, however, we dealt with the case in which the system was no longer disjointed but made up of only one "type" of individuals: the SR is the monopartite version of the SMP. We could therefore say that so far we have dealt with the "1-partite matching problem" ( such as SR) and the "2-partite matching problem" (such as SMP). Here, we introduce the more general "k-partite matching problem", in which the disjointed sets are $ k $: when $ k = 3$ the sets could be called product-consumer-company, when $ k = 4 $ product-worker-company-machines, and so on. \\
The idea of generalizing the SMP to a multi-partite problem derives from Knuth monography: in this work, as already mentioned, he proposed 12 questions and one of these asked if the SMP could be generalized to the case of three separate sets (man, woman and dog). For simplicity, we will often refer to the 3-partite matching problem which we will call "family problem" (also called 3 genders stable marriage, 3GSM \cite{ostrovsky2014s}), but most of the results can be easily generalized for any $ k $.

\subsubsection*{The General Family Problem}
In the family problem (FP) there are three types of categories of $ N $ players: men, women and children (or, in Knuth's version: men, women and dogs). Each agent has a preference list of \textit{pairs} of agents from the other two sets. The goal is to compose "families" of three people (a man, a woman and a child) so that the final matching is stable. The definition of stability is a direct generalization of pairwise stability in SMP (that we can call "triplewise" stability): a matching is stable if there are not three people (composed by agents of different sets) who do not form a family but who mutually prefer each other than to remain in their current family. That is, there must be no "blocking families". \\
In the same way, we can define the "3-dimensional roommates problem" (also called "three persons stable assignment problem" 3PSA \cite{ostrovsky2014s}) in which the agents of the system are all of the same "type", such as students, and must be grouped into triple rooms so that there is no triplet of students who are not in the room together but that would all prefer to stay together than to stay in their current rooms. \\
As Knuth posed the problem in his 12 questions, there was no definition of "ordering preference" or stability. As we shall show studying the many-to-many matching problem, there can be several ways to define preferences (on couples or individuals for example) and stability.\\
To keep everything in close analogy with the SMP for the moment we will keep the notion of stability defined above and the ordering preference on the \textit{pairs} of agents of the other two sets. \\
\\
As we will see in the next paragraph, when the system is k-partite with $ k \ge 3 $, the problem becomes computationally more complex. While polynomial algorithms exist for SMP and SR to find a stable solution (if it exists), this is not true for FP. \\
Furthermore, for $ k \ge 3 $ the theorem of the existence of stability fails: that is, it is not always true that there is a stable solution, and we will demonstrate it at the end of section 5. In reality, as we saw before, this theorem also fails for $ k = 1$ and therefore is valid only for $ k = 2 $, which makes the SMP a very particular model.

\subsubsection*{NP-Completeness of Family Problem}
The next question we ask ourselves is: is there a polynomial algorithm that finds, if it exists, a stable solution to FP or 3PSA? \\
The answer is unfortunately negative as the FP and 3PSA are NP-complete and therefore an efficient algorithm for these two problems is unlikely. The firsts to demonstrate this were Ng and Hirschberg in \cite{ng1991three} and independently also Subramanian in \cite{subramanian1994new}. We will not go into the details of the demonstration but we will show the main steps. The authors in \cite{ng1991three} took advantage of the fact that the "3D-matching problem" is NP-complete to construct a polynomial transformation from this problem to the FP. \\
The definition of 3D-matching problem (3DM) is as follows: given three disjoint sets $ \{A, B, C \} $ and given a set of triple $ T \subseteq \{A, B, C \} $, one wonders if there exists an $ M \subseteq T $ such that $ M $ is a complete matching, meaning that each element of A, B and C appears exactly once in $ M $. This problem turns out to be NP-complete \cite{karp1972reducibility}.
The two main theorems for the proof of Ng and Hirschberg are the following:
\begin{enumerate}
\item Let $ I $ be an instance of 3DM (i.e. a set of preference lists). If with such $ I $ there exists a complete matching $ M $ of 3DM such that $ M \subseteq T $, then the instance $ I_ {FP} $ of the FP, constructed starting from an appropriate polynomial transformation from $ I $, contains stable matching $ M_ {FP} $.
\item If the instance $ I_ {FP} $ has a stable matching, then $ T $ contains a complete matching for instance $ I $.
\end{enumerate}
Since the construction of $ I_ {FP} $ from $ I $ can be done in polynomial time and we know that 3DM is NP-complete, then theorems 1 and 2 together prove that \textit { FP is NP-complete}. Similarly, it can be shown that 3PSA is also NP-complete. \\
This type of result is not surprising in computer science since the boundary between P and NP is often marked between the numbers 2 and 3. Indeed, in our case when $ k = 1, 2 $ we saw that the problem is P-complete (we have a polynomial algorithm $ O (N ^ 2) $), while for $ k \ge 3 $ the problem is NP-complete. \\
As with the SMP with incomplete lists and ties, approximation algorithms also exist for FP and 3PSA, although the complexity of these two problems is even higher. \\
\\
However, at the end of the paper of Ng and Hirschberg, the authors mentioned that their (anonymous) reviewers had pointed out that their conclusions were valid for "inconsistent" preference lists, and wondered if they were also valid for preference lists that are not "inconsistent". To understand the meaning of inconsistent lists, consider the following example: the man $ m $ might prefer the pair $ (w_1, c_1) $ to the pair $ (w_2, c_1) $, and at the same time prefer the pair $ (w_2 , c_2) $ to the pair $ (w_1, c_2) $. This means that the man $ m $ does not consistently prefer the woman $ w_1 $ to the woman $ w_2 $. The question of the reviewers of Ng and Hirschberg is the following: if the preference lists are prevented from being inconsistent, then does the FP (or 3PSA) remain equally NP-complete? \\
Again the answer is affirmative and was demonstrated by Huang in \cite{huang2007two}. The same author has also shown that the problem remains NP-complete even when admitting ties in the preference lists. \\
\\
Having seen that even the most general case of FP is NP-complete, one may wonder if there is a particular case of FP that can be solved in polynomial times. Consider a model in which each player has two separate preference lists for the other two sets. Suppose, for example, that each man first evaluates the women and then the child; that every woman first evaluates the man and then the child, and finally that each child first evaluates the man and then the woman. In this case, one can show that by applying twice the classic GS algorithm, in which men first propose to women and then to children, a stable matching is obtained if they join these two matchings (men-women, men-children) \cite{danilov2003existence}. The problem is that this condition on the preference lists is weak. A single exchange is enough so that the GS algorithm no longer works: it is enough for a single woman to choose first based on the child and then on the man \cite{boros2004stable}.
\newpage

\section{SMP and Biology, Mathematics and Other Applications}

The SMP is applicable to many real-world situations \cite{iwama2008survey, bhatnagar2018new, fenoaltea1, lu2012recommender, medo2008market, lu2008emergence, lu2009role, liao2014firm, lebedev2007using, hitsch2010matching, chakraborti2015statistical, roth1982economics}. Indeed, it appeared to solve a practical problem such as that of satisfactorily placing medical trainees in various hospitals, among which there was a high competition to obtain these medical students.\\
So, since the mid-1900s, the SMP has been widely studied by economists, mathematicians, computer scientists and lately also by physicists.\\ Consequently, the spectrum of applications of this problem has greatly increased and even in 2012 L.S. Shapley and A.E. Roth were awarded the Nobel Prize in Economics for the theory of stable allocations and the practice of market design \cite{persson2012prize}, in which SMP dynamics were exploited. \\
\\
So this section will be multidisciplinary and we will briefly show the main results obtained by different scientific fields that are different from the SMP canonic.\\
We will start from the most "out of the box" application of the SMP: the application to biological systems. In particular, we will see that an "SMP style" process manages to account for the evolutionary dynamics of some communities of microorganisms. \\
We will then describe another unconventional application of SMP: stable resource allocation in the wireless network. We will see that an SMP theory can help solve overloading problems. \\
Finally, we will briefly study some concepts already seen in the previous sections and we will deepen them with the magnifying glass of mathematics. Mathematicians' studies on the SMP go far beyond what we present in this review, but these studies go beyond the scope of this work.

\subsection{SMP and Biology}

Here we want to give an example of SMP's application that goes beyond social, mathematical or IT fields. We want to show that this model can give us clues in various areas of science. Indeed, its simplicity and the concept of "stability" can be useful even in very complex non-social systems: we will therefore deal with SMP and biology.

\subsubsection{Microbial Communities Ecosystem}
Several works exploit the relationship between SMP and biology \cite{ishida2008antibody,goyal2018multiple, goyal2017microbial, dubinkina2019multistability}. \\
In particular, we want to show the latest works carried out by researchers from the University of Illinois \cite{goyal2018multiple, goyal2017microbial}, who developed a mathematical model based on SMP that could help scientists understanding the particular characteristics of microbial communities, allowing them to remain stable despite their strong biodiversity. \\
Microbial communities are groups of microorganisms that can exist in different types of environments such as the soil, the oceans or even in the human body (such as the human gut microbiome). Although these communities involve different types of microorganisms and can be very complex (thousands of different species can coexist in a very small volume), it has been shown that they often exist in multiple stable states that are different from each other because the proportions of the different coexisting species \cite{franzosa2015sequencing}. By "stability" we mean how well a community can manage changes: when a community is stable, for example, it can resist to the modification of one of the nutrients or the invasion of a new species. These changes can also lead to the transition from one stable state to another. Understanding better how the stability of these microorganisms works may be useful to control microbial ecosystems \cite{konopka2009microbial, konopka2015dynamics}: it could be possible to change the status of a soil microbiome by adding microbes or nutrients. At the moment, however, due to its complexity and diversity, it is impossible to manage a microbial community. \\
In \cite{goyal2018multiple} the authors build a predictive model to measure the stability of microbial communities based on the SMP.

\subsubsection{SMP Model of Microbial Community Dynamics}

The main assumption of the model is that, as described for the first time in \cite{monod1949growth}, many microbes tend to use the various nutrients in a specific sequential order. When they are exposed to a medium containing more nutrients, the microbes start using their favourite first. Once the first nutrient is exhausted, they start using their second favourite nutrient until all the nutrients are depleted. It has been shown experimentally that the most prevalent genus in the human gut microbiome behaves in this way \cite{huttenhower2012structure}. \\
The use of the SMP model to describe a dynamic community in which microbes use one nutrient at a time is a conceptual approach that elegantly justifies the existence of multiple stable states, the transition between these states and their resilience. \\
\\
In this case, marriages are represented by one-to-one matching between microbial species and nutrients. In the more general version of the model, all microbes can use all the available nutrients. In the real case, the microbes can use only a subset of all available nutrients and this fact can be easily implemented in the model by exploiting the SMP results with incomplete lists (section 4). \\
So each species of microbes classifies the nutrients from the most preferred - such as glucose for E.coli- to the least preferred. Similarly, the ranking of microbes for nutrients reflects the ability of microbes to use that nutrient. If the microbe $ x $ is the most skilled in using the nutrient $ y $, then the microbe $ x $ will be the first in the nutrient $y$ preference list. In this model, the competitive ability of a microbe for a nutrient is directly proportional to the rate with which this microbe extracts this nutrient from the medium. \\
The final result of the competition between microbes corresponds to a stable state without "blocking pairs", that is, without microbes that can switch to a nutrient $ n $ that they prefer more and, at the same time, be more able to use $ n $ than its current user. \\
The ecosystem will remain in this stable state until it is disturbed by the removal of a nutrient or the addition of a new species of microbes. \\
Another assumption of the model is that the quantity of a certain species of microbes in a stable ecosystem depends on the ranking of the nutrient it is using: if it uses its favourite, it will be more abundant than it would be if it used nutrients with a lower ranking. \\
\\
Figure 14 shows two examples of perturbation of a stable state with two species of microbes ($ M_1 $ and $ M_2 $) and two nutrients ($ N_1 $ and $ N_2 $): one in which a third species of microbes is added $ M_3 $, and one in which a third nutrient is added $ N_3 $. In both examples, there is a transition between one stable state and another, but both additions ($ M_3 $ in the first case and $ N_3 $ in the second) are rejected by the ecosystem. \\
The examples in figure 14 show that:
\begin{enumerate}
\item the invasion of a new microbe increases the competition between the microbes for the various nutrients and, in general, there is a transition to a less "optimal" stable state for the microbes (the sum of the rankings of the "partners" of the various microbes is higher).
\item the addition of a nutrient, on the other hand, reduces competition and brings the system back to a more "microbes-optimal" stable state.
\end{enumerate}

\begin{figure}[!h]
\begin{center}
\includegraphics[width=0.9\textwidth,scale=0.9]{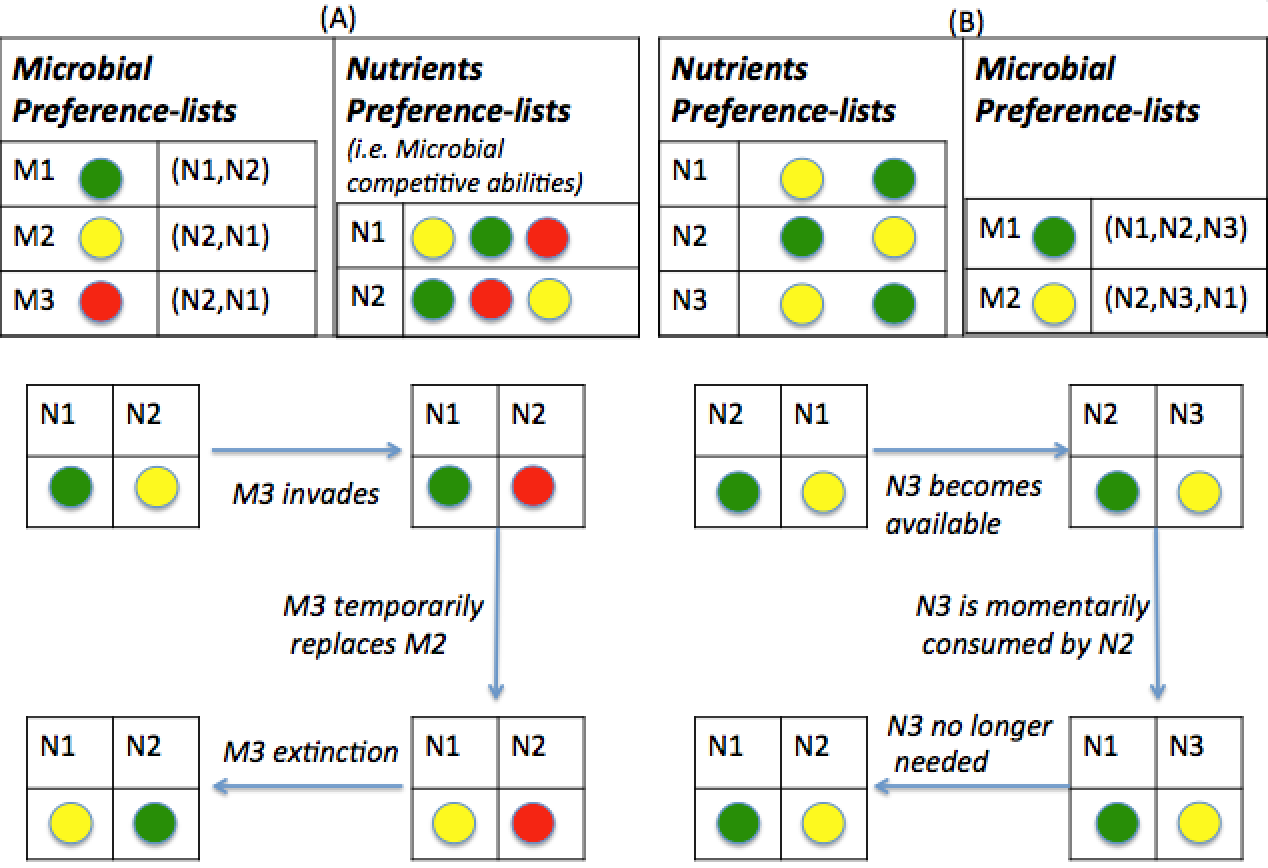}
\caption{\textbf{Microbial community dynamics}: the boxes at the top show the preference lists of microbes and nutrients. In the left panel (A) the system is in an optimal state for the microbes $ M_1 $ and $ M_2 $ and it is shown what happens after the addition of the species $ M_3 $: the system translates into a less optimal state (B) . In the right panel (B) a nutrient $ N_3 $ is added to the ecosystem in order to return to the microbes-optimal (A) state \cite{goyal2018multiple}.}
\end{center}
\end{figure}

In general, as we have seen, the greater the number of microbes and nutrients in the ecosystem, the greater the number of stable states (section 3). Furthermore, the stable states can be ordered starting from the microbes-optimal (corresponding to the GS state in which "the microbes propose") up to the microbes-pessimal, that is, from the state in which the microbes have the lowest "average ranking partner" to the highest one. To move from one state to another, a "marriage" should be broken between a microbe and a nutrient. Likewise, if a nutrient is added into the system, the competition between the microbes increases and the transition between stable states is in the direction of the microbes-pessimal state. Removing a microbe, instead, has the opposite result: the competition decreases and the system shifts to a more optimal state for the microbes. These results have interesting implications: the introduction of an antibiotic in the ecosystem could translate the system towards a more microbes-optimal state and therefore with greater total biomass. Although this conclusion seems rather counter-intuitive, it has been experimentally verified \cite{goyal2017microbial}. \\
\\
More details and results about this model can be found in \cite{goyal2017microbial, goyal2018multiple}. Simplifying, always based on the concepts of the SMP, the authors describe the properties of the network of the stable states of the ecosystem. The nodes (the stable states) are connected if a perturbation on the system connects them. Furthermore, the stable co-existence of large quantities of several different microorganisms is demonstrated through the complementarity of the priorities on nutrients, that is, through the complementarity of the preference lists of the various existing microbes. \\
The basic model explained above can be expanded to make it more realistic through the various variants of the SMP explained in sections 3 and 4. One possibility, for example, is to relax the assumption that there is a one-to-one matching and incorporate the simultaneous use of multiple nutrients by different microbes to have many-to-many matching.\\

\subsection{SMP in Wireless Network}
Smartphones and many other popular handhelds and portable devices today contribute to increasing traffic in the current wireless network. To overcome this problem, new systems such as cognitive radio networks or small cell networks have emerged and, as a consequence, these systems will lead to an increasingly complex wireless architecture.\\
 In such an environment, where the density of the wireless network is continuously growing, it becomes important to manage the problem of resource allocation efficiently. In particular, it is more convenient to focus on the "self-organization" and "self-optimization" approach rather than relying on the traditional centralized mechanism \cite{gu2015matching}. Therefore self-organizing systems, in which small cell base stations and various devices can make resource management decisions quickly, are needed. \\
We now show how the problems of traditional mathematical methods to optimize the resource allocation of the wireless network - such as centralized optimization and game theory - can be solved by applying the fundamental concepts of SMP (figure 15).

\subsubsection{SMP Model for Wireless Resources Allocation Problem}
In general, the wireless resource management problem can be represented as an SMP in which men and women are replaced by \textit{resources} and \textit{users}. The former can represent by different elements such as power or time-frequency; the latter, on the other hand, can represent the various devices or applications for smartphones.\\
The matching can be a one-to-one type, a many-to-one type or a many-to-many type and depending on the type of scenario, each resource and user have a maximum number of agents with which they can be matched. A matching is simply an allocation between resources and users. The goal is to pair users and resources by optimizing the different goals of the players in the system. Each user and resource builds a preference list of the elements of the opposite set. A preference can be defined in terms of a utility function that quantifies the quality of services (QoS) obtained in a given matching \cite{gu2015matching}.\\
As usual, the matching is stable if there are no blockingpairs. If there is no user $ u $ and resource $ r $ that are not coupled together but both would prefer to be.\\
The basic algorithm that guarantees a stable solution is the GS algorithm. It has the advantage of not requiring system agents to know the other agents' preference lists, but any decision is local (based on the preferences of the individual). Hence, the GS dynamic does not require a centralized controller. The GS algorithm has been extended to the case of many-to-many matching by balancing the role of proposers and receivers in the context of wireless networks, for example in \cite{xu2011seen, hamidouche2014many}. \\
There is a great deal of literature in which mathematical methods are proposed to optimize resource allocation in many wireless systems, in particular, centralized optimization and game theory \cite{han2012game, jorswieck2011stable, leshem2011multichannel, naparstek2014distributed, pantisano2013matching, semiari2014self}. These approaches have various limitations including an excessively high computing complexity. \\
Promising techniques based on the SMP have been developed recently to overcome the limitations of previous models \cite{saad2014college, gu2014cheating, chowdhury2019matching}. Some of the advantages of the SMP approach are that one can find different solutions in terms of stability and optimality, reflecting the different purposes of the system and the possibility of implementing efficient algorithms that favour self-organization. \\

\begin{figure}[!h]
\begin{center}
\includegraphics[width=0.8\textwidth,scale=0.8]{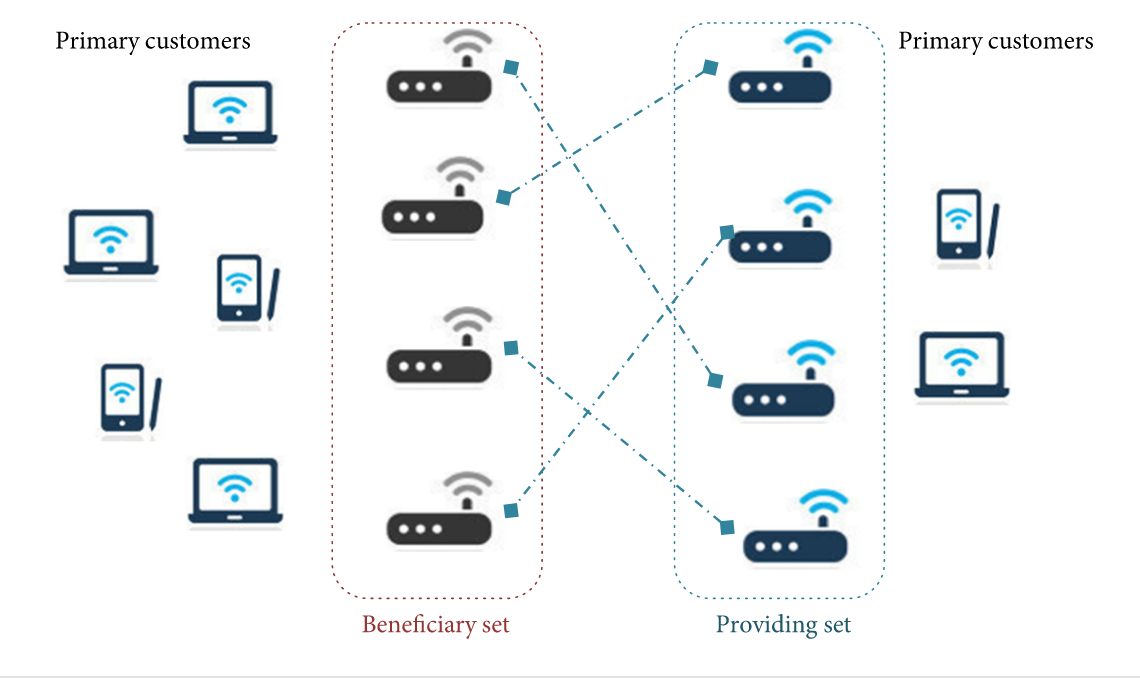}
\caption{\textbf{Matching game in Wireless local area networks}: figure reprinted from \cite{hassine2017access}.}
\end{center}
\end{figure}

Following Gu et al. in \cite{gu2015matching}, we can classify three types of matching model in the context of wireless networks:
\begin{enumerate}
\item \textit{Canonical matching}. This is the classic case where an agent's preferences in the game depend solely on the information available to the agent. This type of matching is particularly appropriate for the cognitive radio network (CRN): it is a two-side system in which licensed primary users PUs (channels) must be accessible to unlicensed secondary users SUs. In a CRN, a centralized system would not be suitable, as PUs and SUs often belong to different operators and cannot be managed centrally. In \cite{naparstek2014distributed, leshem2011multichannel} the CRN is modelled as a one-to-one SMP between PUs and SUs. The preferences of both parties are based on the same utility function calibrated on the rate of transmission. The authors in \cite{naparstek2014distributed, leshem2011multichannel}, through a modified version of the GS algorithm, found a solution that compared to the classic random channel allocation scheme, brought significant improvements for the SUs.
\item \textit{Matching with exeternalities}. In this case, there are \textit{externalities} in the system which form interdependencies between the preferences of the agents. An example concerns small cell networks: when a user is paired with a resource, the preferences of other users will change accordingly, because that resource could cause interference to other resources using the same frequency. Therefore, agent preferences, unlike canonical matching, also depend on the other matchings in the system. \\
A classic example of these externalities is the \textit{peer effect} \cite{lebedev2007using}. The preferences of users on a resource depend on the type and number of users who are connected to that resource. In \cite{saad2014college} the problem related to resource allocation in small cell networks is formulated as a one-to-many matching model in which users can be associated with a single small cell base station (SCBS). At the same time, each SCBS has a maximum number of users with which it can connect. User preference-lists are based on the trade-off between error rate and delay that they can get; while SCBS preference lists are based on load balancing by pushing users on smaller cells. \\
The authors in \cite{pantisano2013matching, bayat2012multiple} developed a new algorithm to overcome the fact that the GS algorithm, due to the peer effect, did not achieve a stable solution. Also in this case, with a matching problem approach, more convenient results are obtained (in terms of utility) compared to traditional methods based on the best neighbour scheme. 
\item \textit{Matching with dynamic}. Finally, this matching class is appropriate for situations in which preferences vary with time according to the conditions of the external environment (time-varying traffic, fast fading or mobility). In this case, the temporal dimension must be taken into account to solve the matching. However, at any moment the problem can be interpreted as one of the two types of matching explained above.
\end{enumerate}

In all types of matching, however, finding a stable solution in the stable marriage sense guarantees system robustness to possible deviations and this creates benefits for both resource owners and users. Having many deviations in the system, that is, many blocking pairs, implies an unstable network operation. \\
\\
Despite all the advantages of matching theory, it has some limitations: first of all, there can be more stable solutions and therefore the problem of which is better appears; moreover, a stable solution is not necessarily the optimal solution. In any case, compared to classic models, the stable marriage approach for the wireless network improves the performance of resource allocation in many fields of application. For a complete overview about matching theory in wireless network one can refer to \cite{han2017matching}. 

\subsection{SMP by Mathematicians}
Mathematicians have also contributed a lot to the study of SMP and especially recently research has been very active. These studies, however, go beyond the purposes of this review and in the following, we will only deepen two topics already covered in the previous sections: in the first paragraph we will deepen some issues concerning the average number of stable solutions in the classical problem; in the second paragraph we will study some themes on the generalizations of the classic model. 

\subsubsection{Maximum Number of Stable Solutions in the Standard Problem}
As we have seen in section 3.2.1, Pittel through probabilistic arguments has solved the integral in equation (1) obtaining an asymptotic formula for the average number of stable solutions in the SMP, i.e. he has obtained the equation (2).\\
However, we saw that this formula did not agree well with the numerical simulations and it was necessary to adopt specific approximations to solve equation (1), obtaining a more correct solution. \\

\begin{figure}[!h]
\begin{center}
\includegraphics[width=0.7\textwidth,scale=0.7]{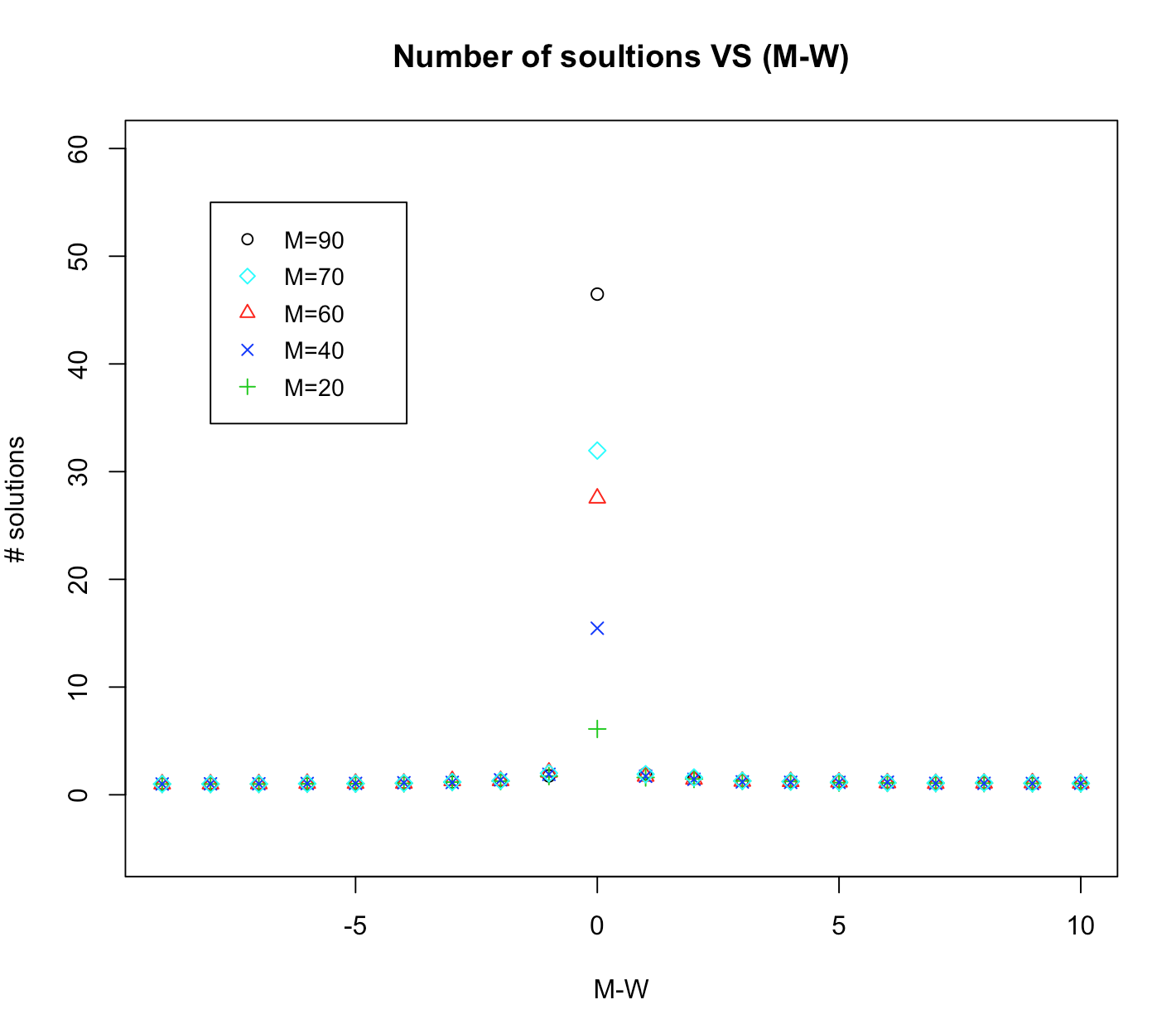}
\caption{\textbf{Number of stable solutions with different number of men and women}: average number of solution as a function of (M-W).}
\end{center}
\end{figure}

Still with regard to the average number of stable solutions, Pittel has obtained a formula for the case in which the number of women is different from the number of men. In particular he has shown that, if $M$ is the number of men and $W$ is the number of women and if we call $ \lambda = log (\frac {M} {M-W}) $, the average number of solutions $ n $ for $ M> W \to \infty $ goes as  \cite{pittel2019likely}:
\begin{equation}
n \approx W \frac{exp(\frac{e^{\lambda}-1-\lambda}{e^{\lambda}-1})}{(M-W)\lambda} \;.
\end{equation}
Consequently, if $\frac{M}{W} \to \infty$, i.e. for big differences between men and women, then we have $ n \to 1 $. In this case, the analytical result is consistent with the simulation results. We have performed a series of numerical simulations, shown in Figure 16: it shows the average number of solutions against the difference between men and women $ M-W $ for five different values of M. Each point in the graph is the average of 100 simulations.\\
\\
Another interesting question that can be asked about the number of stable solutions is the following: since at least one stable matching exists, it becomes natural to ask what the maximum number of stable solutions is, given a certain size $ N $ of the system-which we denote with $f(N)$. In the following paragraphs, we will expose the main results in recent years regarding the upper bound and the lower bound of $f(N)$.

\subsubsection*{$f(N)$ Upper Bound}
The problem of determining the maximum number of stable matching given a system of size $ N $ is a difficult question both from a theoretical and computational point of view. This problem appeared first in 1976 in the Knuth's monograph. In recent years, mathematicians and computer scientists have made numerous efforts to solve this problem, but it remains an open question. \\
Trivially one could say that the maximum number of stable matchings is $ N! $, as it is the maximum number of matching with $2N$ individuals. In 1976 Knuth showed an SMP with size $ N = 4 $ in which there were 10 stable solutions. Subsequently, Eilers showed through an exhaustive computer search that 10 is the maximum number of stable matches for a system of size 4 ($ f(4) = 10 $) and that this system is unique, that is, there is only one way of composing the preference lists to have 10 solutions with $ N = 4 $ \cite{thurber2002concerning}. \\
The first real progress regarding the upper bound of $ f (N) $ was done by Stathoupolos in his master's thesis in 2011, in which he showed that $ f (N) $ is at most of the order $ O (N! / C ^ N) $, for some constant $ c $. More recently, an article by Drgas-Burchardt and witalski \cite{drgas2013number} showed a weaker upper bound of about $ 3N! / 4 $. A further step was made in 2018 by Karlin,  Gharan and Weber \cite{karlin2018simply} who have shown that there is a universal constant $ c $ such that $ f (N) $ is at most $ c ^ N $. To arrive at this conclusion, the authors exploited the theorem previously shown which affirms the existence of a univocal relationship between stable matches and rotations. The best upper bounds before this result were of the order $ 2 ^ {NlogN-O (N)} $.

\subsubsection*{$f(N)$ Lower Bound} 
The first results about $f(N)$ lower bound were achieved with $ N $ being a power of 2: Irving and Leather \cite{irving1986complexity} have shown that if we denote with $ g (N) $ the function that counts the number of stable matching with $ N $ power of 2, then it satisfies the recurrence relation:\begin{equation}
g(N)=3g(N/2)^2-2g(N/4)^4 \;,
\end{equation}
with $ N \ge 4 $ and power of 2, and with $ g (1) = 1 $ and $ g (2) = 2$. Furthermore, they conjectured $ f (N) = g (n) $ for $ N $ power of 2. Later on, Knuth showed that this $ f (N) $ contains at least $ O (2.28 ^ N) $ stable matching. More recently Thurber has extended the work of Irving and Leather to all values of $ N $, not only for powers of 2 \cite{thurber2002concerning}. In the case of $ N $ power of 2 the results of Thurber coincide with those of Irving and Leather, while for all other $ N $ he found a lower bound of $ O (2.28 ^ N / c ^ {logN}) $ for a certain constant $ c $. 

\subsubsection{SMP Expansions}
We now delve into a topic already covered in section 4 concerning the generalizations of SMP studied by mathematicians: we will demonstrate that in the Family problem (FP) the stability theorem is not satisfied, i.e. there may exist no stable solutions.

\subsubsection*{Stability of $k$-Partite Matching Problem}

\begin{figure}[!h]
\begin{center}
\includegraphics[width=0.7\textwidth,scale=0.7]{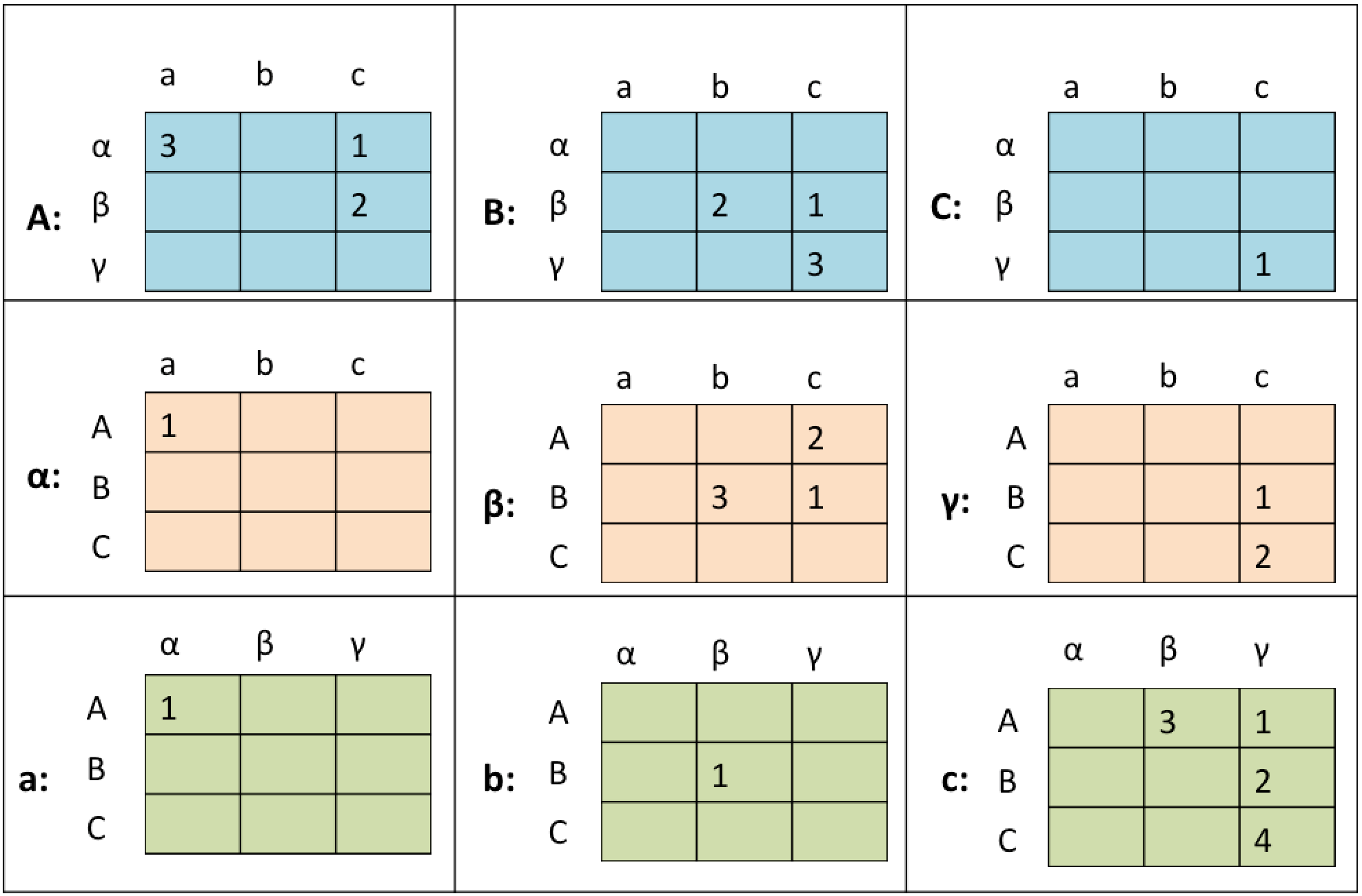}
\caption{\textbf{Alkan's example}: preference-lists of the agents in the Alkan's example \cite{alkan1988nonexistence}.}
\end{center}
\end{figure}

The possibility of the non-existence of stable matching in the FP was first shown by Alkan in 1987 in \cite{alkan1988nonexistence}. He showed through an example that it is not possible that in the FP there could be no stable matching (as Galey and Shapely did for Stable Roommate Problem).\\
Let us analyze Alkan's example. Consider an FP in which there are three men $ \{A, B, C \} $, three women $ \{\alpha, \beta, \gamma \} $ and three children $ \{a, b, c \} $. Each agent builds a preference-list on all possible pairs of agents from the other sets. The possible pairs for an individual are $ N^2 $ which in our example corresponds to 9 possible pairs to put on one's preference-list. The preference-lists of our case are shown in figure 17 (they are partially specified since it is sufficient for our example).\\
Consider the $ A \alpha a $ family, the scores obtained by the three agents are $ \{3,1,1 \} $ respectively. To be stable, a matching must assign man $ A $ to at least his third favourite family. Note, however, that any matching that contains the $ A \alpha c $ family, which is the preferred family of $ A $, is blocked by the $ C \gamma c $ family. Additionally, any matching that contains $ A \beta c $, which is $ A $'s second favourite, is blocked by $ B \gamma c $. Hence, we deduce that all the matchings that do not contain $ A \alpha a $ are unstable.\\
Similarly, it is found that all matchings that do not contain the family $ B \beta b $ are unstable and that all matchings that do not contain the family $ C \gamma c $ are also unstable. So the only possible matching is that which contains the families $ A \alpha a $, $ B \beta b $ and $ C \gamma c $. However, even such a matching is unstable because it is blocked by the family $ A \beta c $.\\
It follows that for the FP specified by the preference-lists in figure 17 there is no stable solution and therefore $ \textit {the FP does not guarantee the existence of a stable matching} $. The same arguments can be used to generalize this result when $ k> 3 $ \cite{alkan1988nonexistence}. \\
Although there may not be a stable solution for the FP, it can be shown that the number of stable solutions goes exponentially with $ N $ and this can be done in a similar way to what was done in the bipartite case \cite{ng1991three}.\\
\newpage

\section{Stable Marriage Problem and Economics}
One of the areas in which the SMP was most successful was the economic field. Many markets lend themselves very well to being modelled through the stable matching theory that we have studied so far. Indeed, in 2012 the Nobel prize for economics was awarded to Shapley and Roth for their theoretical and experimental contribution to the stable matching theory. \\
In this section, we shall revisit the main applications of SMP by economists. After a brief introduction to Game Theory, we will start by telling the story of the 2012 Nobel Prize and then we will describe the main works that led to the winning of this award. In particular, we will see in more detail the famous problem of the allocation of doctors in hospitals, we will study a specific case, based mainly on Roth's work, of allocating students in high schools, we will analyze the SMP applied to the kidney exchange and, finally, we will describe possible extensions of the SMP to different types of markets.\\
However, this is not exhaustive: to have a global vision of current economic research on matching theory, one can look, for example, at \cite{rostek2020matching}.

\subsection{SMP in Game Theory}
In the second half of the 20th century, Game Theory was consolidated as a model of social, biological and economic systems. It was born in 1944 thanks to the work of John von Neuman \cite{von2007theory} and today it is defined as a branch of applied mathematics focused on the description of systems with many agents with conflicting interests. Its study led to various Nobel laureates in economics, for example, that of John Nash, John Harsanyi, and Reinhard Selte in 1994, that of Thomas Schelling and Robert Aumann in 2005 \cite{osborne2004introduction}, and more recently that of Lloyd Shapley and Alvin Roth in 2012.\\

\subsubsection{Introduction to Game Theory}
A game is intended as a set of players each having a set $S$ of possible strategies $s$ and an associated \textit {cost function} $U(s)$ which depends on the strategy used by the player in question and the one used by the other players.\\
 A \textit {state of the game} corresponds to the adoption of a certain strategy for each player. Players are said to be \textit {rational} in the sense that they will try at all times to minimize their cost function. As it normally happens in real life, the final result of the game for a single player depends not only on the strategy adopted by himself but also on that adopted by all the other players, over which that player has no decision-making power.\\
Therefore, in general, players will not have a single optimal strategy but the best of their actions will vary depending on the decisions made by the other players. A classic example is the stock market \cite{challet1997emergence, zhang1998modeling} where one has to decide between buying or selling shares, and the most convenient thing will be to do what fewer players do.\\
\\
The fundamental concept of game theory is the so-called \textit {Nash equilibrium}, from John Nash \cite{gibbons1992primer}. Suppose there is a state $ \prod ^ * = \{s_1 ^ *, ..., s_N ^ * \} $ where the $i$-th player is adopting the strategy $ s ^ * _ i \in S_i $. If that state is such that every agent is playing the best of his response to the decisions of the others, then $ \prod ^ * $ is a Nash equilibrium. These states persist over time since no player is tempted to deviate from his behaviour, as $ s ^ * _ i $ is the strategy that minimizes its cost function, that is, by fixing the strategies of the other players, we have that
\begin{equation}
U_i(s_1^*,...,s^*_i,...,s_N^*) \le U_i(s_1^*,...,s_i,...,s_N^*) \;.
\end{equation}
The definition was accompanied by the proof of Nash's Theorem \cite{gibbons1992primer} which guarantees the existence of these equilibrium states (they can be more than one) in any game if a broader strategy definition than that given so far is included. It is important to note that the stability defined according to Nash is a static concept and does not imply that game dynamics lead to stable states in the Nash sense \cite{lageproblema}.

\subsubsection{SMP and Nash Equilibrium}
As described in section 2, an SMP state is \textit{unstable} if a man $m_i$ and a woman $w_j$ are not married in that state, but they would both prefer to be married rather than stay with their current partner. In real life, this would be enough for the divorce to occur in the marriages to which mi and $w_j$ belong. In this way, they can form a new marriage for themselves, both gaining happiness.\\
So unstable couples lead to the reconfiguration of states if the players are considered acting rationally. A Nash Equilibrium will therefore be a state that does not contain any unstable pair, as no player will find anything better to do than stay with his/her current partner. In other words, such a state will be stable to the individual actions of the players, hence a state of Nash Equilibrium is also called Stable Marriage. The marriage problem, in conclusion, studies the existence and properties of Nash equilibrium states, so it fits very well for economics applications.

\subsection{Stable Matching: a Nobel Prize History}
In the traditional economic analysis, those types of markets in which price plays a fundamental role are extensively studied. In most markets studied by economists, the price "adjusts" to match supply and demand. Indeed, in many cases, these markets function well enough both in theory and in practice. \\
In certain situations, however, the "standard" market mechanisms encounter some problems and there are cases in which the price is no longer the fundamental element as it cannot be used to allocate resources. For example, many schools and universities are prohibited from imposing school fees, or, in the case of human organ transplants, monetary payments are prevented for ethical reasons. In these cases and many others, however, an allocation of resources is required despite the absence of monetary transactions.\\
The problem for economists was to understand how this type of processes work in reality and when the outcome of these processes is efficient. The answer to these questions lies in the theoretical and practical understanding of stable matching theory. The analysis of the mechanisms for allocating resources can be, with a certain degree of approximation, addressed to the analysis of the abstract model of the SMP. \\
\\
To underline the importance of these types of market, the 2012 Nobel prize was awarded to Lloyd Shapley and Alvin Roth thanks to their research on matching theory that extends from abstract models developed in the 1960s, to the empirical works of the 1980s in which practical solutions are applied to real-world problems. Examples include, as we will see in the rest of this section, the assignment of new doctors to hospitals, students to schools or human organs for transplants to recipients. \\
\\
Lloyd Shapley is one of the most important researchers in the field of cooperative game theory. His works not only reinforced the theoretical foundations in this field but also stimulated practical research applied to policymaking. Shapley, in collaboration with Gale, Scarf and Shubik, created the theory of matching markets. In particular with the famous 1962 article \cite{10.2307/2312726}, Gale and Shapley expressed the hope that one day their theoretical work would be finalized in practical applications. So Lloyd Shapley has been recognized for his early theoretical contribution to the matching market theory. The clarity and elegance of the Gale-Shapley paper made this article one of the most important academic "must-read" for economics students across the world.\\
However, the real-world relevance of Shapley's work was recognized only twenty years later, in the early 1980s, with Alvin Roth's work on the market for newly graduated doctors. \\
In Roth and Sotomayor's book (1990) \cite{roth1992two} the main practical works on matching theory are described, including many fundamental results from Roth and coauthors. Roth's research has always been aimed at applying game theory to real situations. In \cite{roth1991game} he describes how laboratory experiments and field observations can connect to game theory and thus make economics a satisfactory experimental science. Regarding matching theory, Roth's work has advanced the understanding of how markets work. Using empirical, experimental and theoretical methods, Roth and his coauthors studied the institutions that improve market performance by understanding the fundamental importance of the concept of stability and therefore encouraging real institutions to create "compatibility". Importantly, these contributions led to the redesign of a large number of real-world markets. \\
\\
In the following, we will study the main applications of SMP in various two-sided markets: doctor allocation, student allocation and kidney exchange. A key feature of these examples is that price is not part of the process. However, the absence of the in the SMP model does not limit its applicability.\\
Shapley and co-workers have examined extensions of the original model which include the price and the salient features of these extensions compared to the original model.\\
A direct application of matching with the price is that of auctions, in which objects must be paired with buyers, and the price is a fundamental element. We will not go into the details of these models but it is useful to know of their existence to understand how vast the spectrum of application of the SMP is. \\
\\
The 2012 Nobel Prize was therefore awarded for contributions to a booming research field in which experimental theory and evidence interact with each other. Although Lloyd Shapley and Alvin Roth have worked independently, the success of their research is due to the combination of Shapley's theoretical works and Roth's practical ones. \\
As we will see in sections 7 and 8, research in this field continues to grow and create new insights on the functioning of social and economic relations (and not only), and it turns out to be a promising field of study for the future.

\subsection{Matching Doctors and Hospitals}
Here we present the main application of the SMP (not only in the economic field). The problem of allocation of doctors to hospitals was the first problem to be studied through matching theory and inspired most of the subsequent works. \\
To give a concrete example we will show the main works of Roth and others on the doctor market in the United States, even if the matching problem theory has also been applied to other countries such as Canada and Britain \cite{gale2012stable, roth2008have}.\\

\begin{figure}[!h]
\begin{center}
\includegraphics[width=0.7\textwidth,scale=0.7]{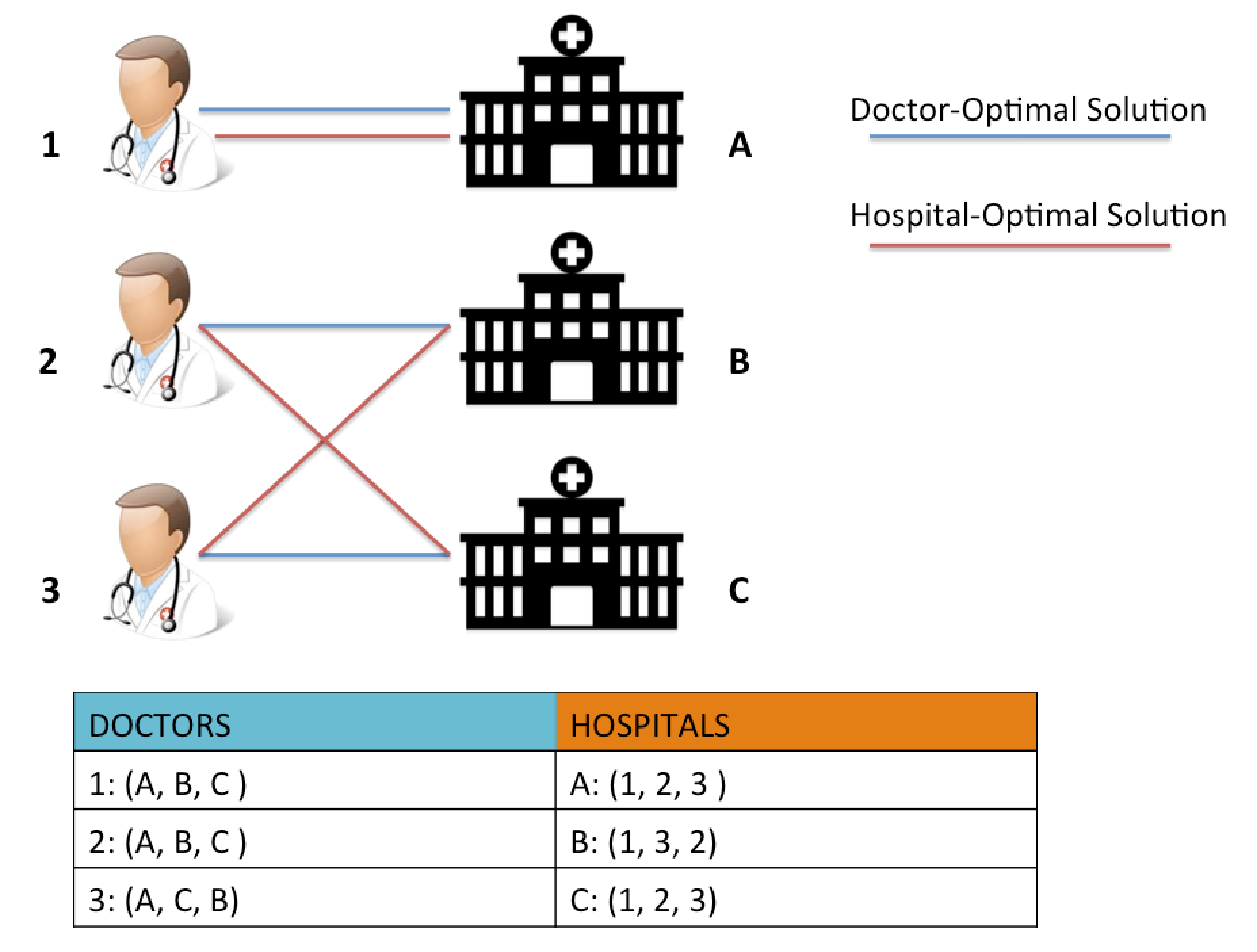}
\caption{\textbf{Matching doctors and hospitals}: example of Doctor-Optimal and Hospital-Optimal solutions with the preference-lists in the table}
\end{center}
\end{figure}

In the United States, medical students are usually hired as interns in hospitals. At the beginning of the last century, this market was very decentralized \cite{shapley2012stable}. Competition for scarce medical students in the 1940s forced hospitals to offer internships to students even many years before graduation. Hospitals were often assigned well before students could show that they were qualified enough or even before students chose which branch of medicine to work in. If an offer was rejected there was a high risk that it was too late to make offers to other candidates.\\
A market structured in this way, as we will see later, clearly produces unstable matching as there is no time to make offers that are beneficial to both parties. To solve this problem, hospitals imposed deadlines to respond to an offer, but this forced students to make decisions too early, without having time to know if they would receive better offers. \\
To solve this problem, in the middle of the last century, it was decided to assign the task of creating the matchings to a central organization: the National Resident Matching Program (NRMP). As we have already seen in section 4, thanks to the work of Roth \cite{roth1984evolution}, it was shown that the NRMP algorithm was no different from the classic GS algorithm. So Roth speculated that the main reason the NRMP worked so well was that it produced stable matchings. He also began studying other medical markets to see which matching algorithms produced stable results and which ones did not and, as can be expected, the algorithms that produced stable matching were the ones that achieved the most success. In the following paragraphs, we will develop the tale just presented in more detail and review the main articles about this topic.

\subsubsection{Market for New Doctor in the US}
The problems concerning the medical market in the United States mentioned above were studied by Roth in \cite{roth1984evolution}. In particular, before the NRMP was established, this market suffered from defects in \textit {unravelling} and \textit {congestion} \cite{shapley2012stable}. Unravelling refers to the fact that offers to students were made long before the students graduated. This did not allow students to show that they were qualified or the offers were so premature that the students did not even know in what field of medicine they would like to practice.\\
As for the congestion problem, it refers to the fact that, when an offer was rejected, it may be too late to make other offers as other students may already have accepted offers from other hospitals.\\
To solve this problem hospitals have imposed deadlines to accept or reject an offer but this forced students to accept or reject without having the possibility to know if the next offers would be better or not. \\
Such unravelling and congestion defects have occurred in many other markets, most notably Roth found them in the following markets \cite{roth1984evolution, roth1994jumping}: entry-level legal, medical market in the US, Canada and UK, business school, Japanese university graduate, clinical psychology internships, optometry and dental residencies in the United States. \\
\\
When indivisible and heterogeneous goods are traded, as in the labour markets mentioned above, the offers must be made individually, not to the global market \cite{shapley2012stable}. The problem of having to coordinate when to propose an offer can cause market decentralization, unravelling and congestion. In \cite{roth1994jumping, roth1997turnaround}, the authors studied through theoretical models how such defects can emerge. \\
Due to these failures of the medical US market, the NRMP was established as a centralized clearinghouse: this institution has the task of creating matching between hospitals and students and the algorithm used was equivalent to the GS algorithm (or deferred-acceptance algorithm), therefore it produced stable matching. As argued by Roth in \cite{roth1984evolution}, stability is the main cause of the success of the NRMP algorithm. Since a stable matching is a Nash equilibrium, stability means that for no pair of the system it is convenient to change. Therefore there are no situations in which, for example, a student is assigned to paediatrics even though he prefers dermatology and, at the same time, that another student is assigned to dermatology department that, in turn,  preferred the previous student. In this case, a feeling of dissatisfaction would arise both on the part of the student and on the part of the hospital. \\
Hence, stability reduces dissatisfaction and has meant that participants in the NRMP were more and more, creating a virtuous circle in which greater satisfaction brought more and more participants.

\subsubsection{New Market Design}
In this paragraph, we see how the insights from the previous paragraph can be exploited to improve the efficiency of labour markets. What we said in the previous discussion is not enough for real-world situations. Some complications are absent in theoretical models. Real institutions have to manage individuals who make mistakes, who do not understand the rules or voluntarily neglect them and who, in general, are not the perfectly rational agents of ideal models. Also, institutions must respect ethical and legal constraints on the transactions that can be executed.\\
As described by Roth and Peranson in \cite{roth1999redesign}, the complexity of the real world forces the NRMP algorithm to be modified. For example, since 1960 the number of married couples who both graduated in medicine were increasing and, often, they both preferred to work in the same geographic region. So they preferred to contact directly the hospitals rather than undergoing the algorithm. \\
Therefore, a couple, from a theoretical point of view, is seen as a compound agent who wants two jobs in the same geographical area violating the "substitutability" assumption, a concept that we will formally define below. Roth in \cite{roth1984evolution} has shown that in a market where some agents are couples, a stable matching may not exist.\\
\\
Another problem was that the NRMP algorithm was an "employer-proposing" algorithm (analogous to the men-oriented GS) and thus favoured hospitals rather than students. \\
In 1995 Alvin Roth was hired on the NRMP Board of Directors to build a new algorithm. The goal of the new design was: "to construct an algorithm that would produce stable matchings as favourable as possible to applicants, while meeting the specific constraints of the medical market" \cite{roth1999redesign}.
The new algorithm designed by Roth and Pernason was an "applicant-proposing" algorithm modified to please couples. The instabilities produced by the presence of couples were resolved sequentially, following the "instability-chaining" algorithm of Roth and Vate \cite{roth1990random}. \\
Since this new algorithm was adopted in 1997, more than 20,000 doctors have been employed in hospitals with it \cite{shapley2012stable}.

\subsection{Student-High School Allocation}
We now focus on the problem of allocating students in schools. There is a lot of literature on this type of problem \cite{abdulkadirouglu2003school, abdulkadirouglu2005new, abdulkadirouglu2017welfare, abdulkadiroglu2006changing, erdil2008s, abdulkadirouglu2009strategy, abdulkadirouglu2003school}. However, we will see in particular the application of the SMP to the Ministry of education (MOE) problem in Singapore \cite{teo2001gale} that is, in turn, mainly based on the Roth works. The analysis of this problem will allow us to introduce a topic of which we have not yet talked about in this review: the possibility for agents in the system (men or women) to act dishonestly (by changing the preference-lists) to obtain a better matching. \\
So, before talking about the practical application of SMP in Singapore schools, following the work of Teo et al. \cite{teo2001gale} we will introduce the concept of "cheating" in SMP and the related algorithms to optimize the problem.

\subsubsection{Optimal Cheating Model}
Consider a two-sided market where a men-oriented GS algorithm is applied. As already explained in section 2, a men-optimal solution is reached with this mechanism. However, women can manipulate their preference-lists to achieve better matching. The goal of this paragraph is to understand under which conditions and how much women benefit from falsifying preference-lists. In \cite{roth1982economics} it has been shown that if one uses the men-oriented GS, no man has the incentive to falsify his preference-list as nobody would get a better matching than what they would get with truthful lists. As for women, on the other hand, Gale and Sotomayor \cite{gale1985ms} have shown that they can distort their preference-lists and force the men-oriented GS to reach a women-optimal solution. In this case, women use their original preference-list but declare "unacceptable" all men under their women-optimal partner. In the following paragraph, we explain what happens if women are forced to use complete preference-lists, that is, without declaring anyone unacceptable. \\
\\
Consider again an SMP where the men-oriented GS algorithm is used. Suppose that the woman $ w $ does not know the preference-lists of the other agents in the game. If you give the woman $ w $ the opportunity to reject proposals and allow her to remain single, a possible algorithm for finding the women-optimal partner for $ w $ is as follows: \cite{teo2001gale}: $ 1) $ during the execution of the men-oriented GS $ w $ refuses all proposals so that in the end $ w $ and a man $ m $ will be single; $ 2) $ Among all the men who proposed to $ w $, call $ m ^ * $ the best proposer. In \cite{teo2001gale} it is shown that $m ^ *$ is the women-optimal partner for $ w $. \\
As already mentioned, the above algorithm only works when the woman $ w $ is allowed to remain single. When this is no longer allowed, the situation becomes more complicated.\\
Again in \cite{teo2001gale}, the authors build an algorithm to determine the optimal strategy for $ w $ (i.e. the falsified preference-list, permuting men, which allows her to achieve a better matching) when she is not allowed to remain single and it is not possible to declare anyone unacceptable:
\begin{enumerate}
\item Run the men-oriented GS algorithm with the real preference-list $ L (m, w) $ for $ w $ (with $ m $ the men-optimal partner). Keep track of all the men who proposed to $ w $.
\item Suppose that a man $ m '$ proposes to $ w $ during the first step. If you put $ m '$ at the top of the $ w $'s list and restart the algorithm, you get a matching where the men-optimal partner of $ w $ is $ m' $. The new list will be called $ L (m ', w) $ and $ m' $ will be called \textit{potential partner}.
\item Repeat the second step to get all the other $ L $ lists with all possible potential partners in the $ L (m, w) $ list. It will be said that the man $ m $ is exhausted.
\item Repeat steps 1, 2 and 3 until all men are exhausted.
\item Call $ m ^ * $ the best of $ w $ potential partners.
\end{enumerate}
It can be shown that $ P (m ^ *, w) $ is the optimal strategy for $ w $. Furthermore, it can be shown that $ w $ is not guaranteed to reach its women-optimal partner. Indeed, there is no guarantee that $ m ^ * $ is better than its men-optimal partner. Think, for example, about the case in which each woman is in the first place exactly for one man, that is, each man has a different woman at the top of his list. In this way, the men-oriented algorithm ends with a men-optimal matching regardless of how women falsify their lists. \\
So, eliminating the possibility for women to declare men unacceptable, makes it more unlikely for them to behave incorrectly during the game. Teo et al., Through simulations, have found that for women the probability of benefiting by cheating is less than $ 10 \% $ and, therefore, not telling the truth does not pay enough. Fortunately, this result is quite comforting for practical applications.

\subsubsection{Singapore Ministry of Education Exercise}
In \cite{teo2001gale} the authors proposed a mechanism based on the GS algorithm to improve the mechanism for assigning students to secondary schools in Singapore. The criticized mechanism worked in this way: before doing the "Primary School Leaving Examination" (PSLE), students had to give the Ministry of education (MOE) a ranking of the 6 schools they preferred; after the PSLE test was run, a student ranking was built based on the scores obtained on that test; at this point, the best students were placed in their first choices and the worst students were left with the last choices or even assigned in schools that are not on their preference-list.\\
One of the main features of this mechanism is that schools have a passive role, that is, students are assigned to schools only based on the PSLE score and not based on the actual preference of the school. This could not adapt well to the new rules imposed by the MOE: the schools would have had more autonomy in the choice of students and the assessment of them would not have been based only on the PSLE, but also on the grades in the individual subjects that different schools assessed differently. In other words, schools could also have had different student preference-lists. \\
Another problem was that the students had to deliver the school preference-lists before taking the test: knowing that many good students would apply for the best schools, it was not convenient for students to build a truthful preference-list. In fact, with this mechanism, if the students put their first six ideal choices, they risked being excluded from all six and being assigned to another school according to criteria decided by the MOE. \\
\\
At this point, it becomes evident that a matching mechanism based on the GS algorithm and the above "cheating model" would be much more convenient for both schools and students. In particular, the MOE exercise lends itself to being an SMP cheating model in which men are schools and women are students. To be more detailed, the model would be a many-to-one matching problem, as each school can contain more than one student. In \cite{teo2001gale} the authors easily map the many-to-one model into a classic one-to-one model. \\
We list below the main advantages of an SMP model for MOE exercise :
\begin{itemize}
\item A general SMP model allows schools to classify students according to different criteria, according to the areas of excellence in which the schools want to invest more.
\item A general SMP model matches all students to schools according to the true preferences of the students.
\item The SMP cheating model allows students, as we have seen in the previous paragraph, to deliver their real preference-lists without incentives to say false, that is, without making strategy based on the results of the PSLE test.
\end{itemize}

\subsection{Kidney Matchings}
The types of matching described so far involve two parts (doctors and hospitals, students and schools) who both make active decisions. However, there are various situations in the real world where one part is active and the other is passive. An example is the matching of the kidneys or other human organs to patients who need transplantation. Here we will deal with this topic.

\subsubsection{Kidney Matching Market before Matching Theory}
Kidney transplantation is a medical treatment that is chosen when kidney disease reaches its terminal state. However, the main problem may be the shortage of kidneys that can be transplanted. As Alvin Roth writes in \cite{roth2008have}, there are more than 70,000 patients on the waiting list for cadaveric kidneys in the United States, but in 2006, for example, fewer than 11,000 transplants of this type had been performed. Also in 2006, approximately 5000 patients died when they were still on the waiting list or had been removed from the list because "too sick to perform the transplant". \\
This type of situation is not exclusive to the United States, in fact even in Great Britain, at the end of 2006, over 6000 people were on the waiting list for cadaveric kidneys, but only about $ 20 \% $ of these patients performed the transplant operation. \\

\begin{figure}[!h]
\begin{center}
\includegraphics[width=0.7\textwidth,scale=0.7]{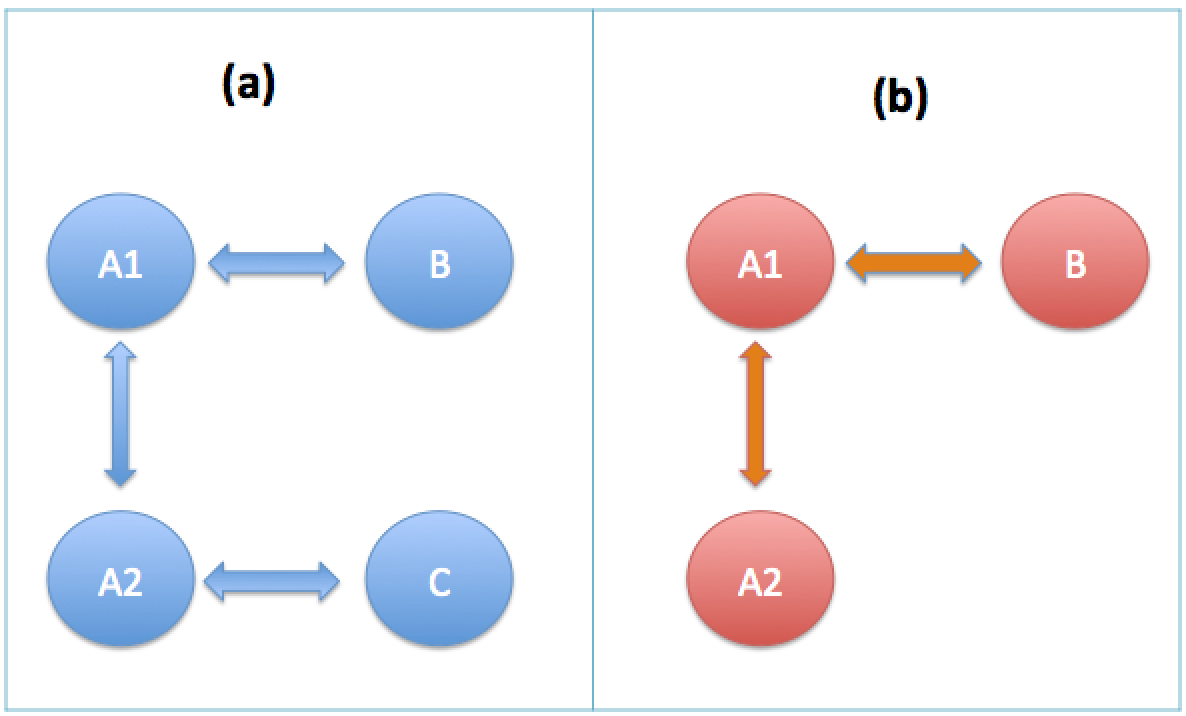}
\caption{\textbf{Kidney matchings}: the double-headed arrows indicate that the connected pairs are compatible, that is, the patient is compatible with the donor. Couples A1 and A2 are in transplant centre A. Couples B and C are in different transplant centres. Transplant Center A only knows its patients, so it can only exchange between them. In (a), transplant centre A also makes exchanges available with other transplant centres. In (b), there could be an exchange between A1 and B, leaving A2 with no exchange. If this were the situation, then transplant centre A could exchange between A1 and A2 without informing the other centres \cite{roth2008have}.}
\end{center}
\end{figure}

Since healthy people have two kidneys and can live on while staying healthy with just one kidney, an alternative option to having a cadaveric kidney transplant is to have a healthy person donate one of his kidneys. Furthermore, in doing so, the patient who undergoes the transplant has a greater chance of long-term success than someone who receives a cadaveric kidney. \\
Unfortunately, however, good health and goodwill are not enough for a person to donate their kidney. The patient and the donor can be biologically incompatible due to the blood type or because the patient's immune system rejects some of the donor's proteins \cite{roth2008have}. In the United States in 2006 there were about 6500 kidney transplants from living donors, while in Britain there were about 590. Hence, unfortunately, the supply of transplantable kidneys was much lower than the demand. Furthermore, in almost all countries it is illegal to buy or sell organs for transplantation because of ethical issues. \\
\\
Economists, in addition to having to deal with political issues to relax this type of law against the sale of organs, must also deal with how to increase the number of transplants subject to existing constraints (both legislative and logistical). This will be the issue that we will deal with in the next paragraph.

\subsubsection{Matching Theory for Kidney Exchange}
As we analyzed in the previous paragraph, the organ transplant market is full of restrictions. Economic transactions cannot exist and biological incompatibilities can exist between the voluntary donor and the patient who needs the transplant. Despite these constraints, there are cases where it is possible to find a solution for the transplant: consider the case in which the patient $ A $ wants to donate to the patient $ a $ but it is incompatible and at the same time the patient $ B $ wants to donate to the patient $ b $ but here too there is incompatibility. If the patient $ A $ is compatible with $ b $ and $ B $ with $ a $, then a transplant is possible. However, this type of bilateral exchange is rare and sometimes impossible for logistical reasons. Roth, Sonmez, and Unver in \cite{roth2005pairwise} have shown that when only bilateral exchanges of this type are feasible, there are efficient outcomes that can be found computationally. \\
As mentioned, some patients could be put on a waiting list in the hope that a compatible donor would be found as soon as possible. Roth, Sonmez, and Unver \cite{roth2004kidney} have introduced a very simple algorithm, called \textit {top trading-cycle}, to allow the presence of waiting lists in the model: the doctor indicates the kidney more appropriate for each patient on the waiting list; if there is a cycle, then the kidneys are exchanged following this cycle.\\
For example if $ A $ wants to donate to $ a $, $ B $ to $ b $ and $ C $ to $ c $, but there are incompatibilities, then if possible one can perform the cyclic exchange $ \{A \to b ; B \to c; C \to a \} $. The algorithm rules allow such a chain, for example, if $ B \to a $ while $ b $ is assigned a "high priority" status in the waiting list. The same authors of this algorithm have built an efficient selection rules chain for this type of exchanges \cite{roth2004kidney}. \\
\\
After this type of research on the organ transplant market, several regional programs have been created in the United States to promote studies towards more efficient and complex kidney exchanges.\\
More recently, attention has shifted towards types of exchanges that include the so-called "altruistic donors", i.e. those people who want to donate an organ but do not have a particular patient in mind \cite{roth2008have}. This type of chains suffers less from logistical problems since transplants do not necessarily have to be performed simultaneously between two patients \cite{roth2006utilizing}.

\subsection{More General Markets}
We now consider an even broader generalization of the SMP: the many-to-many matching problem. In this case, the agents of both sides have a quota $ q \ge1 $. To remain in the theme of marriage, one can think of the metaphor of polygamous marriages in which each man can have multiple wives and each woman can have multiple husbands \cite{baiou2000many}. More useful is to think of two sets composed of firms and workers: each firm wants to hire a set of workers and each worker wants to work for a set of firms \cite{echenique2004theory}.\\
This context, compared to a many-to-one matching market (CAP), is more complex and has many more pitfalls. Moreover, a many-to-one matching market would seem more suitable for describing real situations since normally a firm is thought to contain $ q_f \ge1 $ workers but one worker works at most for one firm ($ q_w = 1 $). However, there are important real-world markets that are many-to-many \cite{echenique2004theory}: think of the famous example of medical interns in the UK where doctors could be assigned to multiple hospitals and one hospital hired many doctors. One can also think about the assignment of teachers to high schools in some countries such as Argentina where about a third of teachers work in several different schools. Finally, the many-to-many environment can be considered as an abstract model of the system of contracts between companies and suppliers. \\
Another important reason why it is necessary to have a consistent theory also for a many-to-many matching market is that most labour markets certainly could have few many-to-many contracts. For example, in the USA more than half of workers work in companies with a non-zero percentage of multiple jobholders \cite{simpson2002link, indexes2014us}. Although this percentage may also be very small, as we will see later, even a single many-to-many contract would make the model very different and therefore not at all approximable to a many-to-one model.

\subsubsection{Many-to-Many Market}
The many-to-many matching problem has been extensively studied by mathematicians, computer scientists and economists. Being very different from the classic SMP, in this review we will not go into too much detail of the model but we will try to show the main results as clearly as possible. \\
We have divided the model into two main variants:
\begin{enumerate}
\item \textit{Reduced many-to-many model (RMM)}: this version of the model is the most similar to the classic SMP and is characterized by the concepts of "stability" and "ordering preference" (in the preference-lists) that we have used so far. For this reason, most of the results of the SMP and the CAP also extend to the RMM. However, this model is very weak for real-world applications.
\item \textit{Generalized many-to-many model (GMM)}: in this case, the definition of "stability" and that of "ordering preference" will be modified to make them more suitable for a many-to-many market. We have already encountered an example of this adaptation in the case of the CAP concerning "ordering preference" when we introduced the concept of "responsive preference". As we will see, in GMM, the results of the classic SMP can only be obtained by implementing appropriate constraints
\end{enumerate}
In the rest of this paragraph, we will study the main results on the existence of stability in the two models (RMM and GMM).

\subsubsection*{Reduced Many-to-Many Matching Model}
In the reduced many-to-many matching model (RMM) there are two finite sets of players: N workers $ W = \{w_1, w_2, ..., w_N \} $ and M firms $ F = \{f_1, f_2 ..., f_M \} $. Each agent in the system has a preference-list of agents from the opposite set which they consider acceptable. In addition, \textit{each} player has a quota ($ q_w $ or $ q_f $) which represents the maximum number of players from the opposite set that can be assigned to him. It is evident that when $ q_w = 1 $ and $ q_f \ge 1 $ it is reduced to CAP while if $ q_w = q_f = 1 $ it is reduced to the SMP. \\
So far, we have not made any hypothesis on the preferences to the sets of players of the opposite set that can be obtained. We are considering, as we have also done with CAP, preferences on individuals and not on sets of individuals. This type of "ordering preference" is called by Mourad Baiou and Michel Balinski \cite{baiou2000many} \textit{"Max-Min individual ordering preference"} and is defined by them in terms of sets of individuals as follows: \\
\begin{itemize}
\item \textit{"Max-Min individual ordering preference"}: a set A of agents is preferred over a set B if the worst individual in set A is preferred over the worst individual in set B. If the worst individuals in the two sets are the same agent, then the preferred set will be the one with more elements.
\end{itemize}
Furthermore, the definition of stability is the usual one that we have used up to this point and which in the literature is called "pairwise stability" \cite{sotomayor1999three}:
\begin{itemize}
\item \textit{pairwise stability}: a matching M is pairwise stable if, for each pair $ \{w_i, f_j \} \notin M $, at least one of the two agents ($ w_i $ or $ f_j $) prefers to remain with its partners in M. In other words, M is pairwise stable if there are no $ w_i $ and $ f_j $ agents who are not partners but if they become partners, dissolving their partnership to remain in their quota and keeping the other partnerships, would both get a higher payoff.
\end{itemize}
With these definitions of ordering preference and stability, it can be demonstrated in a similar way to that done with SMP and CAP that there is always at least one stable solution and that there are worker-optimal and firm-optimal solutions in the usual sense of the SMP. In particular, in \cite{baiou2000many} the authors arrive at these conclusions through a "network approach" in which workers and firms are placed on a grid where workers are in the rows and firms in the columns.\\
The same authors have built an algorithm, called by them "reduction algorithm", which, similarly to the GS algorithm, guarantees the existence of a pairwise stable solution. The algorithm turns out to be of the order of $ O (n ^ 2) $ where $ n = max \{N, M \} $. \\
There are several variations of the RMM model similar to those existing for the classic one-to-one SMP \cite{fleiner2003stable, eirinakis2018stable, bansal2003stable, malhotra2004stability}. For example, in \cite{bansal2003stable} the authors find an efficient algorithm to find the minimum egalitarian stable matching and in \cite{malhotra2004stability} the many-to-many matching problem with ties is studied.

\subsubsection*{Generalize Many-to-Many Matching Model}
In the generalize many-to-many matching problem the situation is a little more complex than in RMM. The preferences of each agent are not considered to individuals, but a \textit{set} of individuals. We give an example from a real situation to understand the importance and greater practical utility of preferences on agent sets rather than on individual agents. Imagine a real estate company $ I $ (with few employees) that has three open positions, i.e. $q_I = 3  $. There are 6 candidates who, according to their curriculum, are ranked in order of preference by the company $ I $ in the following way: $ \{a_1, a_2, a_3, l_4, l_5, c_6 \} $. The first three are architects, then there are two lawyers and finally a business consultant. If the $ I $ company makes its choice according to the \textit{Max-Min individual ordering preference} (defined in the previous paragraph) then it is obvious that $ \{a_1, a_2, a_3 \} $ would be hired. But if $ I $ is a small company, it does not need three architects, it would be better to diversify and hire three candidates with three different roles. In this case it would take $ \{a_1, l_4, c_6 \} $. So this example shows us that individual-based ordering preference is not always a good criterion for building the model. \\
Roth and Sotomayor in \cite{roth1992two} demonstrated the following proposition (already mentioned in the previous paragraph):
\begin{itemize}
\item \textit{if preferences are assessed on sets of individuals arbitrarily, then there is no guarantee that a pairwise stable solution exists neither for the many-to-many matching problem nor for the many-to-one matching problem.}
\end{itemize}
In order for a pairwise stable solution to exist, restrictions must be placed on the ordering preference. Before seeing what are the constraints on the ordering preference that have been most adopted in the literature, let us take another example that shows how much it is disturbing, in this context (i.e. with preferences on the sets of individuals and not on single individuals), the presence also of a single "many-to-many contract" in the system \cite{baiou2000many}: \\
let $ W = \{w, w_1, ... w_{2k} \} $ the workers and $ F = \{f_1, ... f_k, f \} $ the firms. The worker $ w $ is the only worker with a share of $ q_w> 1 $. The workers' preference-lists are:
\[P(w_i, i=1,...,2k)=\{f_1,f_2,...,f_k,f\} \;, \]
\[P(w)=\{(f_1,f), f, f_1\} \;. \]
The firms' preference-lists are:
\[P(f_1)=\{(w,w_1), (w_1, w_2)\} \;, \]
\[P(f)=\{w, w_1, w_2,...,w_k\} \;, \]
\[P(f_i, i=2,...,k)=\{(w_{2i-2}, w_{2i-1}), (w_{2i-1}, w_{2i}), (w, w_{2i})\} \;. \]
 The only possible stable matching in this case is:
 \[ f_i \to (w_{2i-2}w_{2i-1}) \;, \]
 \[ f_1 \to (w, w_1) \;, \]
 \[ f \to w \;. \]
While, if $ w $ is allowed to work for only one firm, then we reduce it to the CAP model and the only stable matching becomes:
 \[ f_i \to (w_{2i-1}w_{2i}) \;, \]
 \[ f \to w \;. \]
This example shows that, if even a single worker can have multiple contracts with different firms, then stability changes for a large number of players in the system. So even if in a labor market the general rule is one-to-one or many-to-one contracts, few many-to-many contracts can make a big difference.\\

\subsubsection{Substitutability Constraint} 
We now introduce the concept of "substitutability" as a constraint in ordering preference in order to find a structure of stable solutions similar to that of SMP also for GMM. Let $ f $ be a firm, then the "substitutability" in the preferences of $ f $ is defined in the following way \cite{kelso1982job, sotomayor1999three}:
\begin{itemize}
\item \textit{Substitutability}: if for firm $ f $ assuming the worker $ w $ is optimal when the available workers are $ w \cup \{S \} $, where $ \{S \} $ is a set of workers and $ S' \subset S $, then $ w $ must be optimal for $ f $ even when the available workers are $ w \cup \{S' \} $ 
\end{itemize}
In other words, substitutability requires that, if $ w $ is chosen by a given set of workers, then it must also be chosen by a smaller set of workers. \\
Therefore, under the conditions of substitutability and pairwise stability Roth in \cite{roth1984stability} has shown that there is always at least one pairwise stable solution in the many-to-many matching problem. Furthermore, the same author has shown that, as in SMP and CAP, there are two particular stable solutions which are one worker-optimal and the other firm-optimal. These solutions can be found with the usual GS algorithm procedure adapted, again by Roth to the case of the many-to-many model. There is also an algorithm for finding all possible stable solutions of the GMM found for the first time by Martinez et al. in \cite{martinez2004algorithm}. \\

\begin{figure}[!h]
\begin{center}
\includegraphics[width=0.9\textwidth,scale=0.9]{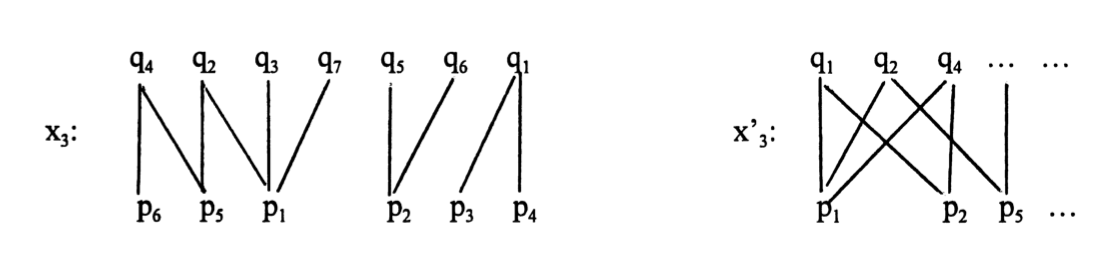}
\caption{\textbf{Corewise stable solution}: $p_1$ prefers $(q_1,q_2,q_4)$ to $(q_2,q_3,q_7)$; $p_2$ prefers $(q_1,q_4)$ to $(q_5,q_6)$; $q_1$ prefers $(p_1,p_2)$ to $(p_3,p_4)$ and finally $q_4$ prefers $(p_1,p_2)$ to $(p_5,p_6)$. Here, the matching $x_3$ is not setwise stable because the agents $(p_1,p_2,q_1,q_4)$ can get a better payoff for all of them by forming a \textit{new} partnership only among themselves, like in the matching $x_3'$ in the (right) figure. However note that the agents $(p_1,p_2,q_1,q_4)$ do not form \textit{all} their partnership only among themselves and so $x_3$ is corewise stable. Figure adapted from \cite{sotomayor1999three}, with permission from Elsevier.}
\end{center}
\end{figure}

Finally, we now mention two other possible definitions of stability which are probably more appropriate for a many-to-many labour market. Pairwise stability, in fact, gives the power to "block" a matching to individual agents (worker-firm) and therefore the stability rule required by this definition is too weak in terms of the duration of the matching \cite{martinez2004algorithm}. \\
The following two more restrictive definitions of stability have been proposed in the literature for a many-to-many market \cite{sotomayor1999three}:
\begin{itemize}
\item \textit{Corewise stability}: a matching is corewise stable if there are no subgroups of agents that, by forming \textit{all} their partnerships only among themselves, can all get a higher payoff.
\item \textit{Setwise stability}: a matching is called setwise stable if there are no subgroups of agents who, by forming their partnerships only among themselves, possibly dissolving some partnerships to remain in their quota and keeping all the others, can obtain a greater payoff.
\end{itemize}
Sotomayor has shown that the definition of setwise stability is a strictly stronger requirement than the other two concepts of stability \cite{sotomayor1999three}. He has indeed shown an example of a case in which a matching is both pairwise stable and corewise stable, but not setwise stable. Figure 20 shows an example of matching which is corewise stable but not setwise stable.
\newpage

\section{SMP Reinterpretation and the Role of Information}
In the above section, we described the wide application of SMP in economics. Here, from the modern point of view of the new theory of information economy \cite{zhang2020matchmakers}, we see that those models have serious flaws and, in the real world applications, one of the major features is not recognized: \textit{information deficiency}.\\
Initially, we will provide a reinterpretation of the original model of the SMP, in particular, we will focus on the Gale-Shapley mechanism and we will see that, using an information-based point of view, women's preference lists are an unnecessary input into the model. This fact has very interesting consequences if we consider SMP as a metaphor for economic society. \\
In the second part of this section, we will see quantitatively how the results of the SMP vary if the concept of partial information is introduced, namely when the preference-lists of the system agents are not complete but have holes that represent the lack of information. The fundamental conclusion is that, if SMP is contextualized as a metaphor for society, the increase in information in society itself increases global happiness. \\
Then we briefly introduce the new theory of information economy. In this way, it will be possible to appreciate how revolutionary the role of information is in the economic and social sphere. For a more detailed and compelling description of this theory, one can refer to Zhang's book \cite{zhang2020matchmakers}. \\
Finally, in the last part, the role of the matchmaker will be described: an external entity that owns all the information available in the system and forces the system agents to match in such a way to obtain the global optimal solution. From a quantitative point of view, we faced this problem at the end of section 3 where the ground state solution of the SMP was analyzed analytically and compared with stable solutions (i.e. in the absence of matchmaker).Here, however, we will place the matchmaker in the context of information economy theory and see what the implications are for society.

\subsection{A New Interpretation for the Gale-Shapley Model}
Let us refer to the original Gale-Shapley algorithm where one side takes the initiative and the other side evaluates whether accepting or rejecting; this was just the choice of a mathematical model, nothing wrong here. But, as an economic model, this raises an important question. Whoever takes the initiative to propose, as we saw, statistically gets much better results than the other side that just sits and waits.  Then one may ask: why does the passive side choose to be passive? Why do not they also play the active role, i.e. both sides propose to each other ?\\
But from this point of view, the original GS model can be re-interpreted as follows. Men and women have to be matched, one side has complete information on the other side (let us say men), whereas the other side (women) has no information on the opposite side. So they just store the information regarding the people that have already proposed to them.\\
Those with information can proceed to rank the other side according to their preference, just like in the GS algorithm. But the women must wait as they do not have any information over men. The women have the same cognitive capability than men but they do not have information so they cannot propose (otherwise these proposals would be completely random). Once a woman receives a proposal, she temporarily keeps him as a candidate, but if she receives two or more candidates, even without any prior knowledge, she is capable to evaluate making a comparison between the two, choosing the best one and rejecting the other. In this way, everything proceeds just like in the GS model. It is important to note that a woman does not have a rank list, but, whenever she receives proposals, she is capable of choosing the best for her so, actually, she never needs the full ranking list as in the original model.\\
Of course, mathematically speaking, the results would be the same as in the original GS model, hence we call the above description as a reinterpretation. Nevertheless, it is useful for a generalization of SMP, i.e. \textit{information deficiency problem}.\\
The re-interpretation sheds lights on the original GS mechanism: here it is clear, missing information impede women to make proposals and hence their passive stance is \textit{not} a choice, but they could not do better otherwise.\\ 
Now, we may ask: if, in the original GS mechanism, women would have the full information, why do not they use it to their advantage? In our opinion, this is contrary to the fundamental principle of rationality.\\
\\
The indirect lesson we learn from the above is that prior information over the other side brings advantages, and missing information puts you at a disadvantage.\\
The real-world applications are full of these circumstances. For example, what happens if both sides have partial information? Following our previous interpretation, each side would choose a hybrid approach, both making and evaluating proposals. Everyone would use as much as possible their information edge to propose, but since he or she misses a substantial part of the information available, they must at the same time evaluate ex post the other side proposals. This means that a person may lack prior knowledge about the other side, but upon receiving a concrete proposal, is capable of evaluating, and comparing (if more than one proposal comes to his/her attention). The best strategy would be to always keep the current best proposal, dropping all the others, and keeping to propose until the current best proposal received is better than the one he/her is about to propose. Similarly, the game will end if the other side accepts his proposal.\\
\\
This modification of the standard SMP has profound consequences: while in mainstream economics, each player is rational but limited by his constraints, here the constraints are the limited prior knowledge on the other side. But rational players must use all the information they have with maximum efficiency. In this light, the original GS model is not rational: if the women have the full information, why should they \textit{choose} to sit and passively wait for the proposals? This is clearly a bad option from the rationality criterion.\\
Of course, one may reinterpret the GS algorithm so that the women, for whichever reason, do not have prior knowledge about the men, but are as intelligent as the men if a proposal or more come to their attention.\\
Acquiring prior knowledge must cost some resources and, if these needed resources are high, one may say that one of the cost-saving measures would be not to evaluate all the men beforehand, but only when the men make proposals, to spend (relatively) much fewer evaluation efforts to check the proposals.\\
But one of the main conclusions is that prior information begets advantages, more prior information begets more advantages, no prior information puts you at a much more disadvantageous position. Nevertheless, the heavily penalized player (from the lack of information) can still get the best he can, even if it would be an optimization under severe limitation to achieve a limited optimization.\\
\\
The lesson learned is the following: "taking initiatives" vs "passively waiting", the former brings more advantages. But it is the amount of information that plays the most important role: if you have the information like the women in the original GS model, they had the full information and not using it is a huge waste of resources (valuable information) that is not condoned by any rationality criterium. On the other hand, with null or inadequate information, taking the initiative to propose is worse off than passively waiting. This is especially true if the quality of information is taken into account (for example the information is not reliable). This latter may induce a player to be overly confident, thinking that he has enough information for taking the initiative to propose. The conclusion is that the degree of initiative-taking should be commensurate with our true information.\\
\\
The case of the GS mechanism is an extreme case in which information is all on the side of men. In the following, we will quantitatively analyze, through a mean-field approximation, what happens to the energies of the agents when the information is partial on both sides. In particular, we will focus on a mechanism in which both parties act as both proposers and judger, meaning that both men and women take the initiative.

\subsection{Stable Marriage Problem with Partial Information}
Computational power is continuously growing with time. According to Moore's law, it is set to double every eighteen months. This increase in computational power will allow for predictions on many systems such as long term weather forecast, financial market predictions and DNA decoding \cite{zhang2001happier}. All of this can be summed up in one sentence: information in society is improving. This will have extraordinary implications in our society and in particular, the increase in information will be the main driving force of economic growth. \\
However, no individual can ever be perfectly informed and it is useful to exploit the metaphor of the SMP to see what happens when each individual has a limited capacity of information. \\
\\
In most of the classic economic models, it is assumed that agents have an infinite availability of information, the same is assumed in the classic SMP. Indeed, all individuals in the system build preference lists by evaluating all individuals of the opposite sex. In real-world, however, it would be very complicated to establish a preference-list for a large number of possible partners because it would be necessary to analyze all the people one by one and the amount of information to be processed would be prohibitive for a normal person.\\
Here we want to modify the classic SMP model so that to include limited information processing power: each man or woman will not be able to build a complete preference-list. Hence, not all individuals will know each other, but the preference-lists will have holes. \\
We will therefore see how the increase (or decrease) of information in the system, defined as the number of individuals present in the preference-lists, modifies the total and individual energy of the agents. There are two extreme cases: one in which both men and women have complete information, and therefore both sides have a motivation to take the initiative, and another in which women do not have their preference-list, i.e. the GS model, and therefore only men benefit from taking the initiative. We will focus on the latter in the next paragraph.

\subsubsection{GS Energy of the Agent with Costly Information} 
We now focus on the energies of men and women in the GS (men-oriented) dynamic when the agents have limited information. First of all, we note that, according to the reasoning made above, since women are passive agents in the GS dynamic, then their preference-lists are superfluous. In fact, at each step, they only have to judge between two men and do not need to know what ranking these two men have on their preference-list. So, once again, the same GS algorithm can be interpreted in such a way that women do not have to waste information processing power.\\
\\
To calculate the energies of men and women in GS dynamics we can follow the reasonings of Caldarelli, Capocci and Laureti in \cite{caldarelli2001sex}. They assume that system agents have limited ability to process information and therefore cannot know all individuals in the system. To limit this information capacity, they consider individuals arranged on a $ k $ -dimensional lattice and assume that everyone can only know individuals at a certain Euclidean distance $ d $. So the "unreachable" partners will be represented as holes in the preference-lists and, under these assumptions, everyone will know only a fraction $ \gamma \sim (d / N)^k $ of the total agents in the system. So the energy of men is equal to the number of proposals divided by $ \gamma $, that is, with the same calculations made in section 3.2.2, we have:
\begin{equation}
x \approx \frac{1}{\gamma}(logN+0.5772) \;.
\end{equation}
Still following the calculations in section 3.2.2 we find that the relationship between the energy of men and women is \cite{caldarelli2001sex}:
\begin{equation}
x \approx \frac{\gamma y}{N[1-(1-\frac{y}{N})^{\gamma}]^2}\;.
\end{equation}
So when the agents have more information, although the competition between them is greater, their energy is lower.\\
\\
So far, we have dealt with the problem using the GS mechanism and we have seen that the benefit to society (of both sides) decreases when the information is less ($ \gamma $ greater). In the following we will see that, as one might imagine, the GS model is not an exception: even when we make both men and women take the initiative, an information deficit harms the whole society and also the side better-informed benefits the most, even if, as we shall see, it is not a zero-sum game.

\subsubsection{Mean-Field Energy with Partial Information} 
Let us assume that men only know $ U<N $ women and that women know $ D <N $ men. For the moment we assume that $ U> D $. The subsets $ U $ and $ D $ are randomly chosen from the total number of individuals of the opposite sex. These hypotheses are more realistic than the classic model since, in reality, no one can know all the people. \\
Our goal is to calculate the average energy of men $ x = X / N $ and women $ y = Y / N $ (where we still consider that each person's energy corresponds to the ranking of the partner on their list). To do this we will use a mean-field approach in the same way as Y.C. Zhang in \cite{zhang2001happier}. \\
Imagine that both men and women are active agents (as they are called in \cite{zhang2001happier}): they are both proposers and, at each step, they make a proposal and evaluate another at the same time. We legitimately consider that the $ U $ women are uniformly distributed in the hypothetical complete list composed of $ N $ women. The list of each man will therefore be "sparse" with an average interval equal to $ N / U $ between a woman and the other. Similar considerations are made for women. \\
So the game begins with every man and woman who propose to someone and evaluate the proposals received at each step. Since the women's lists are sparser ($ D <U $), they will scroll faster and, when the game ends, the men and women have travelled a distance $ x $ and $ y $ respectively in their lists (in general it will be $ y> x $). So on average, every man will make $ xU / N $ before getting married and, since every proposal from a man has a probability $ y / N $ to be accepted, the total probability that a man's proposal is accepted is equal at $ xyU / N ^ 2 $. Similarly, one finds that the total probability that a woman's proposal is accepted is $ xyD / N ^ 2 $. \\
At the end of the trial, there are three types of marriages: those in which only the man has proposed, those in which only the woman has proposed and those in which both have taken the initiative. In the first two cases we mean the man (woman) proposes to a woman (man) who accepts but he (she) is not present in her (his) preference-list. \\
For the matching to exist, it must be that the total probability is equal to 1, therefore we have:
\begin{equation}
\frac{xyU}{N^2}+\frac{xyD}{N^2}(1-\frac{U}{N})=1 \;,
\end{equation}
where the first term contains the sum of the probabilities of marriages in which only the man takes the initiative and in which both take the initiative; the second term, instead, represents only the marriages in which the woman takes the initiative. \\
In addition to this normalization condition, we must add the fact that both men and women scroll their list one step each and that the number of proposals is the same between men and women. So we have to add the condition:
\begin{equation}
xU=yD \;.
\end{equation}
Solving for $ x $ and $ y $ the system, we obtain:
\begin{equation}
x=\frac{N \sqrt{U}}{\sqrt{(U+D-\frac{UD}{N})D}}\;,
\end{equation}
\begin{equation}
y=\frac{N \sqrt{D}}{\sqrt{(U+D-\frac{UD}{N})U}}\;.
\end{equation}
First of all, it should be noted that in the case where $ U = D = N $, the standard SMP theory apply. 
\begin{figure}[!h]
\begin{center}
\includegraphics[width=0.7\textwidth,scale=0.7]{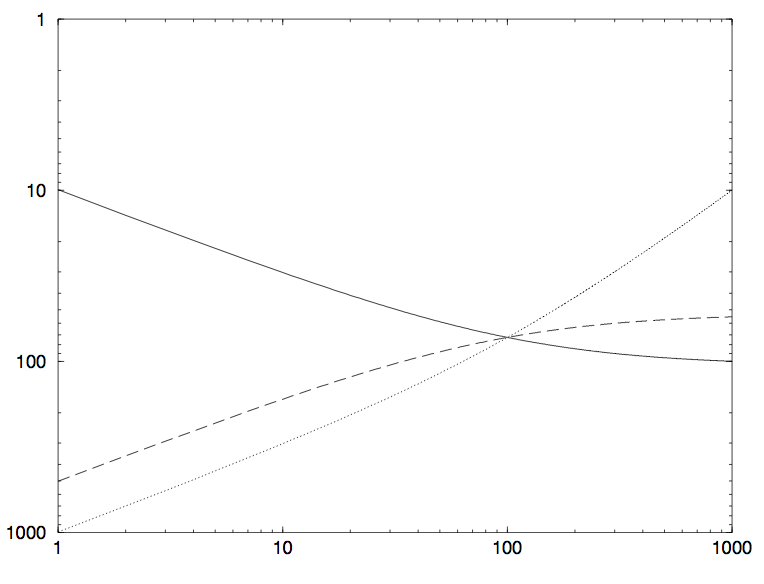}
\caption{\textbf{Average partner's rankings as a function of $D$}: the most up curve from the left correspond to the man's partner ranking, the lowest curve is for the women, and the middle curve is the sum of the two rankings. Here is $U=100$ fixed and $N=1000$. Figure reprinted from \cite{zhang2001happier}, with permission from Elsevier.}
\end{center}
\end{figure}
When $ U> D $ equations (86) and (87) tell us that $ x <y $. In other words, those who have more information also have more benefits. Figure 21 shows the trends of $ x $, $ y $ and $ x + y $ as a function of $ D $, keeping $ U $ fixed. Note that, when $ U> D $, women are much more unhappy than men. By increasing $ D $, the benefit of women increases and that of men decreases. However this decrease does not offset the growth of women's happiness and therefore the total benefits, $ x + y $, grow monotonically with $ D $.\\
All this has important consequences if it is considered the role of information in society \cite{zhang2001happier, zhang2017information, zhang2013broader}, as these results show that the greater the information, the greater the global benefits for society .\\
\\
Another interesting aspect to investigate is whether an initial information difference ca dilate the inequality (in happiness) between the two sides if the information with the same velocity for both men and women is increased \cite{zhang2001happier}.\\
The benefit for men can be defined as $ N-x $, similarly, for women, the benefit is equal to $ N-y $. Now let us grow the amount of information of the same factor for both sides, that is we have that the information of men is $ \xi U $ and that of women is $ \xi D $, where $ \xi> 1 $, and we do increase this parameter. The relative benefit $ R $ can be written as:
\begin{equation}
R=\frac{N-y}{N-x}\;.
\end{equation} 
If, as before, we assume that $ U> D $ we have that R grows monotonically with $ \xi $. So the benefits of the part with less initial information grow faster and this is a socially good result since it means that initial differences do not expand with the uniform growth of information. 

\subsection{Beyond Marriages: the New Theory of Information Economy}

As one can imagine, the information paradigm goes far beyond marriage, as it turns out to be a very important concept in many other areas of social interaction. Indeed, the SMP model serves as a metaphor for other applications. \\
Among the most important social relationships is the commercial one. Consider for example the relationship between firms and consumers. It is generally assumed that a transaction between a firm that sells a product and a consumer who needs this product benefits both parties. In the standard economy, it is assumed that the benefit is equal for both parties and the so-called \textit{economic pie} is the largest. But modern products are much more complex than those of the past, and consumer needs are also more diversified and complicated: the simple law of \textit {supply and demand} based solely on the price of products is no longer adequate \cite{zhang2001happier}. In today's world, finding compatibility between the products sold by firms and consumers takes time and efforts. Consumers cannot know all the potentially useful products and services in the world. Similarly, a firm cannot know all the wishes of all consumers in the world. Therefore, an essential skill in commercial relations is knowing how to find the right counterpart.\\
In this section, we have shown how, through the metaphor of the SMP, the economic pie is as large as possible when the information is the same on both sides, but it is easy to understand that in the real world, in most of the commercial transactions there is \textit{information asymmetry} between firms and consumers. With this asymmetry, the economy is not the largest, but the advent of big data and greater computational power could enlarge this economic pie \cite{zhang2020matchmakers}. \\
\\
Commercial transactions differ from marriages in that, unless it is a polygamous relationship, marriages are one-to-one matchings, while a consumer could use multiple products simultaneously (many-to-one matching). Moreover, at least in principle, people marry only once in their life, while a commercial operation can be repeated several times (in the same restaurant one can go several times in a year).\\
Social interactions are not limited only to marriages or commercial relationships: in today's community the collectivity is becoming more and more important and each of us, in a certain sense, relies on others. Therefore the interconnection between people has never been so intense. Through the internet, one can get in touch with people overseas and, in this way, it is much easier than in the past for people to find each other. Economic and academic institutions are becoming more and more complex and elaborate, allowing previously unthinkable cooperation that could benefit the whole society. \\
In this sense, understanding the importance of information on social relations in the contemporary world is increasingly fundamental.  

\subsubsection{The Role of Information in the Economic Transaction}
As we have just mentioned, consumers and firms look for each other in the market, either independently or helped by third parties (the so-called matchmakers). In standard economics, the information problem is not considered in supply-demand law, conversely, the new theory of information economy focuses on how consumers and firms find each other and on how they manage information deficiency problem. \\
Since the world is increasingly complex, interconnected, and the variety of products is almost unlimited, it is unrealistic to think that consumers will increase their information capacity aligned with time, even if they are increasingly diligent . The new information economy theory goes beyond the mainstream supply-demand law and considers a \textit {cognitive grey area}: if consumers become more and more informed, then they will be able to find more useful products, so it will be easier to match the wishes of consumers with the offer of firms.\\
The new information economy theory postulates that, in addition to the transactions carried out, there is an infinite pool of potential consumers' wants and firms' offers that can be discovered as the information increases.\\
The fundamental conclusion is that the growth of information could enhance economic growth over time without limits. Better transactions can benefit both consumers and firms and therefore the concept of \textit{magic pie} of wealth can be introduced \cite{zhang2020matchmakers}: the classic economic pie is no longer static but can vary in size depending on the quantity of information in the system. The concept of a magic pie implies that economic transactions are not a zero-sum game (the gain of one player does not correspond to the loss of the other) and therefore economic problems should not be treated as optimization problems under constraints.\\

\begin{figure}[!h]
\begin{center}
\includegraphics[width=0.7\textwidth,scale=0.7]{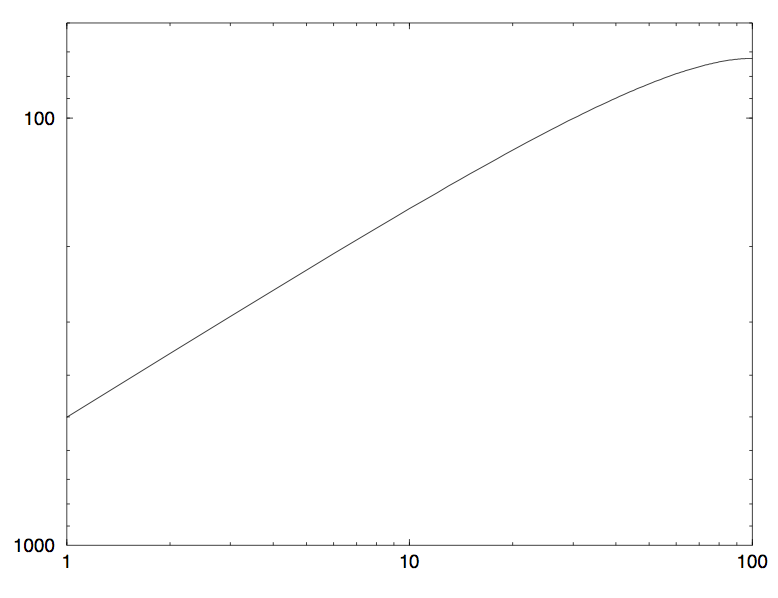}
\caption{\textbf{Sum of rankings of men and women}: $S$ is fixed but $M$ and $K$ vary. Figure reprinted from \cite{zhang2001happier}, with permission from Elsevier.}
\end{center}
\end{figure}

To show how the increase in information benefits both parties, let us go back to the SMP model in which the information of men $ M $ is fixed while we vary the information of women $ K $. As can be seen in figure 21, when women's information capacity surpasses that of men, the benefit of women increases consistently. During this process, the benefit of men decreases although $ M $ is fixed. On the other hand, the loss of men does not compensate for the gain of women and therefore the total benefit, $ S = X + Y $, increases.\\
It is easily shown \cite{zhang2001happier} that the first derivatives of $ S $ with respect to $ K $ and $ M $ are both negatively defined. This implies that this model is not a zero-sum game, and so, exactly as speculated above, a magic pie concept should be introduced. Therefore increasing information, even unilaterally, is beneficial for the whole society. \\
\\
We can also examine the information asymmetry between men and women \cite{zhang2001happier}. Suppose that the total information capacity is fixed, $ M + K = const $. So $ K $ can be raised and $ M $ reduced at the same time. The benefit of women increases while that of men decreases. The most interesting thing is to consider the total sum, or, in economic language, the total size of the economic pie. In our case, it corresponds to the sum of $ N-X $ (men's benefit) and $ N-Y $ (women's benefit).\\
Figure 22 shows that the pie's size is maximum when $ K = M $. The conclusion is that the increase in information can benefit both parties, but this benefit is not always equally divided: if the pie is equally divided then the pie's size is the largest possible.\\
In the following, we will go more in detail about the asymmetry between the two parties and we will analyze the causes and the consequences in the economic society.

\subsubsection{The Fundamental Asymmetry}
The new information economy theory differs from standard economic theories in that, instead of maximizing under fixed constraints, variable constraints are considered here: the desires of consumers and the offer of firms are constraints that can expand over time, the same occurs for their information. At this point, the concept of \textit{fundamental asymmetry} must be introduced \cite{zhang2017information}: it refers to the fact that, even if the constraints of both parties (consumers 'wishes and firms' offer) may vary, it is easier to modify the constraints of firms.\\
When firms are pressed by the market, they have many more opportunities to modify their offer than consumers to expand their wants. Firms can change prices or improve the quality of their products in the short term. While in the long term they can find more efficient ways of production or find new products that anticipate consumers' desires.\\
Note that the driving force that motivates firms to make such efforts is the pressure of consumers. Indeed, it is the increasingly growing demand from consumers that forces firms to renew themselves continuously.\\
Conversely, since each person's income is limited, they can expand their desires more slowly \cite{zhang2017information}. \\
Fundamental asymmetry also manifests itself in another aspect: the tolerance of substitutability \cite{zhang2017information}. If a consumer has a specific desire for a product, it cannot be replaced with another product (if I want a tennis racket, I am not satisfied with a ping pong racket); while for firms it is not a problem to replace customers (if I do not buy that ping pong racket, nothing changes for the seller if another customer buys it). \\
\\
One of the consequences of this asymmetry is that, since firms can expand their offer more easily than consumers can do with their wants, then firms cannot expand their offer without commensurate with consumer demand. Therefore firms should develop business models that exploit fundamental asymmetry to create profit and, at the same time, create benefit for consumers. In \cite{zhang2013broader} for example, it is shown in detail that many new emerging business models have started to follow this trend. \\
Fundamental asymmetry is not only important for business models but also for determining the long term trend of economic evolution. In the short term, a company or its marketing institutions may choose to use smart marketing tools to leverage consumers for profit as this is easier to do. However, according to the fundamental asymmetry, companies have much more space to expand their offers by how much consumers have to expand their desires. Innovative entrepreneurs will be encouraged to follow the prevailing trend aligning their business models and putting more effort into innovation. In the long run, anyone who is following the trend will obtain important rewards, and anyone who goes against the trend could make short-term profits but have improbable long-term prosperity \cite{zhang2017information}.

\subsubsection{The Non-Equilibrium Paradigm}
The most important thesis of the new information economy theory is that information can always grow and improve with the consequence of enhancing economic growth by expanding both supply and demand \cite{zhang2017information}. The consequence of the increase in information is that, not only will there be more transactions, but there will also be more diversification between products. The reasons for this diversification are as follows. First of all, if consumers have more information on the products offered by the firms, they will be able to reveal more diversified desires. This is true also because very often consumers have implicit wants which they become aware of only if prompted by the knowledge of other products. This does not mean that consumers will one day have perfect information since, at the same time, businesses also have an interest in diversifying their products over time. This is because, with increasingly informed consumers, competition in the market intensifies. Greater competition forces firms to develop new frontiers of production where competition is lower \cite{zhang2020matchmakers}.\\
To give a concrete example, consider a situation in which the economy is in a recession, like after a war. In this case, consumer wants are limited only to goods to survive. On the contrary, in the case of an economic boom, consumers are willing to buy even non-essential goods and therefore diversification increases.\\
Each individual has an unlimited number of desires that are not strictly necessary that are hidden in the grey cognitive zone mentioned above and that can be activated through the appropriate stimuli. The economy is expected to grow through a dynamic process in which the wants of consumers are matched with the offer of the firms. Hence, new implicit wants are discovered and therefore the matching process starts again and so on.\\
\\
As we mentioned, standard economic theories are \textit{static} theories that treat economic problems as optimization problems under constraints. On the other hand, the new theory of information economy has its foundations in the concepts of the grey cognitive zone and information improvement. As we have shown above, this growth in information increases the volume of transactions and creates new consumer needs and an innovative offer from firms.\\
Therefore, the new paradigm postulates that each transaction (or each matching) will modify the constraints, creating new opportunities and new risks in the market. According to this new paradigm, we cannot expect to solve any problem neglecting the variability of the constraints. So, simple optimization is not possible in principle. \\
In conclusion, the economy must be treated through a dynamic theory in which there is no set goal to be achieved and there is no predetermined target towards which the economy converges. In this context, information improvement will be the main driving force for an expanding economy. This, in our opinion, is a revolutionary and exciting scenario.

\subsection{The Role of a Matchmaker} 
We now deepen the question: how do people find each other? As we have seen in the course of this review, returning to the marriage metaphor, there are two ways in which people can find their partner. The first possibility is that everyone is looking by themselves so that, by hiring perfectly rational players, they reach a stable solutions of the SMP; the second possibility is that the searches are entrusted to a matchmaker to obtain the ground state. \\
\begin{figure}[!h]
\begin{center}
\includegraphics[width=0.7\textwidth,scale=0.7]{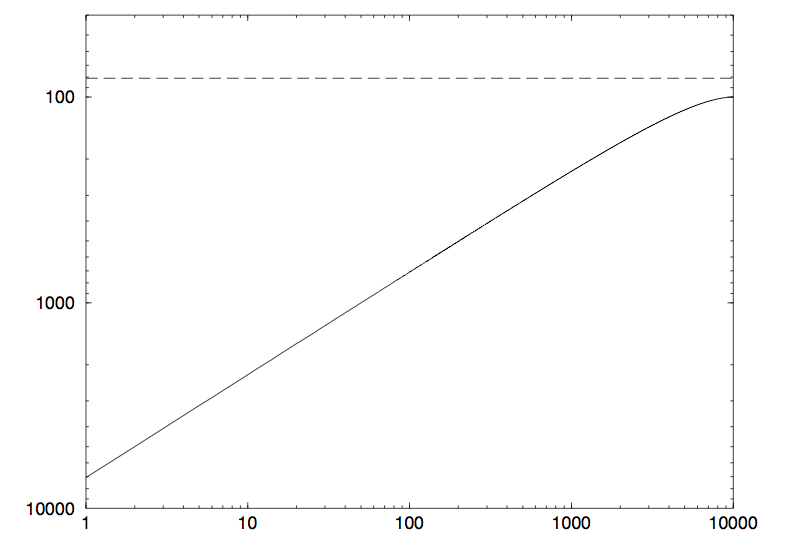}
\caption{\textbf{Happiness of the society as a function of information in the system}: $N=10000$. The dashed curve is the happiness of the society in the matchmaker solution, the solid line is the happiness obtained with the self-searching strategy. Figure reprinted from \cite{zhang2001happier}, with permission from Elsevier.}
\end{center}
\end{figure}
If the first strategy is adopted, each agent must obtain information on all other agents and then there will be $ N ^ 2 $ searches, i.e. the society will spend $ N ^ 2 $ units of effort. If instead the second strategy is adopted, only the matchmaker will have to carry out research and therefore it is clear that the total effort of the society will be $ N $ unit.\\
In addition to the implications for the society's efforts, there are several advantages and disadvantages of adopting one or the other strategy. \\
In general, the matchmaker will try to maximize the \textit {social pie}, which is the sum of the benefits of all the agents in the system minus the commissions to be paid. So the matchmaker has the welfare of society as his goal. However, this is not because of his infinite goodness, but also the matchmaker himself acts selfishly. Indeed, his profits derive directly from the commissions that the agents pay and they are proportional to the total social pie. It can be shown that two groups with partially opposite interests (men and women) and a selfish matchmaker manage to achieve great social pie along with stability as they all act for themselves unintentionally creating benefit for the society \cite{zhang2001happier}. In the next paragraphs, we contextualize the role of matchmaker in the economic market.
\\
In any case, the role of the matchmaker improves the satisfaction of the society that acts selfishly. Taking up the SMP model, Zhang showed in \cite{zhang2001happier} that the gap between stable solutions, obtained through the first strategy mentioned above, and the ground state obtained with the help of a matchmaker increases with decreasing information in the system (figure 24). Obviously, when agents have total information, the gap is reduced to that studied in section 3.5.

\subsubsection{Independent Players and Competition between Matchmakers }
As shown in figure 23, when the information is partial and the players play selfishly, the result they get is very poor. So someone else should assign the partner rather than looking by themselves. The idea of the matchmaker, as we will see in the next paragraph, becomes more and more functional in a world where the connectivity of human social contacts is vast \cite{liljeros2001web}. In fact, for example on the Internet, a few highly connected nodes can contain a huge amount of information \cite{faloutsos1999power}. \\
Let us imagine a society where people agree to give their preference-list to a matchmaker M. Therefore M has all the information available and can find the ground state of the system. Then M offers each player a partner and, if everyone accepts M's proposal, then the society will maximize global happiness. However, a man $ m $ and a woman $ w $ might prefer to marry each other rather than accept M's proposal. This happens for about $ 24 \% $ of men (or women), as we have seen in section 3.5. When this happens, the partners of $ m $ and $ w $ in the ground state are forced to look for a new partner. We will call $ w $ and $ m $ \textit{independent people}. Now, assume that the information is partial and that each person knows only a portion $\alpha $ of the world. When $ \alpha = 1 $, we have the classic case of complete information. In \cite{laureti2003matching} Laureti and Zhang calculated the percentage of independent people $ p (\alpha) $ as a function of the system information, showing that $ p (\alpha) \sim 0.2 \alpha $. So, the more information the more independent people are. When $ \alpha = 1 $ their result agrees with the value of $ 24 \% $ mentioned above. \\
\\
In real life, it is rare for people to employ great efforts to find a replacement for their partner if the partner satisfies them enough \cite{simon1972theories}. For this reason, many people would accept M's proposal, saving time and getting, on average, a good partner.\\
This argument becomes clearer if we imagine that the game is repeated several times: in this case, the \textit {reputation} \cite{laureti2002buyer} comes into play. People would care more about their average utility and could benefit from a long relationship of confidence with the matchmaker \cite{laureti2003matching}. \\
At this point, we must consider the possibility that the matchmaker is corrupt and therefore favours some players at the expense of collective happiness. In any case, the matchmaker can only survive if the average happiness of the people who accept his proposal is higher than that of the people who search by themselves without the help of M.\\
Additionally, when the information given to M is not well used, it becomes more important for people to preserve their privacy. In this situation, it is no longer profitable for M to be bribed and a sort of equilibrium arises. At this equilibrium point, average happiness is higher than that which would be obtained without M \cite{laureti2003matching}. Another aspect to consider is the possibility of competition between multiple matchmakers. All these situations (competition between matchmakers, corruption etc.) can be better studied in a real-life context \cite{lu2009role}. In particular, it is interesting to study the main matchmaker institutions existing in today's economy. In the next paragraph, we will go to deepen this topic.

\subsubsection{Matchmaker Institutions in the Market}
In the contemporary market, whose products are many more and much more diversified, individual consumers cannot cope on their own with the increasingly numerous and complex offer of firms \cite{zhang2020matchmakers}. The task of comparing multiple products and verifying their quality is becoming increasingly difficult. For this reason, more and more information institutions, which we will call matchmakers, are springing up to help consumers find the best products. The most obvious examples are tripadvisor.com, Amazon, eBay or alibaba.com. \\
Information institutions can use the most advanced techniques developed with the advent of big data to recommend relevant products to consumers. For these reasons, matchmakers have a central role in contemporary markets: even if our perception is that of finding products independently, many institutions guide consumers' research, improving their evaluation skills and, at the same time, helping firms to do selective advertising (aimed at specific targets).\\
Matchmaking institutions have an important advantage: a product can be repeatedly viewed over and over again by many different consumers, potentially allowing a customer to benefit from all the information from other consumers by studying that product only once. This way matchmakers become very efficient at bypassing the \textit{information deficiency problem} mentioned above. However, as we mentioned in the previous paragraph, matchmakers are not necessarily impartial between consumers and firms: they often choose to help one side over the other and usually have a great deal of leeway to do so \cite{zhang2020matchmakers}.\\
Due to the limitation of information on both sides, consumers and firms rely on third parties to gather information. This gives matchmakers a great deal of leeway to decide who to help more: the prevailing business models favour firms and this is mainly because firms have a greater motivation to pay matchmakers for their information services. The reason why firms tend to pay matchmakers more willingly can be analyzed in detail in Tibor Scitovsky's  \textit{consumer theory} \cite{scitovsky1995economic}. He suggested that firms are "specialists" while consumers are "generalists". In this way, since the firms are specialists and therefore more "analytic", they can better evaluate the benefits that a matchmaker would bring. Furthermore, consumers are much more numerous than firms and therefore it is easier for matchmakers to establish contracts with the latter \cite{coase1960problem}. \\
Ideally, the matchmaker should maximize the total benefit of both sides. However, there are reasons why this does not happen. The first is a short term reason and concerns the fact that one of the two parties, the firms, pays more willingly. The second is a long term reason and concerns the fact that firms can expand more easily, so matchmakers must align themselves with consumers for matching to be effective.\\
Therefore, for a matchmaker, short term profits conflict with long term prospects and the goal for the future will be to build business models that benefit both parties symmetrically, avoiding the incentives to bribe matchmakers in exchange for short term profits.
\newpage

\section{Recent Studies on SMP Extensions}
The SMP, although it appears to be a simple problem, has aroused a great deal of interest and the diameter of the possible applications is huge. \\
Now, the SMP scheme is an ideal but plausible representation of reality: it manages to capture the essence of the main two-sided markets; the definition of stability naturally fits into the context of the Nash equilibrium in which the agents of the system try to maximize their utility and not the global one; finally, we also saw that if stability is no longer used, i.e. one seeks for the ground state, non-trivial results appear. \\
With the same philosophy of the SMP, it is possible to imagine equally simple but also non-trivial models, which live on different geometries and which therefore schematize different relationships between the agents.\\
Think of $ N $ agents (without distinction between males and females) who, in the same spirit of the SMP, each have a preference-list of the other people in the game. This time, instead of wanting to create a set of stable marriages (or stable double rooms), we want these $ N $ people to sit on a round table and be positioned in such a way that for no one is convenient to change places. This is the problem we will deal with in the second part of this section. In the first part, we will deal with another problem similar in spirit to the SMP, but instead of having men and women, the two parties will be represented by only two negotiators who have to find an agreement among $ N $ possible outcomes, based on their preference-lists. \\
\\
The study of these two problems is recent and was inspired by the simplicity and beauty of the SMP. They are still being studied and their possible extensions are manifold. Both the models that we will treat in this section are simple and ideal models of real situations and we believe they are as promising as the SMP.

\subsection{Negotiation Problem}
Often in social and economic processes, people must reach an agreement with another party. They may have partially overlapping interests and conflicts which lead to a negotiation process for the interested parties.\\
To study such behaviours, three of us propose a simple model of negotiation which has been found not so trivial, called \textit{Negotiation Problem} (NP) \cite{negprob}. Indeed, it belongs to a well-known class of models ranging from spin-glasses to the random matching model, and other models in physics and beyond.\\
The NP, in its basic version, concerns the ideal situation in which two people have to find an agreement among $ N $ possible alternatives to choose, and both players know all the alternatives. For concreteness, think of two friends who must agree on which club to choose to go out in the evening. In this case, the possible alternatives are all the clubs in the city and the two friends, exactly as in the SMP, have a preference-list for these clubs.\\
We assume that the choice of a particular alternative gives a payoff directly correlated with the position in the preference-list of the two players: the highest payoff for the first choice item, the lowest payoff for the worst choice item.\\
Then the goal is to study a \textit{negotiation process} in which the two players do not compete with each other but they selfishly try to maximize their payoff. Note that in this way the game is a non-zero-sum, i.e. the gain of one player does not imply the loss of the other.\\
In a sense, the NP can be seen as a variant of Nash, Rubinstein and von Neumann's models about the well-known bargaining theory \cite{nash1950bargaining,binmore1986nash, rubinstein1982perfect} (or if we want to be more imaginative, like a "two-person version" of the famous \textit{secretary problem} \cite{freeman1983secretary, ferguson1989solved}).\\
\\
In the SMP we mainly studied two ways to create marriages: 1) create a set of stable marriages (in the sense of Nash); 2) create a set of marriages that maximize the total happiness of the system.\\
Similarly in the NP, we can study two ways to approach the problem:
\begin{itemize}
\item Find the alternative on which both players agree, namely the alternative such that if replaced, the payoff of at least one of the two players will decrease. We will call this alternative \textit{negotiated alternative}. 
\item Find the alternative that maximizes the total payoff, i.e. the sum of the payoffs of both players. We will call this alternative \textit{global best alternative}.
\end{itemize} 

\subsubsection*{Negotiation Process and Negotiated Alternative}
Imagine a negotiation process in which each of the two players proposes an alternative to the other, and the one that receives the proposal must accept or reject. The process ends as soon as the two players find an agreement based on their strategies.\\
Finding the optimal strategy can be very complicated and time-consuming, so we consider the simplest possible strategies: players make their proposal in turn and the same alternative cannot be proposed more than once by the same player. Since each of the two players wants to maximize their payoff, they will start by proposing the alternative at the top of their lists, then the second, and so on until a proposal is accepted. We note that in this way the two negotiators know only their preference-lists and have no information on the preferences of the other player. Furthermore, we will show that this strategy, even though it is very simple, is quasi-optimal, in the sense that there is a large probability that, with this strategy, one will find the global best solution (explained in the following paragraph).

\subsubsection*{Global Best Alternative}
Parallel to the negotiation process described in the previous paragraph, we can hypothesize the existence of a matchmaker who tries to maximize the total payoff. To do this he must force the two negotiators to choose both the \textit{global best alternative}, i.e. the alternative that maximizes the total payoff. The matchmaker must have all the information available and must therefore know the preference-lists of both players.

\subsubsection{The Model}
We now formalize the NP. Consider two players $ A $ and $ B $ and a set of $ N $ possible alternatives $ \{o_i \}_{i = 1, ..., N} $. Each player has a preference-list on these alternatives. At each player is assigned a payoff for each alternative. The better the ranking of this alternative in the preference-list, the higher the payoff. To use a more familiar language in physics, instead of talking about "payoff", we will talk about "energy", which will be minimal if the alternative is at the top of the preference-list and maximum if it is the last choice.\\
So, similarly to the SMP, each player will try to minimize their energy, and in the same way, the matchmaker will try to minimize the total energy. The energies of the players $ A $ and $ B $ related to alternative $ i $ will be written as $ \epsilon_i ^ A $ and $ \epsilon_i ^ B $ respectively. \\
In the negotiation process, $ A $ is the player who makes the first proposal (proposes the first alternative), $ B $ will consequently be the second. Players continue the game one turn at a time by alternating their proposals. In this way it is natural to introduce a "time variable" $ t $, which defines the number of proposals that the player $ A $ has made. Since each player first proposes his first choice, then the second and so on, the variable $ t $ represents how much the player $ A $ has gone down in his preference-list. \\
A negotiation process is defined as a set of proposals $ p_t ^ A $ and $ p_t ^ B $ which stops at time $ t ^ * $ when an agreement has been found. The energies associated with the two players at time $ t ^ * $ are defined as $ \epsilon ^ A $ and $ \epsilon ^ B $.

\begin{figure}[!h]
\begin{center}
\includegraphics[width=0.9\textwidth,scale=0.9]{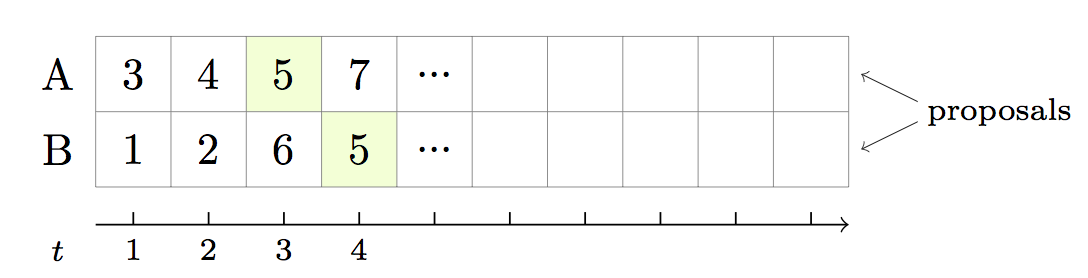}
\caption{\textbf{Example of a negotiation process}: the green box is the negotiated alternative.}
\end{center}
\end{figure}

Figure 24 shows an example of a negotiation process: the $ N $ items are labelled with the numbers from 1 to $ N $ and the two players propose scrolling through their preference-list. Therefore, since in the figure each box represents a proposal, it is clear that the list of $ A $ is $ \{3,4,5,7, ... \} $ and that of $ B $ is $ \{ 1,2,6,5, ... \} $. The first proposal that is repeated twice (not by the same player of course)) is the negotiated alternative defined above. So in this case the chosen alternative is $ o_5 $ at time $ t ^ * = 4 $ and with energies $ \epsilon_5 ^ A = \epsilon ^ A = 3 $ and $ \epsilon_5 ^ B = \epsilon ^ B = 4 $.

\subsubsection*{Interesting Quantities}
Now let us define the most interesting quantities to calculate in the model just described. Given a particular instance of the problem (i.e. given two random preference-lists for players $A$ and $B$) we are interested in the following quantities:
\begin{itemize}
\item Total energy: $s(i)=\epsilon_i^A+\epsilon_i^B$. 
\item Gap function: $\Delta(i)=|\epsilon_i^A-\epsilon_i^B|$.
\item Negotiated alternative: the alternative $ z $ that is chosen after a negotiation process at time $ t ^ * $.
\item Global best alternative: the alternative $ l $ that minimizes the total energy $ s (l) $.
\end{itemize}

With these quantities it is possible to compare the solution of the negotiation process (the negotiated alternative) with the global best alternative. However, what interests us most is not the result of a particular instance of the problem, but the average result we get given a statistical distribution of the preference-lists. The simplest choice that can be made is to take random permutations $ \pi_N $ of the preference-lists, i.e. given $ N $ alternatives there are $ N! $ possible preference-lists and therefore the probability of one in particular is $ 1 / N! $. So the interesting quantities to study become:
\begin{itemize}
\item Average negotiation energy: $\overline{\epsilon^A+\epsilon^B}=\frac{1}{(N!)^2} \sum_{\epsilon^A \in \pi_N} \sum_{\epsilon^B \in \pi_N} s(z) $.
\item Average minimum energy: $\overline{s}=\frac{1}{(N!)^2} \sum_{\epsilon^A \in \pi_N} \sum_{\epsilon^B \in \pi_N} min(\epsilon^A_i+\epsilon^B_i)$ with $i=1,...,N$.
\item Average negotiation gap function: $\overline{\Delta_{neg}}=\frac{1}{(N!)^2} \sum_{\epsilon^A \in \pi_N} \sum_{\epsilon^B \in \pi_N} \Delta(z)$.
\item Average global best gap function: $\overline{\Delta_{glo}}=\frac{1}{(N!)^2} \sum_{\epsilon^A \in \pi_N} \sum_{\epsilon^B \in \pi_N} \Delta(l)$.
\end{itemize}
In these definitions the quantities $ (\epsilon_1 ^ A, ..., \epsilon_N ^ A) $, $ z $ and $ l $ are now stochastic variables that depend on the configuration of the permutation of the preference-lists. Note also that given the symmetry between the players $ A $ and $ B $, it is possible to consider permuting only the list of one of the two players leaving the other one fixed.

\subsubsection{Negotiation Process Results}
We begin to evaluate the interesting quantities related to the \textit {negotiation process}. First we calculate the probability $ P(\epsilon ^ A, \epsilon ^ B) $ that the game ends with energies $ \epsilon ^ A $ and $ \epsilon ^ B $. If a negotiation process stops at time $ t ^ * $, it means that all proposals made in $ t <t ^ * $ have been rejected. From the perspective of one of the two players, this means that each of his proposals in $ t <t ^ * $ has, for the other player, energy greater than $ t ^ * $. Otherwise, that player would have made such proposal before $ t ^ * $ and the process would have already ended. So the probability that the i-th proposal is greater than $ t ^ * $ corresponds to $ 1- \frac {t ^ *} {N-i + 1} $, where normalization $ N-i + 1 $ takes into account the fact that, at time $ t = i $, the possible alternatives to be proposed are less than the total (since the first $ i-1 $ items have already been proposed). Similarly, the probability that the alternative with energies $ \epsilon ^ A $ and $ \epsilon ^ B $ is accepted at time $ t ^ * $ corresponds to $ \frac {1}{N-t ^ * + 1} $. Putting it all together, the probability that the game ends with energies $ \epsilon ^ A $ and $ \epsilon ^ B $ is:
\begin{equation}
P(\epsilon^A,\epsilon^B)=\frac{1}{N-t^*+1} \prod_{i=1}^{t^*-1}\left(1-\frac{t^*}{N-i+1} \right) \;.
\end{equation}
It is now convenient to write $ t ^ * $ in terms of energies: $ t ^ * = max (\epsilon ^ A, \epsilon ^ B) $. Recalling that, for small $ x $, it holds $ e ^ x \sim 1 + x $, we have that, for large $ N $, the equation (89) becomes
\begin{equation}
P(\epsilon^A,\epsilon^B) \sim \frac{1}{N} \exp\left\lbrace \frac{-max(\epsilon^A,\epsilon^B)^2}{N}\right\rbrace \;.
\end{equation}

At this point we can use the probability distribution (90) to average the total energy $ \epsilon ^ A + \epsilon ^ B $. We can write

\begin{equation}
\overline{\epsilon^A+\epsilon^B}=\sum_{\epsilon^A=1}^{N} \sum_{\epsilon^B=1}^{N} P(\epsilon^A,\epsilon^B) (\epsilon^A+\epsilon^B) = \sum_{\epsilon^A=1}^{N} \sum_{\epsilon^B=1}^{N} \frac{1}{N} \exp\left\lbrace \frac{-max(\epsilon^A,\epsilon^B)^2}{N}\right\rbrace (\epsilon^A+\epsilon^B)    \;.
\end{equation}

To solve the equation (91) we pass to the limit $ N \to \infty $, therefore we pass to the continuous transforming the summation into an integral. The final result is:

\begin{equation}
\begin{aligned}
\overline{\epsilon^A+\epsilon^B} \sim \frac{3\sqrt{\pi}}{4} \sqrt{N} \;.
\end{aligned}
\end{equation}

\begin{figure}[!h]
\begin{center}
\includegraphics[width=0.9\textwidth,scale=0.9]{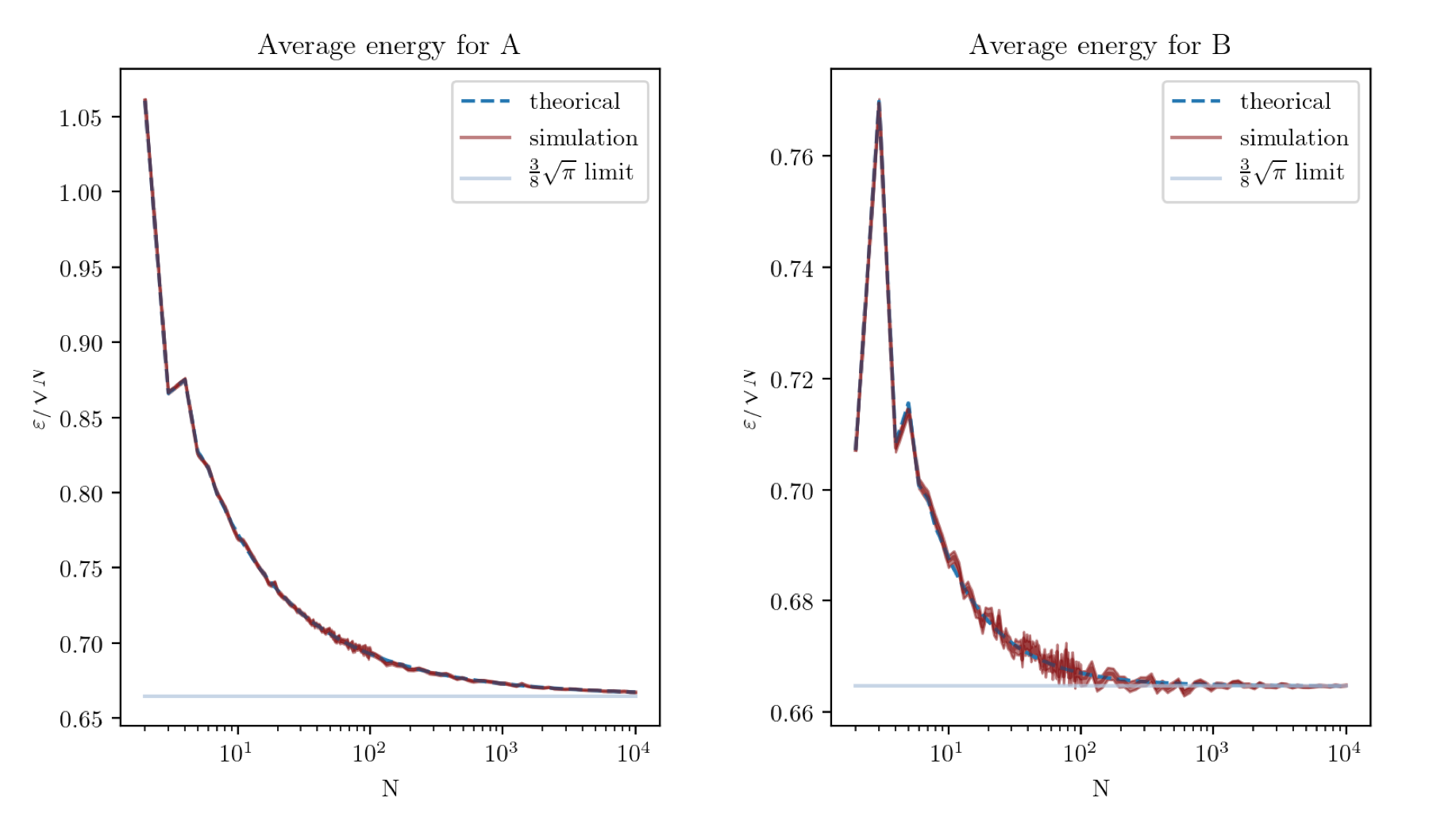}
\caption{\textbf{Average energy in the negotiation process}: theoretical (equation 8.1) and simulation results for the average energy in the negotiation process. The energy is normalized by a factor $N^{1/2}$ and is plotted against the number of alternatives N. The asymptotic limit is indicated by the grey line. The simulations are done by finding the negotiating solution over 100000 iterations and the shaded area is the confidence level within the standard error.}
\end{center}
\end{figure}

Now we also calculate in a similar way the \textit{average negotiation gap function}, $ \overline{\Delta_ {neg}} $. \\
As before we use the probability distribution (90) to average the gap function $\Delta(z)=|\epsilon^A-\epsilon^B|$:
\begin{equation}
\overline{\Delta_{neg}}=\sum_{\epsilon^A=1}^{N} \sum_{\epsilon^B=1}^{N} P(\epsilon^A,\epsilon^B) |\epsilon^A-\epsilon^B|=\sum_{\epsilon^A=1}^{N} \sum_{\epsilon^B=1}^{N} P(\epsilon^A,\epsilon^B) (max(\epsilon^A,\epsilon^B)-min(\epsilon^A,\epsilon^B)) \;.
\end{equation}
By again adopting the limit for large $ N $, we have

\begin{equation}
\overline{\Delta_{neg}} \sim \frac{1}{2}\Gamma\left(\frac{1}{2}\right)\sqrt{N}=\frac{\sqrt{\pi}}{4}\sqrt{N} \;.
\end{equation}

Note that the average of the total energy is three times larger than the average of the energy difference between the two players, i.e. 
\begin{equation}
\overline{\Delta_{neg}}=\frac{\overline{\epsilon^A+\epsilon^B}}{3}\;.
\end{equation}

We did some approximations to obtain the previous results. The most important approximation in the thermodynamic limit ($ N \to \infty $) is the exchange between integral and summation: in general, given a test function $ f (x) $ the error resulting from changing the summation with the integral is as follows
\begin{equation}
\text{error}= \sum_{i=1}^{N} f(i)-\int_{1}^{N} f(x) \,dx=\sum_{i=1}^{N} \int_{i}^{i+1} [f(i)-f(x)] \,dx \;.
\end{equation}
In our case the quantities we want to calculate are the moments $ \overline {x} ^ q \sim \int_ {1} ^ {N} \frac {x ^ {q + 1}} {N} e ^ {- x ^ 2 / N} \, dx $. So our test functions boil down to the family $ f (x) = \frac {x ^ {q + 1}} {N} e ^ {- x ^ 2 / N} $ for each $ q> 0 $. So in our case, by the changing of variable $ x = \omega \sqrt {N} $, equation (96) is solved as follows
\begin{equation}
\begin{aligned}
\text{error}=N^{q/2} \sum_{\omega_i}^{\omega_N} \int_{\omega_i}^{\omega_{i+1}} \left( \omega_i^{q+1} e^{-\omega_i^2}- \omega^{q+1} e^{-\omega^2}\right)= \\
= N^{q/2} \left(\sum_{\omega_i}^{\omega_N} e^{-\omega_i^2}[2\omega_i^{q+2}-(q+1)\omega_i^q]\frac{\Delta\omega_i^2}{2} +O(1/N)  \right) \sim N^{q/2} \left(\frac{C}{N} +O(1/N)  \right) \;,
\end{aligned}
\end{equation}
where we have defined 
\[\Delta\omega_i=\omega_{i+1}-\omega_i=1/\sqrt{N} \;, \]
\[C=\sum_{\omega_i}^{\omega_N} e^{-\omega_i^2}(2\omega_i^{q+2}-(q+1)\omega_i^q) \;.  \]
Note that the constant $ C $, due to the exponential term, does not strongly depend on $ N $. Thus, for large $ N $, the corrections are of order $ N ^ {q / 2-1} $. \\
\\
In the following, we will study other aspect and some small variants of the negotiation process.

\subsubsection*{Switching the Players}

By convention in the previous paragraphs, we referred to the player $ A $ as the first player to propose during the process. However, the player $ A $ may want to change this rule. Note that the first to perform his turn suffers a considerable disadvantage with respect to the other player. If $ A $ makes the first move, $ B $ has the advantage of accepting only the proposals that have an energy $ \epsilon^B_i <t $. \\

\begin{figure}[!h]
\begin{center}
\includegraphics[width=0.6\textwidth,scale=0.6]{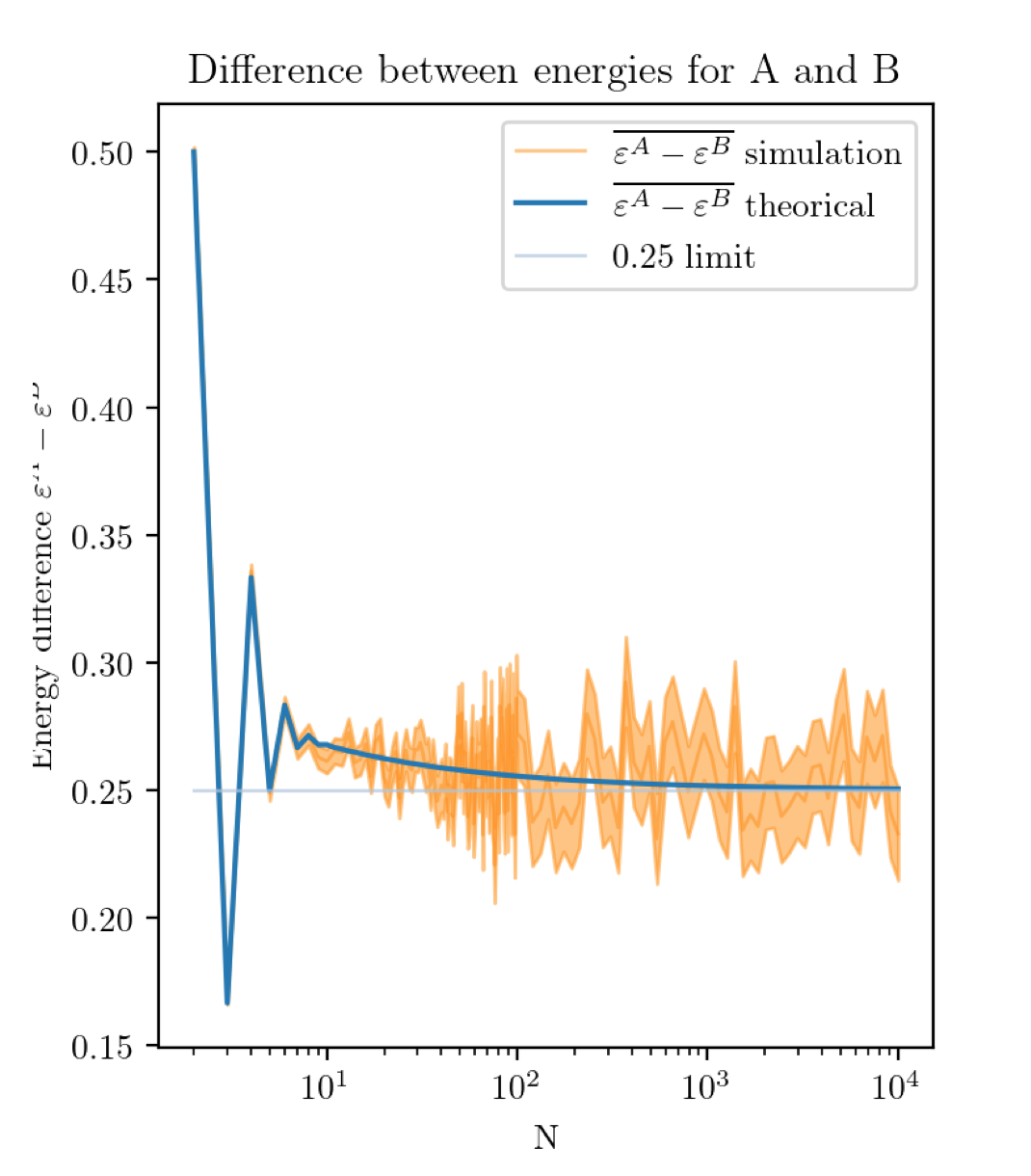}
\caption{\textbf{Difference between energies of A and B}: average energy difference $\overline{\epsilon^A-\epsilon^B}$. There are both the theoretical result and the simulation result.}
\end{center}
\end{figure}

The difference in energy between the starting player (A) and the opposite player (B) can be calculated exactly like before, i.e. on average we have

\begin{equation}
\overline{\epsilon^A-\epsilon^B}=\sum_{\epsilon^A=1}^{N} \sum_{\epsilon^B=1}^{N} P(\epsilon^A,\epsilon^B) (\epsilon^A-\epsilon^B) \;.
\end{equation}
Note that this equation differs from equation (93) as there is no modulus in the difference between the two energies. By adopting the same approximations we obtain

\begin{equation}
\overline{\epsilon^A-\epsilon^B}=\int_{0}^{\sqrt{N}} \sqrt{N}\left(1-e^{-x/\sqrt{N}} \right)e^{-x^2}x^2  \,dx=\frac{1}{4} \;.
\end{equation}

The result is interesting in that, for large $ N $, the difference in energy between the starting player and the second player is constant, and it corresponds to 0.25. \\
\\
So, what if $ B $ instead of $ A $ starts (with the same preference-lists)? As we will see, for large $ N $, the game would end in the same way regardless of who starts, i.e. the two players would agree on the same alternative whether $A$ starts or $ B $ starts. In some cases, however, it may happen that the result changes based on who makes the first proposal. Figure 27 shows an example.
\begin{figure}[!h]
\begin{center}
\includegraphics[width=0.9\textwidth,scale=0.9]{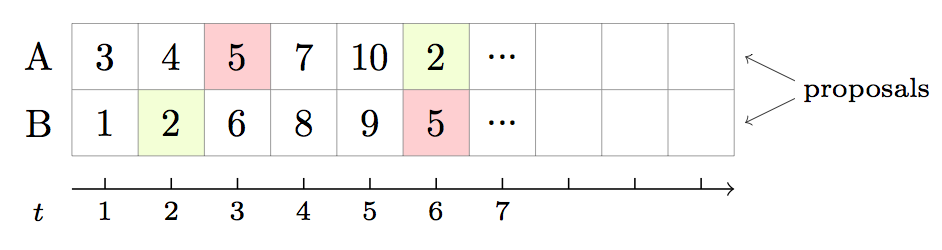}
\caption{\textbf{First mover disadvantage}: example of a negotiation process where $A$ starts (green box) and where $B$ starts (red box).}
\end{center}
\end{figure}

As one can see from the figure, if $ A $ is the first player to propose in the process, it ends with alternative $ o_2 $ and with energies $ \epsilon ^ A = 6 $ and $ \epsilon ^ B = 2 $; if $ B $ is the first player, the process ends with alternative $ o_5 $ and with energies $ \epsilon ^ A = 3 $ and $ \epsilon ^ B = 6 $. \\
When does the process end with different results? This happens when $ A $ proposes to $ B $ in $ t <t ^ * $ so that $ \epsilon_i ^ B = t ^ * $, where $ t ^ * $ is the time the process ends when $ A $ is the first player. In other words whenever the proposals of A and B can be accepted at the same ending time $t^*$. Otherwise, the $ B $ proposal in $ t ^ * $ would be rejected by $ A $, and $ B $ would accept the immediately subsequent $ A $ proposal and the process would end in the same way as when $ A $ makes the first move. \\
The probability $ P (t ^ *) $ that $ A $ \textit{does not} make a proposal such that $ \epsilon_i ^ B = t ^ * $ corresponds to the probability that the proposal of $ A $ in $ t ^ * $ is such that $ \epsilon ^ B <t ^ * $ (so that $ B $ accepts), multiplied by the probability that all proposals of $ A $ for $ t <t ^ * $ are such that $ \epsilon ^ B_i> t ^ * $, that is

\begin{equation}
P(t^*)=\frac{t^*}{N-t^*+1} \prod_{i=1}^{t^*-1}\left(1-\frac{t^*}{N-i+1} \right) \;.
\end{equation}
For large $N$ this probability goes as 
\begin{equation}
P(t^*) \sim \frac{t^*}{N} \exp \left(\frac{-(t^*)^2}{N} \right) \;.
\end{equation}
At this point, being (101) the probability that the result of the process is always the same, it is easy to show that it becomes more and more important for large $N$.\\
In conclusion, the probability that the solution depends on who starts the game is negligible for large $N$.\\
\\
Therefore, for small $ N $, the result may not be symmetrical, but this asymmetry is negligible in the thermodynamic limit.\\
One can also study the case of an "asymmetric" negotiation process in all respects. Consider the situation in which one of the two players proposes more alternatives at a time than the other player. The result that was found is that, called $ r $ the number of alternatives a player proposes at a time, his energy grows as $ \sim \sqrt {rN} $.

\subsubsection*{Second Negotiation Solution}
Now consider the case in which for some reason the first negotiated alternative is skipped. In the example of the friends who have to choose the place to go, this can happen because of an error or forgetfulness by one of the two. Once the first negotiation alternative has been skipped, the two players continue with the negotiation process until another agreement is found, i.e. a second negotiation alternative. \\
Again we want to calculate the average total energy $ \overline {\epsilon ^ {A, 2} + \epsilon ^ {B, 2}} $ and the average negotiation gap function $ \overline {\Delta_ {neg, 2} } = \overline {| \epsilon ^ {A, 2} - \epsilon ^ {B, 2} |} $. To do this we must first calculate the probability distribution of the second stopping time $ t ^ * _ 2 $. The reasoning is similar to what was done previously but in this case we must impose the constraint that the first stopping time is $ t ^ * _ 1 $. To satisfy this condition one just has to multiply by the term $ \frac {t ^ * _ 1} {N-t ^ * _ 1 + 1} $ which, for large $ N, $ becomes $ \frac {t ^ * _ 1} {N} $. So the probability distribution of $ t ^ * _ 2 $ given $ t ^ * _ 1 $ corresponds, for large $ N $, to
\begin{equation}
P(t^*_1,t^*_2) \sim \frac{t^*_1t^*_2}{N^2} e^{-(t^*_2)^2/N} \theta(t^*_2-t^*_1) \;,
\end{equation} 
where $ \theta (t ^ * _ 2-t ^ * _ 1) $ is the Heaviside function and is 1 for $ t ^ * _ 2> t ^ * _ 1 $ and 0 otherwise: it serves to guarantee that $ t ^ * _ 2 > t ^ * _ 1$.\\
We can evaluate $ \overline {t ^ * _ 2} $ in the following way. Without making explicit accounts, note that the time $ t ^ * _ 2 $ corresponds to the sum of two energies, in fact it is equal to $ t ^ * _1 + (t ^*_ 2-t ^ * _1) $. So the result is identical to what is done in equation (92): 
\begin{equation}
\overline{t^*_2}=\frac{3 \sqrt{\pi}\sqrt{N}}{4} \;.
\end{equation}
To evaluate the average total energy it is convenient to reason in this way: we know that $ t ^ * _ 2 = max (\epsilon ^ {A, 2}, \epsilon ^ {B, 2}) $, then the probability that $ \epsilon ^ {A, 2} $ is the maximum is $ P (\epsilon ^ {A, 2} = t ^ * _ 2) = 1/2 $; if instead $ \epsilon ^ {A, 2} $ is not the maximum, its expected value is $ \frac {t ^ * _ 2} {2} $, and therefore it holds $ P (\epsilon ^ {A, 2} = t ^ * _ 2/2) = 1/2 $. So on average we have
\begin{equation}
\overline{\epsilon^{A,2}}= \frac{1}{2}\overline{t^*_2}+\frac{1}{2}\frac{\overline{t^*_2}}{2}=\frac{3}{4}\overline{t^*_2}=\frac{9}{16}\sqrt{\pi N} \;. 
\end{equation} 
Since there is symmetry, on average, between the two players, looking at the results in (92), one obtains
\begin{equation}
\overline{\epsilon^{A,2}+\epsilon^{B,2}}=\frac{3}{2}\overline{\epsilon^A+\epsilon^B} \;,
\end{equation}
and this tells us that if you make a mistake and skip the first negotiation solution, the total energy increases by $ 50 \% $, and therefore also the negotiation time on average increases by the same amount. \\

\begin{figure}[!h]
\begin{center}
\includegraphics[width=0.5\textwidth,scale=0.5]{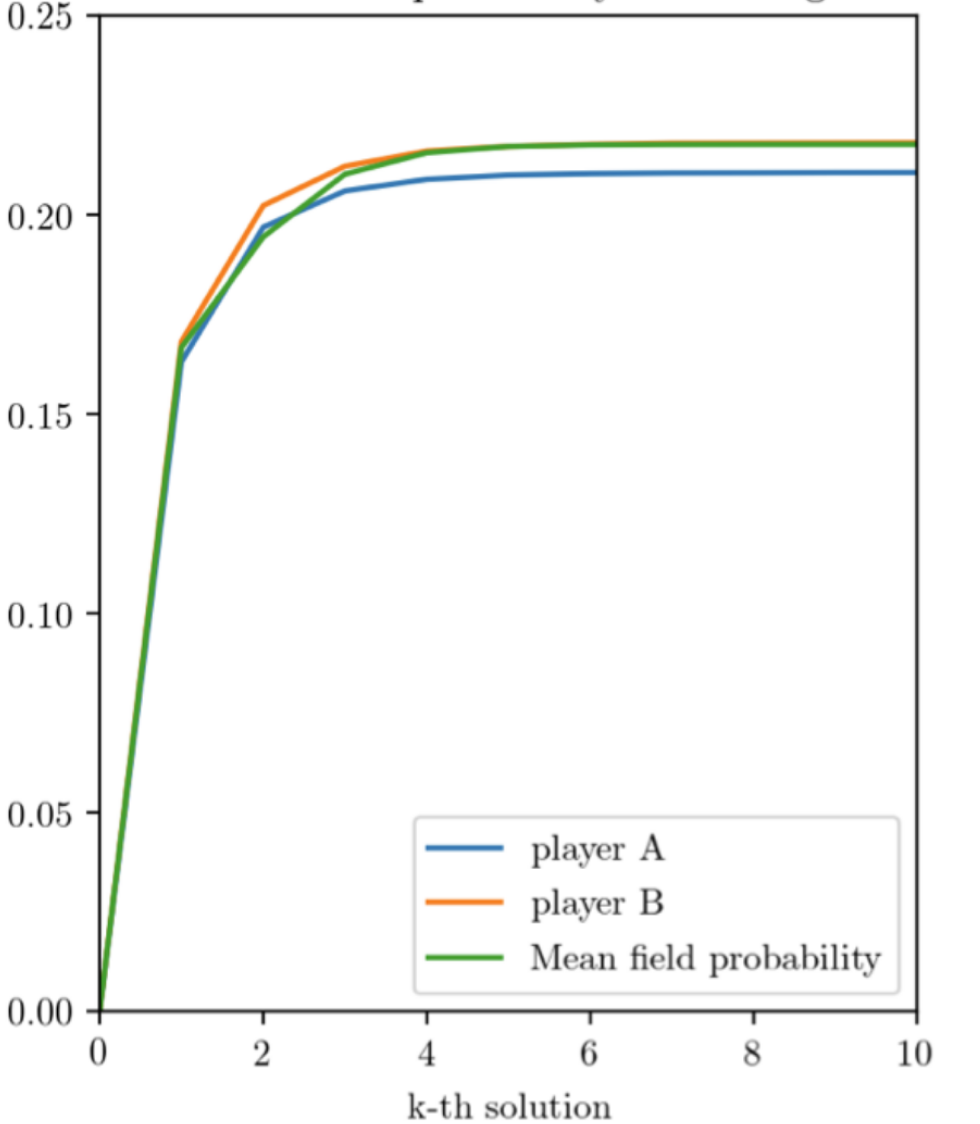}
\caption{\textbf{Cumulative probability of winning}: simulations of the probability of gaining a better deal than the negotiation solution in function of the number of refusals. Done over 100000 iterations, it is compared to the mean field approximation results.}
\end{center}
\end{figure}

To calculate now the second negotiation gap function just note that in general $ \overline {\Delta_ {neg}} = \overline {t ^ *}/2 $, in fact the gap function can assume all the values between 1 and $ t ^ * - 1 $ with the same probability. So we get the following relationship
\begin{equation}
\overline{\Delta_{neg,2}}=\frac{t^*_2}{2}=\frac{3}{2}\overline{\Delta_{neg}} \;.
\end{equation}

The above arguments can be extended to study the n-th negotiation solution. In other words, the average values of $ t ^ * _ 3, t ^ * _ 4, ..., t ^ * _ k $ can be calculated. We found the interesting relation
\begin{equation}
\overline{t^*_{k+1}}=\left(1+\frac{1}{2k} \right)\overline{t^*_k} \;.
\end{equation}
When $k$ becomes larger we have
\begin{equation}
\overline{t^*_k} \sim \overline{t^*} \sqrt{N} \sim N \;,
\end{equation}
hence, the condition that $ t_k $ cannot be greater than $ N $ implies that there is a maximum $ k = cN $ beyond which there are no more negotiation solutions.\\
\\
One can also study the probability that a player, say $ B $, will get a better alternative by rejecting the negotiation solution with $ \epsilon ^ B $ energy. To do this, one can calculate the fraction of times that $ \epsilon ^ B / \epsilon ^ {B, 2} <1 $. We used a mean-field approximation in which the random variables of the problem can be replaced with their average, finding $ P_1 = \epsilon ^ B / \epsilon ^ {B, 2} = t ^ * / 2t ^ * _ 2 = 1/3 $. If, on the other hand, player $ A $ purposely refuses the first negotiation solution, one has $ P_1 = \epsilon ^ A / \epsilon ^ {A, 2} = 1/6 $. \\
Now, if player B keeps tries to reject subsequent negotiation solutions on purpose, hoping to get better alternatives, we get the following iterative relationship

\begin{equation}
P_{k+1}=(1-P_k)\frac{t^*}{4t^*_k} \;,
\end{equation}
which goes to 0 quickly as $ k $ increases. Its cumulative probability has a finite limit, that is
\begin{equation}
P_{\infty}=\sum_{k=1}^{\infty}P_k \sim 0.217 \;,
\end{equation}
this means that $ B $'s strategy of denying negotiation solutions has a $ \sim 80.7 \% $ chance of failing. Figure 28 shows the results of the numerical simulations.

\subsubsection{Matchmaker Process Results}
Let us forget about the negotiation process and assume the existence of a matchmaker that forces the two players to choose the alternative that minimizes the total energy. In practice we want to find the alternative $ l $ that minimizes the sum $ s (l) = \min[s (i)] = \epsilon_l ^ A + \epsilon_l ^ B $. As usual we want to find the value of $ s $ averaged over all the possible configurations of the preference lists, i.e. $ \overline {s} $. \\
\begin{figure}[!h]
\begin{center}
\includegraphics[width=0.6\textwidth,scale=0.6]{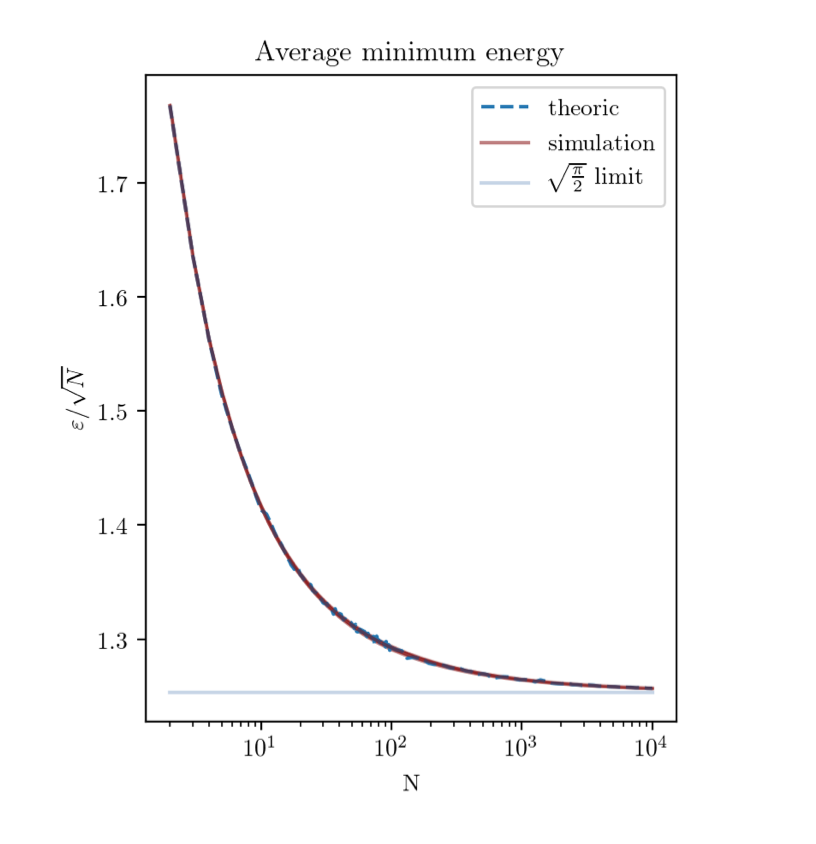}
\caption{\textbf{Average minimum total energy}: normalized average energy of the global minimum against the number of alternatives $N$. Simulations are done over 100000 samples and the theoretical result is found by numerical computation from the given probability distribution. The grey line shows the $\sqrt{\pi/2}$ limit.}
\end{center}
\end{figure}
To do this we reason in this way: if $ q $ is the minimum, we must then calculate the probability that the minimum is exactly $ q $ and then find the average value for all possible values that $ q $ can assume. To find the probability that $ q $ is the minimum we have to find the probability $ P (s (i = 1, ..., N) \ge q) $ that $ \epsilon ^ A_i + \epsilon ^ B_i \ge q $ for $ i = 1, ..., N $. Given the symmetry of the problem, we can sort the alternatives of one player and consider the possible permutations for the other player, that is, the problem becomes calculating the probability that $ i + \pi (i) \ge q $ for $ i = 1, .. ., $ N. \\
It holds that $ P (s (i) \ge q) = 1-P (s (i) \le q) $ and therefore we have that 
\begin{equation}
P(s(i=1,...,N) \ge q)= \prod_{i=1}^{q}\left(1-P(s(i) < q)\right) \;.
\end{equation}
Now, fixed $ i $, the probability that $ i + \pi (i) <q $ corresponds to $ P (i + \pi (i) <q) = \frac {qi-1}{N-i + 1} $ so we can rewrite (111) as (starting $ i $ from 0 instead of 1)
\begin{equation}
P(s(i=1,...,N) \ge q)=\prod_{i=0}^{q-1}\left(1-\frac{q-i-2}{N-i} \right) \;,
\end{equation}
if we consider the limit for very large $ N $ we have
\begin{equation}
P(s(i=1,...,N) \ge q) \sim \exp\left(\frac{-q^2}{2N} \right) \;,
\end{equation}
where the factor 2 in the denominator of the exponent appears because in this limit it holds $ \prod_ {i = 0} ^ {q-1} (q-i) \to q ^ 2/2 $. At this point, to find $ \overline {s} $ we just need to add up all the possible $ q $. Using the scaling $ q \to x \sqrt{N} $ and within the limit $ N \to \infty $ we obtain
\begin{equation}
\overline{s}=\sum_{q=2}^{N+1}P(s(i=1,...,N) \ge q) \sim \sqrt{N}\int_{0}^{\infty} e^{-x^2/2} \, dx=\sqrt{\frac{\pi}{2}}\sqrt{N} \;.
\end{equation} 
Note that the matchmaker improves the negotiation process solution by about $ 6 \% $. \\
\\
Now let us calculate the \textit{average global best gap function}, $ \overline{\Delta_ {glo}} $. In this case, the situation is a bit more complicated than the negotiation process. Given a minimum value of the total energy $ s (l) $ there are different possible values that the individual energies of the two players can take. So let us make the approximation that the minimum is unique and so $ \Delta_ {glo} $ is simply a random number between 1 and $ s (l) -1 $. We get
\begin{equation}
\overline{\Delta_{glo}}=\frac{\overline{s}}{2}=\sqrt{\frac{\pi}{8}}\sqrt{N} \;.
\end{equation} 
In this case, the gap function is larger than the one of negotiation. Since the total cost is lower with respect to the negotiation, then the inequality between $ A $ and $ B $ is greater.
\\
Now, in the same spirit of equation (109), we found a similar iterative equation for the n-th global minimum:
\begin{equation}
\overline{s_{k+1}}=\left(1+\frac{1}{2k} \right)\overline{s_k} \;.
\end{equation}

\begin{figure}[!h]
\begin{center}
\includegraphics[width=0.8\textwidth,scale=0.8]{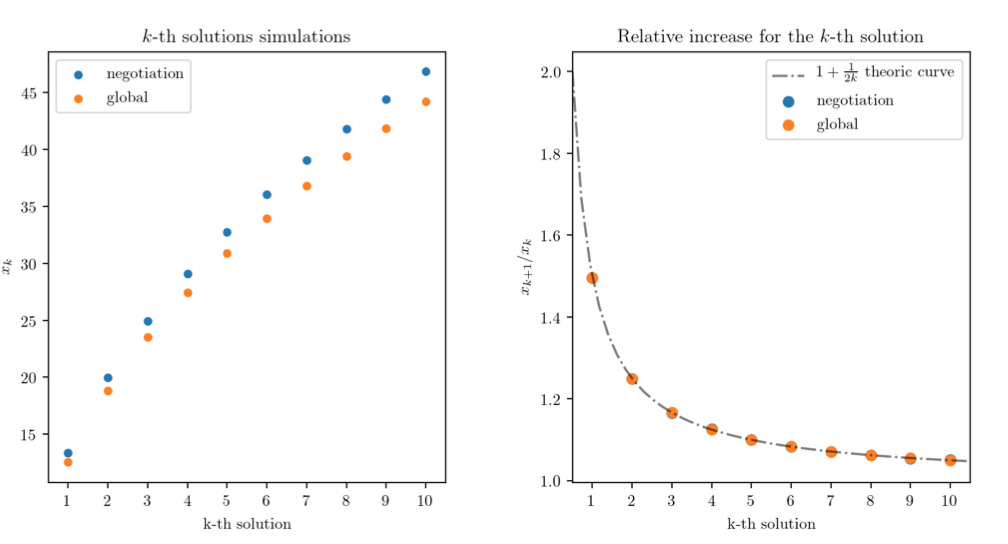}
\caption{\textbf{k-th solution}: average energies for the nested solutions for the negotiation and the global ones. On the left the average energies in function of the k-th solution; on the right the fractional increase for both systems. We can see it follows the law $1+1/2k$.}
\end{center}
\end{figure}

\subsubsection{Overlapping Probability}
Here we want to calculate what is the probability that through the negotiation process the two players obtain the ground state. In other words, we answer the question: how likely is the solution with minimum total energy through the negotiation process? \\
Like before, we put the alternatives of one of the two players in order and consider the possible permutations of the other player. In practice we want $ l + \pi (l) = \epsilon ^ A + \epsilon ^ B $, where $ l $ is the global best alternative and $ \epsilon ^ A $ and $ \epsilon ^ B $ are the two energies in the negotiation process. For the \textit{negotiation alternative} to be equal to the \textit {global best alternative} these three conditions must apply:
\begin{enumerate}
\item For $0<i<l$ we need that $i+\pi(i) \ge \epsilon^A+\epsilon^B$, otherwise $\epsilon^A+\epsilon^B$ would not be the minimum. If $\epsilon^A>\epsilon^B$ then $l=\epsilon^B$. 
\item If $\epsilon^A>\epsilon^B$, for $l\le i<\epsilon^A$ we need that $\pi(i) \ge \epsilon^A$, otherwise $\epsilon^A+\epsilon^B$ would be grater than $i+\pi(i)$.
\item If $\epsilon^A>\epsilon^B$, for $\epsilon^A \le i<\epsilon^A+\epsilon^B$ we need that $i+\pi(i) \ge \epsilon^A+\epsilon^B$, otherwise $\epsilon^A+\epsilon^B$ would not be the minimum.
\end{enumerate}
In the limit for large $ N $, we have that the probability for which the first condition holds is equal to
\begin{equation}
P_1=\prod_{i=1}^{\epsilon^B-1} \left(1-\frac{\epsilon^A+\epsilon^B-i-1}{N-i+1} \right) \sim \exp\left\lbrace \frac{-\epsilon^B(2\epsilon^A+\epsilon^B)}{2N}\right\rbrace \;.
\end{equation} 
The probability that the second condition holds is equal to
\begin{equation}
P_2= \frac{1}{N-\epsilon^B+1}\prod_{i=\epsilon^B+1}^{\epsilon^A-1} \left(1-\frac{\epsilon^A}{N-i+1} \right) \sim \frac{1}{N}\exp\left\lbrace \frac{-\epsilon^A(\epsilon^A-\epsilon^B)}{N}\right\rbrace \;,
\end{equation} 
where the first factor before the product ensures that $ \epsilon ^ B = l $ is accepted, that is, the alternative chosen is the one that produces the minimum total energy.\\
Finally, the probability that the third condition holds is equal to
\begin{equation}
P_3=\prod_{i=\epsilon^A}^{\epsilon^A+\epsilon^B-1} \left(1-\frac{\epsilon^A+\epsilon^B-i-1}{N-i-1} \right) \sim \exp\left\lbrace \frac{-(\epsilon^B)^2}{2N}\right\rbrace \;.
\end{equation}  
Multiplying $ P_1 $, $ P_2 $ and $ P_3 $ gives the probability $ P (z = l) $ that the \textit {negotiation item} $ z $ is equal to the \textit {global best alternative} $ l $. For large $ N $ one gets
\begin{equation}
P(z=l)=P_1P_2P_3 \sim \frac{1}{N}\exp \left\lbrace \frac{-\epsilon^B(2\epsilon^A+\epsilon^B)}{2N}+ \frac{-\epsilon^A(\epsilon^A-\epsilon^B)}{N}+ \frac{-(\epsilon^B)^2}{2N}  \right\rbrace \;.
\end{equation}
After a bit of algebra, the final result is 
\begin{equation}
P(z=l) \sim \frac{1}{N} \exp \left\lbrace -\frac{(\epsilon^A)^2+(\epsilon^B)^2}{N}  \right\rbrace \;.
\end{equation}
Finally, adding up all possible $ z $, i.e. adding up all the possible values of $ \epsilon ^ A $ and $ \epsilon ^ B $, we obtain the average value of the probability that $ z = l $. So passing to the continuum, in the limit $ N \to \infty $, and integrating we have
\begin{equation}
\sum_{\epsilon^A,\epsilon^B}P(z=l) \sim \frac{1}{N}\int_{0}^{\infty} e^{-(\epsilon^A)^2/2} \, d\epsilon^A \int_{0}^{\infty} e^{-(\epsilon^B)^2/2} \, d\epsilon^B = \frac{\pi}{4} \;.
\end{equation}
Hence, even without a matchmaker the probability that the two players reach the solution with minimum total energy with a negotiation process is quite high, indeed it is $ \pi / 4 \sim 78.5 \% $.

\subsubsection*{Matchmaker vs Negotiation Process}
As we have just seen, only in $ 22 \% $ of the cases the negotiation solution is different from the global minimum solution. In the ground state, the gap between the energies of the two players increases compared to that of the negotiation solution. In particular, the energy of one of the two players increases while the other decreases. In other words, in the global best solution, one of the two players has to sacrifice himself. \\
Let $ \epsilon ^ A $ and $ \epsilon ^ B $ be the energies of the players in the negotiation solution, while $ \epsilon ^ A_g $ and $ \epsilon ^ B_g $ the energies of the players in the global minimum solution. While during the negotiation process the players scroll down through the possible alternatives, their energies cannot be strictly greater or strictly less than their respective energies in the global minimum solution. They cannot be larger because the negotiation solution would have been equal to the global minimum solution; they cannot be smaller because the global minimum solution would not be the solution with minimum energy.\\ Therefore only two scenarios are possible:
$$\text{$\epsilon^A \ge \epsilon^A_g$ and $\epsilon^B \le \epsilon^B_g \;,$}$$
or
$$\text{$\epsilon^A \le \epsilon^A_g$ and $\epsilon^B \ge \epsilon^B_g \;.$}$$
So there is a player who, in the global minimum solution, earns more and one who earns less. But the gain of one compensates for the loss of the other. Indeed, assuming that $ B $ is the loser in the global minimum solution, we have
$$loss(B)=\epsilon^B_g - \epsilon^B \le gain(A)=\epsilon^A - \epsilon^A_g \;,$$
and rearranging we have:
\begin{equation}
\epsilon^A_g + \epsilon^B_g \le \epsilon^A + \epsilon^B \;,
\end{equation}
that is true by definition.

\subsubsection{Negotiation Problem with $m$ Players}
Everything we have studied above can be generalized to the case of $ m $ players. Very often in the real world, there are more than two agents who have to find agreement. A recent example is the case of the countries of the European Union during the Covid-19 pandemic, where different countries with different needs had to find an economic agreement to heal the crisis caused by the virus. It is therefore superfluous to underline the importance of expanding the classic NP to an NP with $ m $ agents. In the following, we will first study the solution of the negotiation process. It is analogous to the case of two agents but now there are $m$ players proposing each turn. Later we will study the global minimum solution in which we look for the solution that minimizes the sum of the energies of all $ m $ players. For simplicity, we will consider only the case of the thermodynamic limit (large $ N $).

\subsubsection*{$m$-Agents Negotiation Process}
In the process of negotiating with $ m $ players, one can repeat arguments similar to the previous one. So the goal is to calculate the probability that the process ends with energies $ \epsilon ^ 1, ..., \epsilon ^ m $, where $ \epsilon ^ i $ is the energy of player $ i $. We found that
\begin{equation}
P(\epsilon^1,...,\epsilon^m) \approx \frac{1}{N^{m-1}} \exp \left(-\frac{t^m}{N^{m-1}} \right) \;,
\end{equation}
where we have defined $ t = max (\epsilon ^ i) $, and it is the instant at which the negotiation process ends. As we said above, we have obtained equation (124) in the thermodynamic limit. If we want to do the exact calculation, the problem becomes very complex. To show this, we calculate the exact joint probability in the case of negotiation between 3 players: $A$, $B$, and $C$. The joint probability $P(\epsilon^A,\epsilon^B,\epsilon^C)$ is the probability of finding the negotiation solution with energies $\epsilon^A$, $\epsilon^B$ and $\epsilon^C$. We consider the case $t^*=\epsilon^A$, the other cases are similar. Hence the joint probability is equivalent to find the probability that both player $B$ and player $C$ accept the proposal of player $A$ after $t^*-1$ rejections. Let us denote by $p_1$ the number of times that $\epsilon_{i}^B<\epsilon^A$ and $\epsilon_{i}^C>\epsilon^A$, for all possible configurations of the preference-lists. Similarly, we denote by $p_2$ the number of times that $\epsilon_{i}^B>\epsilon^A$ and $\epsilon_{i}^C<\epsilon^A$ and, finally, we denote by $p_3$ the number of times that $\epsilon_{i}^B>\epsilon^A$ and $\epsilon_{i}^C>\epsilon^A$. Then we can write the joint probability as
\[
P(\epsilon^A,\epsilon^B,\epsilon^C)=\sum\limits_{p_1+p_2+p_3=\epsilon^A-2} \frac{\binom{\epsilon^A-1}{p_1}(p_1)!\binom{N-\epsilon^A+1}{p_2+p_3}(p_2+p_3)!}{\binom{N}{\epsilon^A-1}(\epsilon^A-1)!} \frac{\binom{\epsilon^A-1}{p_2}(p_2)!\binom{N-\epsilon^A+1}{p_1+p_3}(p_1+p_3)!}{\binom{N}{\epsilon^A-1}(\epsilon^A-1)!} \;.
\]

Returning to the general case of $m$ players, from equation (124) we found that the average energy per person is
\begin{equation}
\overline{\epsilon^i} \approx \frac{m+1}{2m} \Gamma \left(1+\frac{1}{m} \right)N^{1-1/m} \;,
\end{equation}
and as one can note, when $ m = 2 $, one gets the result of the classic NP. Moreover, when the number of agents increases, the average energy per players becomes larger: this is because when more and more people have to find an agreement it is more complicated to find an alternative that satisfies everyone. 

\subsubsection*{$m$-Agents Matchmaker Process}
We can calculate the global minimum solution with $ N $ alternatives and $ m $ players in the large $ N $ limit as follows. We first calculate the probability that all energies are greater than a certain threshold $ l $. Treating each energy independently we get
\begin{equation}
P(s>l)=(1-F(l))^N \;,
\end{equation}
where $s=min \left(\sum_{i=1}^{m}  \epsilon^r_i \right)$ and $i=1,...,N$. Furthermore, $ F (x) $ is the cumulative distribution of the sum of $ m $ independently and identically distributed variables (i.i.d).\\
When $ N $ becomes large, only values of $ F (l) $ close to zero will contribute significantly. Therefore for $ N $ i.i.d variables we can perform the approximation $ F (l) \sim l ^ m / N ^ mm! $. By using again the exponential approximation we have
\begin{equation}
P(s>l) \approx \exp \left(-\frac{l^m}{m!N^{m-1}} \right) \;.
\end{equation}
With calculations similar to the classic case we finally obtain that the average energy per player is
\begin{equation}
\frac{\overline{s}}{m} \approx \frac{(m!)^1/m}{m} \Gamma \left(1+\frac{1}{m} \right) N^{1-1/m} \;.
\end{equation}
Also in this case, if $ m = 2 $, the results of the classic NP are reproduced.

\subsubsection{Ground State with Replica Method}
As we did to calculate the SMP ground state using statistical mechanics arguments, also for the NP we want to use the replica method to calculate the ground state. The thermodynamics of the problem is equivalent to the \textit{random energy model} \cite{derrida1981random}. In any case in this paragraph, we will show the main highlights for obtaining the solution with minimum energy that we have already obtained before with probabilistic arguments.\\
As before, the fundamental quantity to calculate is the partition function, which in this case is written as:
\begin{equation}
Z_{\beta}=\sum_{i=1}^N e^{-\beta l_i} \;,
\end{equation}
where $ l_i $ is the sum of the energies of the two players for the item $ i $. For the moment we say that the $ l_i $ are random numbers taken from a $ \rho (l) $ distribution. We are interested to compute the minimum energy, that is obtain the zero temperature limit ($1/T=\beta \to \infty$) of the free energy $F$:
\begin{equation}  
\min_{i=1,..,N} (l_i)=\lim\limits_{\beta \rightarrow \infty} -\frac{1}{\beta}\ln (Z_{\beta}) =\lim\limits_{\beta \rightarrow \infty} F(\beta) \;.
 \end{equation}
In particular we are interested in the configurational average of the minimum, i.e. his average over the distribution $\rho(l)$. To do this we will use the replica method again, in particular we will use the following relation to make calculations feasible:
\begin{equation}  
F(\beta)=\lim_{n\rightarrow 0} - \frac{\overline{Z_\beta^{n}}-1}{\beta n}=\frac{\partial}{\partial n} \overline{Z_\beta^{n}}|_{n=0} \;.
\end{equation}

Solving the limit $ n \to 0 $, we obtain the following expression for the free energy:
\begin{equation}
F(\beta)=-\frac{1}{\beta}\left[\int_{-\infty}^{0} \frac{dz}{z}\, \left(1+\phi(z)\right)^N-\int_0^\infty \frac{dl}{l} e^{-l} \right] \;,
\end{equation}
where $\phi(z)=\sum_{p=1}^\infty \frac{g(p\beta)}{p!}z^{p}$ and $g(p\beta):=\mathbb{E}(e^{-\beta lp})$.\\

It is interesting to note that the procedure above leads to the integral form of the logarithm. In particular we have that:
\begin{equation} 
\mathbb{E}\left(\log(Z)\right)=\int_0^{\infty} \frac{dl}{l} \left[e^{-l}-
\mathbb{E}\left(e^{-l Z}\right)\right]=\int_0^{\infty} \frac{dl}{l} \left[e^{-l}-
\mathbb{E}\left(e^{-l e^{-\beta y}}\right)^N\right] \;. 
\end{equation}
Working from this expression we substitute $l=e^{w}$ and obtain:
\begin{equation}
\mathbb{E}\left(\log(Z)\right)=
\int_{-\infty}^{\infty} dw \left[e^{-e^{w}}-
\left(1-\int_0^\infty dy \rho(y)\left[1-e^{-e^{-(\beta y-l)}}\right]\right)^N\right] \;.
\end{equation}

Finally, passing to the limit $\beta \to \infty$ we obtain
\begin{equation} 
s_{min}=\lim_{\beta\rightarrow \infty}F_\beta=\int_0^\infty dq\left(1-\int_0^{q} dy\rho(y)\right)^N:= \int_0^\infty dq\left(1-F(q)\right)^N:=\int_0^\infty dq P(s>q) \;,
\end{equation}
where we used the definition of equation (126). Indeed it is the calculation of the minimum value from N realizations extracted with the distribution $\rho(x)$.\\
For $N$ large we can use the approximation that for $\rho(q)\sim\frac{q^{r-1}}{N^r {r-1}!}$ then $F(q)\sim \frac{q^{r}}{N^rr!}$. So we have
\begin{equation}
s_{min}=\int_0^\infty dq\left(1-\int_0^{q} dy\rho(y)\right)^N\sim 
\int_0^\infty dq\exp\left\{-N\int_0^{q} dy\rho(y)\right\}\sim \int_0^\infty dq\exp\left\{-\frac{q^{r}}{N^{r-1}r!} \right\} \;,
\end{equation}
which is exactly the same of equations (127) and (128).

\subsection{Seating Problem}
We now introduce another model inspired by the philosophy of the SMP: the \textit {seating problem}. As we have already mentioned in the introduction to this section, the problem is to position $ N $ individuals to sit in a round table, hence the name of the problem.\\
Each individual has a preference-list of all other individuals in the game. The cost function of a single individual is given by the sum of the rankings (in his preference-list) of the individual sitting on his right and the one sitting on his left. Once again, the most interesting questions concern the stability and optimization of energy (cost function). Stability means Nash equilibrium, i.e. in a stable state of the seating problem, there is no pair of players in which both want to swap places. In other words, no individual should agree to swap places with another, as this would lead to an increase in their energy and therefore in their unhappiness. Concerning optimization, however, one looks for the minimization of the total energy of the system, i.e. the sum of the energies of the individual players regardless of the stability of the system. \\
\\
The seating problem can be seen as a roommate problem (section 4) in which some rooms are "intersected" in a sequential manner, i.e. each individual belongs to two rooms at the same time and each room must have 2 tenants. Although the analogy with the roommate problem just made may seem unclear and too speculative, it helps us to understand how broad the field of applicability of this model can be: as we have seen throughout this review the SMP and its generalizations have been very successful in modelling most of the two-sided markets, the seating problem presents itself as an analogous model in which, however, couples are strongly interacting, and this is closer to a situation in the real world.\\
The research on the seating problem is still open. Despite its apparent simplicity, it is extremely complex both from a theoretical and a computational point of view. While we are writing, no formal stability results exist yet and the problem of finding the solution with the lowest total energy is shown to be NP-complete. Despite this, we found the value of the global minimum analytically, and this is what we will focus on in the rest of this section. However, we have first to formalize the problem from a mathematical point of view.

\subsubsection{The Model}
In the seating problem, we have to arrange $ N $ people in a round table taking into account their mutual preferences. Assume that each player assigns all the other players an energy between 0 and 1. In particular, we say that $ l_ {ij} $ is the energy of player $ i $ when he is sitting next to player $ j $: the lower the energy and the happier that individual is.\\
The total energy of a single player is given by $ l_ {ij} + l_ {ik} $, which is the sum of the player sitting on his right, $ j $, and the one sitting on his left, $ k $.\\
A system setup is defined by each player's placement, so we will say that $ P (i) $ is the player on the table's $ i $ site. If such a table is modeled by a 1-dimensional chain composed of $ N $ elements, then a player's energy can be written as
\begin{equation}
C_{P(i)}=l_{P(i)}l_{P(i+1)}+l_{P(i)}l_{P(i-1)} \;,
\end{equation}
where $i=0,1,...,N-1$ and with the periodic boundary condition $P(N)=P(0)$.

\subsubsection{SP Ground State}
We now want to find the solution that minimizes the total energy, that is, we want to find the configuration in which the energy is given by
\begin{equation}
E_{min}^{SP}= \min_P \left(\sum_{i=0}^{N-1} C_{P(i)} \right)=\min_P \left(\sum_{i=0}^{N-1} l_{P(i)}l_{P(i+1)}+l_{P(i)}l_{P(i-1)} \right) \;.
\end{equation}
It can be shown that finding such a solution is an NP-complete problem and the complexity of the system characterizes it as a frustrated system. For this reason, the statistical mechanics and the replica method, already used to find the ground states of the SMP and the NP, are useful again. Fortunately, the problem can be mapped to another problem, namely the \textit {Traveling Salesman Problem} (TSP). The TSP is one of the most famous combinatorial problems, and, as we mentioned in section 3.1, it also turns out to be NP-complete \cite{gharan2011randomized}. Despite this, there is an analytical solution for the ground state through the replica method and it is due to Mezard and Parisi \cite{mezard1986replica}. As we will see, there is a correspondence between the TSP and the Seating Problem (SP), therefore by exploiting the results in \cite{mezard1986replica} we can obtain the ground state for the SP.  

\subsubsection*{Mapping the SP in the TSP}
The TSP consists in finding the tour that passes through $ N $ city (only once) that minimizes the distance travelled. In other words, our salesman must find the minimum path that starts from $ a $ and returns to $ a $, passing through all $ N $ cities on his map only once.\\
Formally, we have $ N $ cities and each pair of cities is separated by a Euclidean distance $ d_ {ij} $. The "energy" of the system can be defined as the total distance traveled by the salesman passing through all the cities only once. If $ d_ {P (i) P (i + 1)} $ is defined as the distance between the city in site $ i $ and that in site $ i + 1 $, then the energy can be written as 
\begin{equation}
E^{TSP}=\sum_{i=0}^{N-1} d_{P(i)P(i+1)} \;.
\end{equation}
The ground state is the solution that minimizes this distance and it is
\begin{equation}
E_{min}^{TSP}= \min_P \left(\sum_{i=0}^{N-1} d_{P(i)P(i+1)} \right) \;.
\end{equation}
The SP can be seen as a non-Euclidean TSP \cite{saalweachter2008non} (in which there is no correlation between the distances of the cities) and asymmetrical \cite{laporte1987generalized}, i.e. in general $ d_ {ab} \ne d_ {ba} $. In any case, comparing equations (138) and (140), we note that with the change of variables
\begin{equation}
l_{P(i)}l_{P(i+1)}+l_{P(i)}l_{P(i-1)}=d_{P(i)P(i+1)} \;,
\end{equation}
there is a correspondence between the SP and the TSP. In other words, knowing the solution of the minimum of the TSP it is easy to obtain that of the SP.\\
\\
In the thermodynamic limit, one can analytically find the value of the minimum energy of the TSP, and therefore of the SP. As usual, the quantity we are interested in is the minimum averaged over a distance distribution $ \rho (l) $, that is
$$ E=\int \rho(l)E \,dl \;. $$
To calculate this quantity, we can approximate the probability distribution as $ \rho (l) = \frac {l ^ re ^ {- l}} {(r + 1)!} $ For the SP, and $ \rho (d) = \frac {d ^ {2r + 1} e ^ {- d}} {(2r + 2)!} $ for the TSP \cite{mezard1986replica}. Then one can use the replica method again. In particular, first the partition function of the system has to be calculated:
\begin{equation}
Z(\beta)=\sum_{P} \exp \left(-\beta \sum_{i=0}^{N-1}C_{P(i)} \right) \;,
\end{equation}
and then one has to calculate the configurational average of the logarithm of the partition function, i.e. $ \overline{log (Z)} $. Finally, the following relation is used to calculate the configurational average of the energy
$$\overline{E_{min}}=\lim_{\beta \to \infty} -\frac{1}{\beta}\overline{log(Z)} \;.$$
Following the arguments developed in \cite{mezard1986replica} three of us have shown that the SP ground state with the $ l $ uniformly distributed, that is with $ r = 0 $, is
\begin{equation}
\overline{E_{min}^{SP}(r=0)}=\overline{E_{min}^{TSP}(r=1)} \sim 1.817 \sqrt{N} \;.
\end{equation}
So, in this case, the same thing encountered with the SMP occurs: the case $ r = 1 $ corresponds to having a triangular probability distribution for the energies. Indeed, the energy of the SP mapped in that of the TSP is nothing more than the sum of two random variables uniformly distributed.\\
Note that, to carry out the calculations in the TSP, Mezard and parisi used the approximation that the distances between the cities are not independent: this is obviously not possible in a Euclidean space. In the SP, on the other hand, we performed the calculations, for the first time, without the need of any approximation, in fact, in the definition of the problem, the preference-lists of the players are not correlated and are independent with each other.

\section{Conclusions}
In this review, we have told about SMP in the various disciplines in which it has been studied, showing theoretical results, its applications and different models inspired by it.\\
We have focused more on the studies conducted by physicists, showing that this perspective adds interesting non-trivial results and, in our opinion, necessary for a deep understanding of SMP.\\
In fact, in addition to studying the stability of the system through the conventional techniques of game theory, we also emphasized the importance of studying the ground state. Finding the solution that minimizes the energy of the system, and therefore that maximizes the overall benefit, is a problem characterized by a high degree of complexity, which can be studied through the powerful tools of statistical mechanics and frustrated systems. Finding the ground state of the system is a problem that is only apparently disconnected from that of stability: knowing the globally best solution allows us to quantify in terms of energy the efficiency of stable solutions. In other words, the ground state constitutes a benchmark, or a comparison parameter, with all possible matching strategies. For example, in the third section, we showed that the energy gap between the stable optimal solution and the global optimal solution is significant, indeed the ground state is better in energy than $ 19 \% $. In section 7, instead, we have seen that this gap becomes larger when the agents of the system do not have complete information. Finally, in the last section we have seen that, in the case of a negotiation between two players, the stable solution and the ground state coincide $ 78 \% $ of the time.\\
The gap between stable solutions and the ground state is not only theoretical speculation but is of great importance for practical applications. As we saw in section 7, for example, it is crucial to recognize in which situations the role of the matchmaker becomes necessary. In the case of economic transactions on online platforms, where agents are irrational and have partial information, a matchmaker is essential if business and consumers want to obtain the maximum global benefit. In the case of a negotiation between two rational players, instead, the role of the matchmaker risks being an unnecessary additional cost for the society.\\
Studying the ground state and stable solutions of SMP, however, are not the only possibilities. As we have seen in sections 4, 5 and 6 there are many variations and applications of the problem still unexplored by physicists. With the approach of physicists, it is possible to study more deeply and quantitatively other mechanisms and dynamics of matching that would be interesting both from the theoretical point of view and for the possible applications. Indeed, the two-sided systems are innumerable in the real world and require a more systematic study.\\
We hope that in this review we have clearly shown the beauty of the Stable Marriage Problem and all its possible ramifications. The possibilities and the unknown lands are still countless in this area and we hope to have enticed the reader to explore them.

\section*{Acknowledgement}
The authors would like to thank Fei Jing, Matus Medo, Sergei Maslov, Guiyuan Shi, and Ruijie Wu who supported this work with relevant discussions, and Flaminia Fenoaltea who helped in the writing phase. This work was partially supported by the Swiss National Science Foundation (grant no. $200020\_182498/1$).
\newpage

\printbibliography
\end{document}